\documentclass[aps,prd,twocolumn,nofootinbib,preprintnumbers,floatfix]{revtex4-2}

\usepackage[utf8]{inputenc}
\usepackage[T1]{fontenc}
\usepackage{lmodern}
\usepackage{amsmath,amssymb,amsfonts,amsthm,mathtools}
\usepackage{bm,bbm}
\usepackage{mathrsfs}
\usepackage{braket}
\usepackage{physics}
\usepackage{graphicx}
\usepackage{booktabs,array}
\usepackage{xcolor}
\usepackage{tikz}
\usepackage{hyperref}

\usetikzlibrary{arrows.meta,decorations.markings}
\usetikzlibrary{decorations.pathmorphing}

\hypersetup{
  colorlinks=true,
  linkcolor=blue,
  citecolor=blue,
  urlcolor=blue
}

\newcommand{\cO}{\mathcal{O}}

\newcommand{\ws}{\mathrm{ws}}
\newcommand{\bdry}{\mathrm{bdry}}

\begin{document}
\title{Gluon GTMD at strong coupling:\\
fixed-spin saddle factorization and Reggeization}

\author{Kiminad A.~Mamo}
\affiliation{Department of Physics, University of Connecticut, Storrs, CT 06269-3046, USA}
\email{ska25005@uconn.edu}

\author{Ismail Zahed}
\email{ismail.zahed@stonybrook.edu}
\affiliation{Center for Nuclear Theory, Department of Physics and Astronomy,
Stony Brook University, Stony Brook, New York 11794-3800, USA}

\date{\today}

\begin{abstract}
Generalized transverse-momentum-dependent parton distributions (GTMDs) are the most complete two-parton correlation functions in QCD, encoding the joint spatial and momentum structure of hadrons. Through appropriate projections and limits they yield generalized parton distributions (GPDs), transverse-momentum-dependent distributions (TMDs), parton distribution functions (PDFs), and phase-space (Wigner) distributions. We construct conformal moments of unpolarized gluon GTMDs at strong coupling using gauge/string duality. For fixed even conformal spin $j$, we distinguish the local boundary limit at $b_T=0$ from the finite-separation regime $b_T>0$, where the planar semiclassical amplitude is governed by a minimal worldsheet. There the GTMD moment factorizes into a universal staple-worldsheet soft factor and a stripped spin-$j$ Witten amplitude carrying target dependence. The cusp of the renormalized minimal area generates the rapidity-logarithmic Collins-Soper structure. We derive universal ultraviolet and infrared endpoint reductions. As $b_T\to0^+$, the finite-separation sector matches onto the local conformal moment through a universal overlap kernel. At large $b_T$, after cusp/perimeter subtraction, it factorizes into target projections and infrared transfer kernels. The ultraviolet endpoint is universal within the leading saddle, whereas the infrared tail depends on the holographic completion: soft-wall, gap-matched hard-wall, and repulsive-wall backgrounds generate algebraic, exponential, and Gaussian falloffs, respectively. Analytic continuation in $j$ yields the low-$x$ Regge regime governed by the holographic Pomeron spectral curve. The framework describes hadron tomography, transverse structure, rapidity evolution, and Reggeization for GTMD moments and provides a unified starting point for holographic studies of observables relevant to the Electron-Ion Collider.
\end{abstract}

\maketitle

\section{Introduction}

GTMDs occupy the apex of the hierarchy of QCD parton correlation
functions. Through suitable projections and kinematic limits they
generate GPDs, TMDs, PDFs, and Wigner distributions, thereby encoding
the combined momentum-space and spatial tomography of hadrons, that
forms a central component of the physics program of the Electron-Ion
Collider.

Understanding the multidimensional structure of hadrons in terms of
partonic degrees of freedom is a central goal of contemporary QCD.  Collinear
parton distribution functions encode longitudinal momentum structure, while
transverse-momentum-dependent distributions (TMDs) resolve intrinsic
transverse momentum \cite{Collins:1981uk,Collins:1984kg,Ji:2004wu,Collins:2011zzd,
Bacchetta:2006tn}, and generalized parton distributions (GPDs) encode
longitudinal momentum and transverse spatial correlations through momentum
transfer \cite{Ji:1996ek,Radyushkin:1996ru,Diehl:2003ny,Belitsky:2005qn}.
Generalized transverse-momentum-dependent distributions (GTMDs) unify these
limits into a single phase-space description of partons in hadrons
\cite{Ji:2003ak,Belitsky:2003nz,Meissner:2007rx,Meissner:2009ww,
Lorce:2011dv,Lorce:2011kd,Lorce:2013pza,Kanazawa:2014nha}.

At the operator level, GTMDs are gauge-invariant nonlocal correlators of
field-strength operators connected by staple-shaped Wilson lines.  The
staples introduce rapidity divergences, soft factors, and Collins-Soper
evolution beyond ordinary ultraviolet renormalization
\cite{Collins:2011zzd,Echevarria:2012js,Echevarria:2016mrc}.  At weak
coupling these structures are organized through perturbative factorization
and matching, where GTMDs reduce in appropriate limits to TMDs, GPDs, and
collinear PDFs \cite{Bertone:2022nxa,Bertone:2025gtmd}.

At strong coupling, gauge/string duality provides a complementary
nonperturbative description: boundary correlators are represented by bulk
Witten diagrams \cite{Maldacena:1997re,Gubser:1998bc,Witten:1998qj,
Freedman:1998tz,DHoker:1999jke}, while Wilson lines are represented by
classical string worldsheets \cite{Maldacena:1998im,Rey:1998ik,
Drukker:1999zq}.  The same framework geometrizes confinement and Wilson-loop
area laws \cite{Brandhuber:1998bs,Kinar:1998vq} and relates cusped Wilson
lines to anomalous dimensions and rapidity evolution
\cite{Korchemsky:1987wg,Kruczenski:2002fb,Kruczenski:2007cy,
Makeenko:2006ds}.  The broader GTMD and Wigner-distribution literature
includes light-front quark models, spectator and bag-model calculations,
gluon GTMD studies, and small-$x$/CGC analyses
\cite{Lorce:2011dv,Lorce:2011kd,Kanazawa:2014nha,Rajan:2016tlg,
Maynard:2026gtmd,Tan:2024zjs,Chakrabarti:2025gtmd,
Hagiwara:2016kam,Zhou:2016rnt,Bhattacharya:2018lgm,
Bhattacharya:2024pi0,Benic:2026gtmd,Bhattacharya:2026qnd}.  A fixed-spin
strong-coupling formulation of gluon GTMDs has remained largely unexplored.

The purpose of this work is to formulate gluon GTMD conformal moments at
strong coupling and to identify the geometric structures controlling their
transverse behavior and high-energy evolution.  We organize the calculation at
fixed even conformal spin~$j$, corresponding to twist-two gluon operators.  At
fixed spin, the holographic exchange is a single bulk spin-$j$ field, which
allows the Wilson-line geometry to be separated from the hadronic radial
structure.

The leading-saddle factorization is piecewise in the strict local and finite-separation sectors
\begin{equation}
\begin{aligned}
F_j^g(\xi,t,0;\mu,\zeta)
&=
\widetilde F_j^{g,{\rm bdry}}(\xi,t;\mu),
\\[3pt]
F_j^g(\xi,t,b_T;\mu,\zeta)
&=
S(b_T;\mu,\zeta)\,
\widetilde F_j^{g,{\rm ws}}(\xi,t,b_T;\mu)
\\[-1pt]
&\quad+
\mathcal O(N_c^{-2},\lambda^{-1/2}),
\qquad b_T>0 .
\label{eq:intro_piecewise_factorization}
\end{aligned}
\end{equation}
Here $S$ is the universal vacuum worldsheet soft factor determined by the
staple geometry, while $\widetilde F_j^{g,{\rm ws}}$ is the stripped bulk
exchange amplitude containing the target dependence.  The notation
\begin{equation}
F_j^g
=
F_{j,\bdry}^g+F_{j,\ws}^g+\cdots
\label{eq:intro_saddle_decomposition}
\end{equation}
is only a saddle-sector bookkeeping convention: the boundary term is the
strict local/contact Witten diagram at $b_T=0$, and the worldsheet term is the
finite-separation saddle for $b_T>0$.  The ellipsis denotes string-loop,
multi-saddle, and fluctuation corrections.

The stripped finite-separation amplitude has a fixed-spin Witten-diagram
representation in confining holographic backgrounds.  Its ultraviolet endpoint
is universal within the saddle construction and
%\begin{equation}
\(b_T\rightarrow0^+,\)
%\end{equation}
\begin{equation}
\widetilde F_j^{g,{\rm ws}}(\xi,t,b_T)
\longrightarrow
\mathcal K_j^{\rm UV}(b_T)\,
\widetilde F_j^{g,{\rm bdry}}(\xi,t),
\label{eq:intro_uv_reduction}
\end{equation}
with $\mathcal K_j^{\rm UV}$ a worldsheet overlap kernel.  For
$\mathrm{Re}[4+\gamma_c(j)]>0$,
\begin{align}
\lim_{b_T\to0^+}\widetilde F_j^{g,{\rm ws}}(\xi,t,b_T)&=0,
\nonumber\\
\widetilde F_j^{g,{\rm full}}(\xi,t,0)&=\widetilde F_j^{g,{\rm bdry}}(\xi,t).
\label{eq:intro_bzero_distinction}
\end{align}
Thus the endpoint of the finite-separation worldsheet saddle is not the
perturbative small-$b_T$ OPE of a fully subtracted GTMD.  The strict local
matrix element is represented by the separate boundary Witten diagram; the
comparison with perturbative matching is given in Sec.~\ref{SEC_VII}.

In the infrared endpoint 
\(b_T\rightarrow\infty,\)
and after cusp/perimeter subtraction and assuming a finite-depth central saddle,
the amplitude reduces to
\begin{equation}
\widetilde F_j^{g,{\rm ws}}(\xi,t,b_T)
\longrightarrow
\mathcal K_j^{\rm IR}(b_T)\,
\widehat{\mathcal T}_j(\xi).
\label{eq:intro_ir_reduction}
\end{equation}
The soft-wall transfer gives an algebraic tail, while gap-matched hard-wall
and repulsive-wall completions give exponential and Gaussian tails,
respectively.  These large-$b_T$ tails are infrared transfer-kernel
completions; the universal statement is the radial endpoint logic and, in
particular, the ultraviolet overlap kernel.

The low-$x$ regime is obtained by analytic continuation of the fixed-spin
moments into the complex $j$-plane, as in the BPST Pomeron and conformal
Regge theory \cite{Brower:2006ea,Brower:2010wf,Costa:2012cb,Brower:2013}.
The rightmost strong-coupling singularity is
\begin{equation}
j_0^g
=
2-\frac{2}{\sqrt{\lambda}},
\end{equation}
and the resulting GTMD exhibits the characteristic BPST diffusion in rapidity
and logarithmic transverse separation.  Reggeization introduces no additional
nonperturbative input; it analytically continues the same fixed-spin
amplitudes that determine the finite-$b_T$ moments.

The practical output is a conformal-moment-space recipe for future GTMD
parametrizations.  One may use holographic string parametrizations of GPD
conformal moments as input \cite{Mamo:2024vjh,Mamo:2024jwp}, dress each
moment with the finite-separation kernels derived here, and perform the
inverse Mellin-Barnes transform in complex spin to return to $x$-space
\cite{Mueller:2005ed}.  This supplies the transverse dressing and
rapidity/Regge organization needed to promote existing string-GPD fits into
candidate holographic-string GTMD parametrizations for processes such as
exclusive double quarkonium, exclusive $\pi^0$, and exclusive heavy vector or
axial-vector meson production \cite{Bhattacharya:2018lgm,Bhattacharya:2024pi0,
Bhattacharya:2026qnd}.

This paper is organized as follows.  In Sec.~\ref{SEC_II} we define
gauge-invariant gluon GTMDs and their fixed-spin conformal moments.  In
Sec.~\ref{SEC_III} we derive the strong-coupling soft-factor separation.  In
Sec.~\ref{SEC_IV} we develop the fixed-spin holographic representation.  In
Sec.~\ref{SEC_V} we derive the transverse endpoint reductions.  In
Sec.~\ref{SEC_VI} we Reggeize the fixed-spin moments.  In Sec.~\ref{SEC_VII}
we compare with perturbative GTMD factorization, and in Sec.~\ref{SEC_VIII}
we summarize the resulting framework.

\section{Gauge-invariant Definition of Gluon GTMDs}
\label{SEC_II}

We use a notation close to the modern GTMD definitions in
Refs.~\cite{Meissner:2009ww,Lorce:2013pza,Echevarria:2016mrc,Bertone:2022nxa,Bertone:2025gtmd}.  In particular, the kinematics will be set by
\begin{equation}
\begin{gathered}
\bar P=\frac{P+P'}{2},
\qquad
\Delta=P'-P,
\qquad
t=\Delta^2,\\
\xi=-\frac{\Delta^+}{2\bar P^+}.
\end{gathered}
\label{eq:secII_kinematics}
\end{equation}
For a staple direction \(n\) and transverse displacement \(\bm z_\perp\),
the unsubtracted  gluon GTMD correlator is
\begin{widetext}
\begin{align}
W_g^{ij}(x,\xi,\bm k_\perp,\bm\Delta_\perp;n)
&=
\frac{1}{x\bar P^+}
\int\frac{dz^-\,d^2\bm z_\perp}{(2\pi)^3}
\exp\!\left[ix\bar P^+z^- -i\bm k_\perp\!\cdot\!\bm z_\perp\right]
\nonumber\\
&\quad\times
\left.
\matrixel{P'}{\mathrm{Tr}\!\big[
F^{+i}(-z/2)\,U_{\rm st}[-z/2,z/2]\,
F^{+j}(z/2)\,U_{\rm st}[z/2,-z/2]
\big]}{P}
\right|_{z^+=0} .
\label{eq:secII_standard_gtmd_correlator}
\end{align}
\end{widetext}
where $x\bar P^+$ refers to the longitudinal momentum of parton-x. 
All reference to the resolution $\mu$ will be assumed.
The leading-twist unpolarized scalar projection is obtained by contracting
the transverse indices.  The object used below is its mixed
transverse-coordinate representation,
\begin{equation}
F_g(x,\xi,t,\bm b_T)
\equiv
\int\frac{d^2\bm k_\perp}{(2\pi)^2}\,
\exp\!\big(i\bm k_\perp\!\cdot\!\bm b_T\big)
\,x\bar P^+\,\delta_{ij}W_g^{ij},
\label{eq:secII_bspace_gtmd}
\end{equation}
using the transverse Fourier convention stated in Appendix~A.  Equivalently, it can be written as the
matrix element of the bilocal operator
\begin{equation}
\begin{aligned}
\mathcal O_g(y)
&=
\mathrm{Tr}\,
F^{+i}(0)\,U_{\rm st}(0,y)\,
F^+_{\ i}(y)\,U_{\rm st}(y,0),
\end{aligned}
\label{eq:secII_bilocal_operator}
\end{equation}
with \(y=(0,y^-,\bm b_T),\) and the standard understanding that the staple is regulated away from
the light cone before Collins-Soper evolution is taken.  Thus
\(\bm b_T\) in this paper is the transverse separation conjugate to
the average partonic transverse momentum \(\bm k_\perp\), while \(t\) or
\(\bm\Delta_\perp\) describes the off-forward momentum transfer.

The natural gauge link for the gluon operator is in the adjoint
representation.  In Eq.~\eqref{eq:secII_standard_gtmd_correlator} we write
the color trace in fundamental matrix notation,
\(F^{\mu\nu}=F^{a\,\mu\nu}T^a\), with
\(\mathrm{Tr}(T^aT^b)=\delta^{ab}/2\), because this is the most compact
notation for the field-strength bilocal.  The actual Wilson line may be
written as
\begin{equation}
U_{\rm adj}^{ab}[C]
=
2\,\mathrm{Tr}\!\left[T^aU_F[C]T^bU_F^\dagger[C]\right],
\label{eq:secII_adjoint_largeN_identity}
\end{equation}
\[\mathrm{Tr}_{\rm adj}U_{\rm adj}
=
|\mathrm{Tr}_F U_F|^2-1,\]
so that at large \(N_c\) an adjoint eikonal source may be represented by the
planar two-fundamental-string, or Casimir-rescaled, saddle.  In the present
leading-saddle treatment this is a normalization convention. The effective
representation dependence is absorbed into the worldsheet tension, the
soft-factor normalization, and the fixed-spin coupling \(g_j(\lambda)\).  No
finite-\(N_c\) representation independence or separate adjoint-string dynamics
is assumed.  Screening, recombination, and multi-string effects belong to the
omitted \(1/N_c^2\), multi-saddle, and worldsheet-fluctuation sectors.

\begin{table*}[t]
\centering
\caption{Notation for the fixed-spin saddle sectors and endpoint kernels.}
\label{tab:notation_fixed_spin}
\small
\begin{tabular}{ll}
\toprule
Symbol & Meaning \\
\midrule
\(F_j^g\) & full fixed-spin GTMD moment, organized as \(F_{j,\bdry}^g+F_{j,\ws}^g+\cdots\) \\
\(F_{j,\bdry}^g\), \(\widetilde F_j^{g,{\rm bdry}}\) & strict \(b_T=0\) boundary/GPD conformal moment \\
\(F_{j,\ws}^g\) & finite-separation worldsheet sector for \(b_T>0\), including the soft factor \\
\(\widetilde F_j^{g,{\rm ws}}\) & stripped finite-separation worldsheet Witten diagram \\
\(\widehat{\mathcal T}^{(c)}_j\) & target projection in the infrared endpoint ordering \\
\(\mathcal K_j^{\rm UV}\) & universal small-\(b_T\) worldsheet overlap kernel \\
\(\mathcal K_j^{\rm IR,SW/HW/RW}\) & model-dependent cusp-subtracted infrared transfer kernels \\
\(\Delta_c(j),\nu_j,\alpha_j\) & \(\nu_j=\Delta_c(j)-2\), \(\alpha_j=\Delta_c(j)-(j-2)\) \\
\(A_{j,t},B_j\) & soft-wall Tricomi indices in the transfer kernel \\
\bottomrule
\end{tabular}
\end{table*}

The partonic support region \(-1\le x\le1\) uses the standard C-even gluon
extension to negative \(x\); the negative-\(x\) part is fixed by crossing of
the gluon operator rather than by an independent antiquark-like distribution,
and only even conformal spins contribute in the channel used here.
The conventional GTMD correlator contains the prefactor
\(1/(x\bar P^+)\) in Eq.~\eqref{eq:secII_standard_gtmd_correlator}.  We
absorb this conventional factor into the scalar quantity
\(F_g\) in Eq.~\eqref{eq:secII_bspace_gtmd}.  At finite skewness, the
fixed-spin object diagonal in conformal partial waves is not a plain
Mellin moment but the gluon conformal, or Gegenbauer, moment.  Following
the gluon Gegenbauer normalization used in
Ref.~\cite{Mamo:2024vjh}, the C-even spin-\(j\) moment in the present
normalization is
\begin{align}
\mathbb F_j^g(\xi,t,b_T)
&=
\mathcal N_j^g
\int_{-1}^{1}dx\,
\xi^{\,j-2}
C_{j-2}^{5/2}\!\left(\frac{x}{\xi}\right)
\nonumber\\
&\quad\times
F_g(x,\xi,t,b_T),
\qquad
j=2,4,\ldots .
\label{eq:secII_conformal_moment}
\end{align}
where
\begin{equation}
\mathcal N_j^g
=
\frac{\Gamma(5/2)\,\Gamma(j-1)}
{2^{j-2}\Gamma(j+1/2)} .
\label{eq:secII_gluon_conformal_norm}
\end{equation}
The normalization follows from the Gegenbauer orthogonality relation
\begin{equation}
\int_{-1}^{1}dy\,(1-y^2)^2
C_n^{5/2}(y)C_m^{5/2}(y)
=
 h_n^{(5/2)}\delta_{nm},
\label{eq:secII_gegenbauer_orthogonality}
\end{equation}
with
\begin{equation}
h_n^{(5/2)}
=
\frac{\pi\,2^{-4}\Gamma(n+5)}
{n!\,(n+5/2)\Gamma(5/2)^2},
\qquad n=j-2 .
\label{eq:secII_gegenbauer_norm_hn}
\end{equation}
Equivalently, the normalized kernel obeys
\begin{equation}
\lim_{\xi\to0}
\mathcal N_j^g\,\xi^{\,j-2}
C_{j-2}^{5/2}\!\left(\frac{x}{\xi}\right)
=
x^{j-2},
\label{eq:secII_forward_kernel_limit}
\end{equation}
so that the conformal moment reduces exactly to the Mellin moment
\begin{align}
F_j^{g,{\rm ws}}(0,t,b_T)
&=
\int_{-1}^{1}dx\,x^{j-2}F_g(x,0,t,b_T),
\nonumber\\
j&=2,4,\ldots .
\label{eq:secII_mellin_moment}
\end{align}
Below we use \(F_j^g\) for the fixed-spin conformal moment and write the
Mellin form explicitly when the \(\xi\to0\) limit is taken.

The power \(x^{j-2}\) in Eq.~\eqref{eq:secII_mellin_moment} is fixed by
the local twist-two gluon operator selected by the longitudinal operator
product expansion,
\begin{align}
\mathcal O^{(g)}_{+\cdots+;j}(0)
&\sim
\mathrm{Tr}\!\big[
F^{+i}(0)(iD^+)^{j-2}F^+_{\ i}(0)
\big]
\nonumber\\
&\quad+
\text{trace subtractions and mixing terms}.
\label{eq:secII_twist_two_operator}
\end{align}
Equivalently, in the normalization of
Eq.~\eqref{eq:secII_standard_gtmd_correlator}, the same local operator
is selected by an \(x^{j-1}\) moment of the conventional scalar GTMD.
Only even \(j\) contribute in the C-even gluon channel considered here.
The convention is the same as in the string-based all-skewness GPD
parametrization of Refs.~\cite{Mueller:2005ed,Mamo:2024vjh,Mamo:2024jwp}.

In the large-\(N_c\), strong-coupling holographic limit used below, the
exchanged closed spin-\(j\) field should be understood as the singlet
gluonic/Pomeron-channel representative of the twist-two sector.  Pure-gluon,
quark-singlet, and finite-coupling operator-basis effects are not resolved
as separate leading-saddle exchanges.  They are encoded, to the extent that
they are retained in the model, in the effective spectral curve, the radial
normalization, and the coupling \(g_j(\lambda)\).

At fixed \(j\), the longitudinal nonlocality collapses into the local
operator \eqref{eq:secII_twist_two_operator}.  The remaining nonlocality
is transverse and is carried by the staple Wilson line and the separation
\(b_T\), which become the classical worldsheet data in the
holographic description.

\section{Strong-coupling Factorization of Fixed-spin GTMDs}
\label{SEC_III}

We now derive the factorization of the fixed-spin gluon GTMD conformal moment directly from its dual string representation, using the standard AdS/CFT relation between boundary correlators and bulk Witten diagrams \cite{Maldacena:1997re,Gubser:1998bc,Witten:1998qj,Freedman:1998tz,DHoker:1999jke}.
The key point is to distinguish  between the external hadronic states and the bilocal gluon operator.
In this section and for simplicity, the incoming and outgoing hadrons are represented in the bulk by closed-string source operators, while the gauge-invariant bilocal gluon operator is represented by a nonlocal defect tied to the staple-shaped Wilson-line contour on the boundary, and coupled to the bulk field dual to the twist-two gluon operator \cite{Berenstein:1998ij,Miwa:2006vv,Alday:2011ga,Alday:2011pf,Buchbinder:2012vr}, see also \cite{Gimenez-Grau:2023fcy,carmi2026aspectswittendiagramsholographic} for recent discussions of AdS Witten diagrams with defect.

For notational simplicity we represent the external hadrons by effective scalar closed-string vertices.  This is only a Witten-diagram bookkeeping device: the scalar vertex carries the same normalizable bulk wave function as the physical hadron mode and no claim is made that the hadron is a bosonic-string tachyon.  Thus the incoming and outgoing hadron states are created by
\begin{eqnarray}
V_T(P)
&=&
\int d^2\sigma\,
\sqrt{g}\,
\Phi_P\!\left(X(\sigma)\right),
\nonumber\\
V_T(P')
&=&
\int d^2\sigma\,
\sqrt{g}\,
\Phi_{P'}\!\left(X(\sigma)\right),
\end{eqnarray}
where $\Phi_P$ and $\Phi_{P'}$ are the bulk wavefunctions valued on the worldsheet, dual to the external hadrons on the worldsheet.

The bilocal gluon operator with staple Wilson line is treated differently.
At fixed even spin $j$, its conformal moment is represented by coupling the classical worldsheet ending on the staple contour $C$ to the bulk spin-$j$ field $H^{(j)}_{M_1\cdots M_j}$ dual to the twist-two gluon operator.
Schematically, the corresponding insertion is
\begin{align}
V^{(j)}_{\rm bilocal}[C]
&=
\int_{\Sigma:\,\partial\Sigma=C} d^2\sigma\,
\sqrt{g}\,
\mathcal{J}^{M_1\cdots M_j}[X(\sigma);C]
\nonumber\\
&\quad\times
H^{(j)}_{M_1\cdots M_j}(X(\sigma)).
\end{align}
where $\mathcal{J}^{M_1\cdots M_j}$ is the worldsheet current induced by the bilocal gluon operator and by the staple geometry.
All dependence on the transverse separation $b_T$ and on the rapidity geometry is carried by this insertion.

The fixed-spin GTMD conformal moment is therefore represented by the string amplitude
\begin{align}
F^g_j(\xi,t,b_T;\mu,\zeta)
&\sim
\int_{\partial\Sigma=C}\mathcal{D}X\,
e^{-S_{\rm NG}[X]}
\nonumber\\
&\quad\times
V_T(P')\,
V^{(j)}_{\rm bilocal}[C]\,
V_T(P).
\label{eq:secIII_string_amplitude}
\end{align}
Here $S_{\rm NG}$ is the Euclidean Nambu-Goto action for the holographic Wilson-line worldsheet \cite{Maldacena:1998im,Rey:1998ik,Drukker:1999zq,Alday:2007hr,Alday:2010vh},
\begin{equation}
S_{\rm NG}[X]
=
\frac{\sqrt{\lambda}}{2\pi}
\int d^2\sigma\,
\sqrt{\det g_{ab}},
\end{equation}
with induced metric
\begin{equation}
g_{ab}
=
G_{MN}(X)\,\partial_a X^M \partial_b X^N.
\end{equation}

In the strong-coupling limit $\lambda\gg1$, the path integral is dominated by the classical worldsheet $\Sigma_{\rm cl}$ whose boundary is the staple contour:
\begin{equation}
X^M(\sigma)=X^M_{\rm cl}(\sigma)+\delta X^M(\sigma),
\qquad
\partial\Sigma_{\rm cl}=C.
\end{equation}
Expanding around the saddle gives
\begin{equation}
S_{\rm NG}[X]
=
S_{\rm NG}[X_{\rm cl}]
+
\frac{1}{2}\,\delta X\,\mathcal{O}_{\rm ws}\,\delta X
+\cdots,
\end{equation}
so that worldsheet fluctuations are suppressed by $\lambda^{-1/2}$.
To leading semiclassical order,
\begin{align}
F_j^{g,{\rm ws}}(\xi,t,b_T;\mu,\zeta)
&=
e^{-S_{\rm NG}[\Sigma_{\rm cl}]}\,
\mathcal{A}^{(j)}_{\rm exch}(\xi,t,b_T)
\nonumber\\
&\quad+
\mathcal{O}(\lambda^{-1/2}),
\end{align}
where
\begin{equation}
\mathcal{A}^{(j)}_{\rm exch}
=
\left.
V_T(P')\,V^{(j)}_{\rm bilocal}[C]\,V_T(P)
\right|_{X=X_{\rm cl}}
\end{equation}
is the bulk exchange amplitude evaluated on the classical worldsheet background.

At planar order, the relevant process is single closed-string exchange between the hadron sources and the bilocal worldsheet source, as in the standard tree-level Witten-diagram expansion of holographic correlators \cite{Witten:1998qj,Freedman:1998tz,Costa:2011mg,Costa:2014kfa,Hijano:2015zsa,Dyer:2017zef}.
Equivalently, in the supergravity limit this is exchange of the bulk spin-$j$ field $H^{(j)}$.
Integrating out $H^{(j)}$ at tree level yields
\begin{eqnarray}
&&\mathcal{A}^{(j)}_{\rm exch}(\xi,t,b_T)
=\nonumber\\
&&\int_0^{z_{\rm IR}}dz
\int_0^{z_{\rm IR}}dz'\,
J^{(j)}_{\rm had}(\xi,t;z)\,
G_j(z,z';t)\,
J^{(j)}_{\rm op}(b_T;z'),\nonumber\\
\end{eqnarray}
up to corrections from multi-string exchange, which are suppressed by powers of $1/N_c^2$.
Here $J^{(j)}_{\rm had}$ is generated by the two hadron vertices,
\begin{equation}
J^{(j)}_{\rm had}(\xi,t;z)
\sim
\Phi_{P'}(z)\,
\mathcal{V}^{(j)}_{\rm had}(\xi,t;z)\,
\Phi_P(z),
\end{equation}
while $J^{(j)}_{\rm op}$ is generated by evaluating the bilocal worldsheet current on the classical surface,
\begin{equation}
J^{(j)}_{\rm op}(b_T;z)
\sim
\int_{\Sigma_{\rm cl}} d^2\sigma\,
\sqrt{g_{\rm cl}}\,
\mathcal{J}^{(j)}[X_{\rm cl}(\sigma);C]\,
\delta\!\left(z-z_{\rm cl}(\sigma)\right).
\end{equation}

It follows that, to leading order in the simultaneous large-$N_c$ and
large-$\lambda$ expansion, the finite-separation branch of the piecewise
saddle decomposition in Eq.~\eqref{eq:intro_piecewise_factorization} is
obtained by multiplying the stripped Witten diagram by the universal vacuum
worldsheet factor.  The shorthand \[F^g_j=F^g_{j,\bdry}+F^g_{j,\ws}+\cdots\] will
be used only as a sector decomposition: the boundary sector is a strict
local/contact Witten diagram at $b_T=0$, not a smooth term in the finite-$b_T$
worldsheet saddle.  For $b_T>0$, the finite-separation worldsheet sector is
\begin{equation}
F^g_{j,\ws}
=
S\,\widetilde F_{j}^{g,{\rm ws}}
+
\mathcal{O}(N_c^{-2},\lambda^{-1/2}),
\qquad b_T>0 .
\label{eq:secIII_ws_factorization}
\end{equation}
with
\begin{equation}
S(b_T;\mu,\zeta)
=
\exp\!\left[-S_{\rm NG}[\Sigma_{\rm min}(C)]\right]
\end{equation}
and
\begin{align}
\widetilde F_{j}^{g,{\rm ws}}(\xi,t,b_T;\mu)
&=
\int_0^{z_{\rm IR}}dz
\int_0^{z_{\rm IR}}dz'\,
J^{(j)}_{\rm had}(\xi,t;z)
\nonumber\\
&\quad\times
G_j(z,z';t)\,
J^{(j)}_{\rm op}(b_T;z').
\label{eq:secIII_stripped_ws}
\end{align}
The subtraction of the perimeter divergences associated with the isolated Wilson lines is understood in the definition of $S$.
The soft-factor bookkeeping used below is fixed as follows.  The symbol
\(S^{1/2}\) denotes the vacuum factor associated with one staple in one GTMD
matrix element, while \(S\) denotes the product of the two staple factors in
the unsubtracted worldsheet convention.  A square-root or full-soft
subtracted GTMD is obtained only after redistributing this same vacuum factor
as in Eq.~\eqref{eq:secIII_subtracted_scheme_map}.  Thus the formula below is
written for one staple, and the full unsubtracted soft factor contains twice
its cusp logarithm.
For a regulated pair of staple directions with relative rapidity
\(\chi\), the self-contained worldsheet evaluation summarized in
Appendix~\ref{app:soft_factor} gives, per staple,
\begin{widetext}
\begin{align}
-\ln S^{1/2}(b_T;\chi)
=
\Gamma_{\rm cusp}^{\rm ws}(\lambda)\,
\mathfrak c_M(\chi)\ln\frac{L}{\epsilon}
%\nonumber\\
+
\frac{\sqrt\lambda}{2\pi}
\left[
\frac{b_T^2}{2}\,\mathcal F_{\rm ws}^{(M)}(\chi)
\ln\frac{\rho_c}{\epsilon}
\right.
%\nonumber\\
\left.
+A_{\rm strip}^{\rm ren}(b_T;z_{\rm IR})
+A_{\rm IR}
+\cdots
\right].
\label{eq:secIII_soft_factor_main}
\end{align}
\end{widetext}
Here
\begin{equation}
\Gamma_{\rm cusp}^{\rm ws}(\lambda)
\equiv
\frac{\sqrt\lambda}{2\pi}
\label{eq:secIII_ws_cusp_convention}
\end{equation}
is the coefficient in the single-staple, \(S^{1/2}\), Nambu-Goto
normalization used here.  In this convention \(\mathfrak c_M(\chi)\) denotes
the total regulated cusp geometry assigned to one staple factor.  It should
not be compared without these labels to full Wilson-loop, per-geometric-cusp,
two-staple, or scaling-function definitions, which differ by factors of two
depending on whether one quotes one cusp, one staple, \(S^{1/2}\), or the full
soft factor \(S\).  Throughout the main text, \(S^{1/2}\) denotes the vacuum
factor associated with one staple in a single GTMD matrix element, \(S\)
denotes the product of the two staple factors in the unsubtracted worldsheet
convention, and therefore \(-\ln S\) contains
\(2\Gamma_{\rm cusp}^{\rm ws}\mathfrak c_M(\chi)\ln(L/\epsilon)\) at leading
saddle.  The dimensionless function
\(\mathfrak c_M(\chi)\) is the finite-angle cusp function of the regulated
worldsheet scheme.  Only its large-rapidity behavior is needed for
Collins-Soper evolution,
\begin{equation}
\mathfrak c_M(\chi)=\chi+\mathcal O(\chi^0),
\qquad
\partial_\chi\mathfrak c_M(\chi)\to1
\quad (\chi\to\infty).
\label{eq:secIII_cusp_large_chi}
\end{equation}
We therefore do not identify the finite-\(\chi\) strong-coupling minimal
surface with the familiar perturbative eikonal finite-angle form.  The
strip and infrared terms are rapidity independent at leading saddle and
encode model-dependent large-\(b_T\) physics.  The mapping between
\(\chi\) and the rapidity scale \(\zeta\) is regulator dependent, but all
such scheme dependence is confined to the soft factor.

The convention in Eq.~\eqref{eq:secIII_ws_factorization} is the
unsubtracted worldsheet convention: the vacuum Wilson-line factor
multiplies the target-dependent stripped amplitude.  A perturbatively
subtracted GTMD convention redistributes this vacuum factor.  Algebraically,
for a scheme that removes a power \(\alpha\) of the vacuum soft factor,
\begin{widetext}
\begin{align}
F_{j}^{g,{\rm sub}[\alpha]}(\xi,t,b_T;\mu,\zeta)
=
S(b_T;\mu,\zeta)^{-\alpha}
F_{j}^{g,{\rm unsub}}(\xi,t,b_T;\mu,\zeta)
=
S(b_T;\mu,\zeta)^{1-\alpha}
\widetilde F_{j}^{g,{\rm ws}}(\xi,t,b_T;\mu).
\label{eq:secIII_subtracted_scheme_map}
\end{align}
\end{widetext}
with \(\alpha=1/2\) or \(\alpha=1\) corresponding to common square-root
or full-soft conventions.  The strong-coupling statement is the
separation of the universal Wilson-line area from the target Witten
diagram, not a claim about a unique perturbative subtraction scheme.  The
rapidity variable \(\zeta\) is fixed by the regulated staple directions.
Any kinematic dependence of the rapidity regulator at nonzero skewness is
included in the chosen \(\zeta\); the remaining explicit \(\xi\) and
\(t\) dependence resides in \(\widetilde F_{j}^{g,{\rm ws}}\).

\begin{table*}[t]
\centering
\caption{Soft-factor bookkeeping used in the text.  The coefficient column
refers to the coefficient of \(\ln(L/\epsilon)\) in \(-\ln\) of the indicated
soft factor at leading worldsheet saddle.  The function \(\mathfrak c_M\) is
normalized as the total regulated cusp geometry assigned to one staple factor
in the \(S^{1/2}\) convention.}
\label{tab:soft_factor_conventions}
\small
\begin{tabular}{llll}
\toprule
Object & Content & Coefficient & Convention \\
\midrule
\(S^{1/2}\) & one staple factor &
\(\Gamma_{\rm cusp}^{\rm ws}\mathfrak c_M(\chi)\) &
single matrix element \\
\(S\) & two staple factors &
\(2\Gamma_{\rm cusp}^{\rm ws}\mathfrak c_M(\chi)\) &
unsubtracted worldsheet \\
\(F_j^{g,{\rm sub}[\alpha]}\) & scheme dependent &
\(2(1-\alpha)\Gamma_{\rm cusp}^{\rm ws}\mathfrak c_M(\chi)\) residual &
soft redistribution \\
\bottomrule
\end{tabular}
\end{table*}

This factorization has a precise origin.
The factor $S$ is the vacuum weight of the classical worldsheet determined solely by the staple contour and therefore depends only on the Wilson-line geometry, the renormalization scale, and the rapidity regulator.
All hadronic information enters only through the exchange amplitude $\widetilde F_{j}^{g,{\rm ws}}$ for the finite-separation sector and through $\widetilde F_{j}^{g,{\rm bdry}}$ for the strict boundary sector.
In particular, the dependence on the skewness $\xi$ and the momentum transfer $t$ resides in $J^{(j)}_{\rm had}$ and in the bulk propagator $G_j$, whereas the dependence on the transverse geometry of the bilocal operator resides in $J^{(j)}_{\rm op}$.

To summarise: Eqs.
\eqref{eq:intro_piecewise_factorization} and
\eqref{eq:secIII_ws_factorization} are leading-saddle statements at planar
large $N_c$ and semiclassical large $\lambda$.  They are obtained  at fixed even
spin $j$ before Regge continuation and for $b_T>0$.  The spin-$j$ insertion is
treated as a probe on the classical staple worldsheet.  Perimeter and cusp
subtractions are included in the soft factor.  The infrared endpoint assumes a
cusp-subtracted central worldsheet saddle at finite holographic depth.  The
soft-wall, hard-wall, and repulsive-wall large-$b_T$ tails are therefore data
of the chosen infrared transfer-kernel completion, not universal strong-coupling
predictions.
%\end{minipage}}

\section{Fixed-spin Holographic Representation}
\label{SEC_IV}

We now make explicit the two ingredients entering the stripped
finite-separation worldsheet amplitude: the hadron-side source and the
operator-side worldsheet source.  The full GTMD moment is the
piecewise saddle-sector object in Eq.~\eqref{eq:intro_piecewise_factorization}.
Thus the strict \(b_T=0\) value is supplied by
\(F^g_{j,\bdry}=\widetilde F_{j}^{g,{\rm bdry}}\), whereas for \(b_T>0\)
\begin{equation}
F^g_{j,\ws}(\xi,t,b_T;\mu,\zeta)
=
S(b_T;\mu,\zeta)\,
\widetilde F_{j}^{g,{\rm ws}}(\xi,t,b_T;\mu).
\label{eq:secIV_full_fixed_spin_factorization}
\end{equation}
In this section \(\widetilde F_{j}^{g,{\rm ws}}\) denotes the stripped
finite-separation worldsheet moment; the boundary sector is not included in
it.

At fixed even conformal spin \(j\), the bulk field exchanged between the
hadron sector and the bilocal operator sector is the closed spin-\(j\)
field dual to the twist-two gluon operator
\cite{Brower:2006ea,Costa:2011mg,Costa:2014kfa,Mamo:2019mka,Mamo:2024jwp}.  The
stripped amplitude has the Witten-diagram form
\begin{eqnarray}
&&\widetilde F_{j}^{g,{\rm ws}}(\xi,t,b_T)
=\nonumber\\
&&\int_0^{z_{\mathrm{IR}}} dz
\int_0^{z_{\mathrm{IR}}} dz'\,
J^{(j)}_{\mathrm{had}}(\xi,t;z)\,
G_j(z,z';t)\,
J^{(j)}_{\mathrm{op}}(b_T;z') .\nonumber\\
\label{eq:secIV_stripped_witten}
\end{eqnarray}
The upper limit \(z_{\rm IR}\) denotes the infrared scale of the confining
background.  In a hard-wall model \(z_{\rm IR}=z_0\)
(simplified repulsive wall), while in a soft-wall
background \(z_{\rm IR}=\infty\)
\cite{Karch:2006pv,Polchinski:2001tt,Polchinski:2002jw,Brodsky:2014yha}.

\subsection{Hadron-side source at fixed spin}

The hadron-side source is generated by the pair of effective closed-string
vertices associated with the incoming and outgoing hadrons.  We use one
notation for this object throughout the paper:
\begin{equation}
J^{(j)}_{\mathrm{had}}(\xi,t;z)
\equiv
\rho_j(z;\xi),
\label{eq:secIV_hadron_source}
\end{equation}
with the momentum-transfer dependence carried by the exchanged propagator,
or equivalently by the bulk-to-boundary transfer kernel
\(\mathcal H_j(K,z)\).  The skewness polynomial and the external
normalizable modes are included in \(\rho_j(z;\xi)\).  In the simplest
factorized convention,
\begin{equation}
\rho_j(z;\xi)
=
P_j(\xi)\,\rho_j^{(0)}(z),
\label{eq:secIV_rho_skewness_convention}
\end{equation}
where \(P_j(\xi)\) is the conformal polynomial at spin \(j\) and
\(\rho_j^{(0)}(z)\) is the radial density built from the external
normalizable modes.  More general target models may replace
\(P_j(\xi)\rho_j^{(0)}(z)\) by a non-factorized \(\rho_j(z;\xi)\), but the
same symbol denotes the full hadron-side radial weight in every formula
below.

For fermionic hadrons in the soft-wall model, a typical radial density is
\begin{equation}
\rho_j^{(0)}(z)
=
\frac{R^5}{2}
e^{-\kappa^2 z^2}
\left(
\tilde n_R^{\,2} z^{2\tau+j-5}
+
\tilde n_L^{\,2} z^{2\tau+j-3}
\right).
\label{eq:secIV_rho_fermion}
\end{equation}
This source contains all hadronic information at given twist $\tau$ and
skewness \(\xi\).  In particular, the universal Wilson-line geometry does
not enter Eq.~\eqref{eq:secIV_rho_fermion}.
The endpoint factorization derived below is spin-agnostic at the level of
the radial overlap; Eq.~\eqref{eq:secIV_rho_fermion} is only the later
specialization to a spin-\(1/2\) target in the soft-wall model, while the
effective scalar vertices of Sec.~\ref{SEC_III} serve only as
Witten-diagram bookkeeping.
Similar soft-wall/holographic hadron densities and gluon gravitational
form factors have been used in holographic analyses of gluon GPDs and
diffractive processes \cite{Mamo:2019mka,Mamo:2024jwp}.

\subsection{Operator-side source from the four-cusp worldsheet}

The operator-side source is generated by the bilocal gluon insertion
evaluated on the classical four-cusp worldsheet ending on the staple
contour, following the same logic used for Wilson loops with
local-operator insertions and null-polygon worldsheets
\cite{Berenstein:1998ij,Miwa:2006vv,Alday:2007hr,Alday:2010vh,Alday:2011pf,Buchbinder:2012vr}.
Near the AdS boundary the bulk spin-\(j\) field scales as
\begin{equation}
H^{(j)}\sim z^{\Delta(j)},
\end{equation}
where \(\Delta(j)\) is the conformal dimension of the twist-two gluon
operator of spin \(j\).

For the four-cusp saddle, the near-boundary radial profile is
\begin{align}
z_{\rm ws}(u,v;b)
&=
b\,z_b(u,v),
\qquad
z_b(u,v)
=
\frac{1}{\cosh u\,\cosh v},
\nonumber\\
(u,v)&\in\mathbb{R}^2,
\qquad
b\equiv |\bm b_T| .
\label{eq:secIV_ws_profile}
\end{align}
The fixed-spin projection of the bilocal insertion carries the worldsheet
factor
\begin{equation}
\left(\sqrt{2}\kappa_c z_{\rm ws}\right)^{-(j-2)} .
\end{equation}
The overall coefficient of the worldsheet source is written as
\(\widetilde g_5^{\,2}g_j(\lambda)\).  Here \(\widetilde g_5\) is the
effective five-dimensional closed-channel normalization used for the
spin-\(j\) propagator, while \(g_j(\lambda)\) is the dimensionless
string-tension coupling of the spin-\(j\) vertex.  In the unrescaled AdS
normalization one may parameterize \(g_j(\lambda)=c_j\lambda^{j/4}\), with
\(c_j\) fixed by the spin-\(j\) two-point normalization.  The product
\(\widetilde g_5^{\,2}g_j\) is kept symbolic because it is convention
and matching dependent; importantly it carries no \(z_{\rm ws}\) or
\(b_T\) dependence and therefore cannot change the transverse endpoint
powers.  Appendix~\ref{app:spinj_vertex_power} gives the detailed
normalization and power-counting check.
Thus a convenient representation of the operator-side source is
\begin{eqnarray}
&&J^{(j)}_{\rm op}(b_T;z')
=
\widetilde g_5^{\,2}g_j(\lambda)
\int_{-\infty}^{\infty}du
\int_{-\infty}^{\infty}dv\,
\left(\sqrt{2}\kappa_c z_{\rm ws}\right)^{-(j-2)}
\nonumber\\
&&\hspace{3.0cm}\times
\delta\!\left(z'-z_{\rm ws}(u,v;b)\right) .
\label{eq:secIV_operator_source}
\end{eqnarray}
In Eq.~\eqref{eq:secIV_operator_source}, the notation \(du\,dv\) denotes
conformal worldsheet coordinates after the induced measure and local
Jacobian factors have been absorbed into the renormalized current
\(\mathcal J_{\rm ws}^{(j)}\) and the coupling \(g_j(\lambda)\).  Thus no
separate factor of \(\sqrt{g_{\rm cl}}\) has been dropped; it is included
in the definition of the operator-side source.  In conformal gauge the
near-boundary four-cusp solution is scale invariant: after
\(z_{\rm ws}=b\,z_b(u,v)\), the induced area element and the local Jacobian
are functions of \(u,v\) only in the renormalized current convention.  They
therefore cannot generate an additional power of \(b\).  The only
transverse power comes from the explicit spin-\(j\) source factor and the
near-boundary radial wave, as derived in Appendix~\ref{app:spinj_vertex_power}.
When this source is contracted with a near-boundary spin-\(j\) mode,
\(\psi_j(z_{\rm ws})\sim z_{\rm ws}^{\Delta_c(j)}\), the physical
worldsheet power is
\begin{equation}
z_{\rm ws}^{\Delta_c(j)}
z_{\rm ws}^{-(j-2)}
=
z_{\rm ws}^{\Delta_c(j)-(j-2)} .
\label{eq:secIV_physical_power}
\end{equation}
This combination is the transverse ultraviolet power which appears in the
transverse-separation kernels.

The absence of an additional factor \(z_{\rm ws}^{2(j-2)}\) in
Eq.~\eqref{eq:secIV_operator_source}, relative to the Witten diagram for DVCS \cite{Mamo:2026vuq,Mamo:2026fjh}, is intentional.  The current
\(\mathcal J_{\rm ws}^{M_1\cdots M_j}\) is defined with curved/tangent
bulk indices and the BPST spin-\(j\) field is normalized in the same
closed-channel convention as the propagator \(G_c\).  With these
conventions the only universal source prefactor is
\(z_{\rm ws}^{-(j-2)}\).  A formula with an extra
\(z_{\rm ws}^{2(j-2)}\) corresponds to a different boundary-frame
normalization of the worldsheet current; using that factor together with
the present propagator convention would double count the vielbein/metric
conversion.  Appendix~\ref{app:spinj_vertex_power} gives the detailed
power-counting check, including the graviton limit.

\subsection{Endpoint moment summary}

It is useful already at this stage to state the three fixed-spin moment
reductions that follow from the same Witten diagram, using in particular the following notations
%We introduce
\begin{widetext}
\begin{gather}
b\equiv |\bm b_T|,
\qquad
\bar b\equiv \sqrt{2}\,\kappa_c b,
\qquad
K^2=-t,
\qquad
a_{t,c}\equiv -\frac{t}{4\kappa_c^2},
\\
\Delta_c(j)=2+j+\gamma_c(j),
\qquad
\alpha_j\equiv \Delta_c(j)-(j-2)=4+\gamma_c(j),
\qquad
\nu_j\equiv \Delta_c(j)-2=j+\gamma_c(j).
\label{eq:secIV_delta_alpha_nu}
\end{gather}
\end{widetext}
The strict boundary conformal moment is denoted by
\begin{equation}
\widetilde F_{j}^{g,{\rm bdry}}(\xi,t)
\equiv
\widetilde F_j^{g,{\rm full}}(\xi,t,0).
\end{equation}
In the infrared entries below, the transfer kernels refer to the
cusp/perimeter-subtracted central worldsheet region.  Their coefficients and
scales are model and saddle data of the chosen confining completion; only the
radial endpoint logic is universal at leading saddle.

The three endpoint reductions are
\begin{widetext}
\begin{align}
{\rm UV:}\qquad
\widetilde F_{j}^{g,{\rm ws}}(\xi,t,b_T)
&\xrightarrow[b_T\to0^+]{}
\mathcal K^{\rm UV}_j(\lambda,b_T)\,
\widetilde F_{j}^{g,{\rm bdry}}(\xi,t),
\qquad
\mathcal K^{\rm UV}_j
\sim
\bar b^{\,4+\gamma_c(j)},
\nonumber\\
{\rm IR\;SW:}\qquad
\widetilde F_{j,{\rm IR},{\rm SW}}^{g,{\rm ws}}(\xi,t,b_T)
&\xrightarrow[b_T\to\infty]{}
\mathcal K^{\rm IR,SW}_j(\lambda,b_T,t)\,
\widehat{\mathcal T}^{(c)}_j(\xi),
\qquad
\mathcal K^{\rm IR,SW}_j
\sim
C_j(t)\,\bar b^{2-j-a_{t,c}},
\nonumber\\
{\rm IR\;HW:}\qquad
\widetilde F_{j,{\rm IR},{\rm HW}}^{g,{\rm ws}}(\xi,t,b_T)
&\xrightarrow[b_T\to\infty]{}
\mathcal K^{\rm IR,HW,gap}_j(\lambda,b_T)\,
\widehat{\mathcal T}^{(c)}_j(\xi),
\qquad
\mathcal K^{\rm IR,HW,gap}_j
\sim
\widetilde C_j^{\rm HW}
(\kappa_{\rm IR}b)^{\frac{7}{2}-j}e^{-\kappa_{\rm IR}b}.
\label{eq:secIV_endpoint_moment_results}
\end{align}
\end{widetext}
The repulsive wall result will be quoted in the end, with its full derivation given in Appendix~\ref{app:RW_GTMD_endpoint}.
The target-side projection appearing in the two infrared reductions is
\begin{equation}
\widehat{\mathcal T}^{(c)}_j(\xi)
\equiv
\int_0^\infty dz\,
\rho_j(z;\xi)\,
\left(\sqrt{2}\kappa_c z\right)^{\Delta_c(j)} .
\label{eq:secIV_target_projection}
\end{equation}
The ultraviolet endpoint factorizes onto the boundary conformal moment,
while the two infrared endpoints factorize onto the target projection
\(\widehat{\mathcal T}^{(c)}_j(\xi)\).  The soft-wall infrared endpoint
gives an algebraic deep-bulk transfer, whereas the gap-matched hard-wall
completion gives a confining exponential with physical scale
\(\kappa_{\rm IR}=M_{\rm gap}\widehat z_{\rm IR}\).  The compact
\(\bar b^{7/2-j}e^{-\bar b}\) form is the specialization
\(\kappa_{\rm IR}=\sqrt2\kappa_c\), or the convention in which the gap factor
is absorbed into the fitted infrared scale.

\subsection{Small-\texorpdfstring{\(x\)}{x} and small-\texorpdfstring{\(b_T\)}{b} consequence}

The fixed-spin endpoint formulas above are the ingredients that are
analytically continued in section~\ref{SEC_VI}.  In the DGLAP-like
small-skewness region, \(\xi\simeq0\), the \(x\)-dependent GTMD is
reconstructed from its conformal/Mellin moments by an inverse Mellin transform in
\(j\).  At strong coupling the rightmost singularity of the analytically
continued gluon moment is the BPST branch point
\cite{Brower:2006ea,Brower:2010wf,Costa:2012cb,Brower:2013},
\begin{equation}
j=j_0^g,
\qquad
j_0^g
=
2-\frac{2}{\sqrt{\lambda}} .
\label{eq:secIV_bpst_intercept}
\end{equation}
Thus the leading small-\(x\) power is
\begin{equation}
x^{1-j_0^g}
=
x^{-\left(1-\frac{2}{\sqrt{\lambda}}\right)} .
\label{eq:secIV_smallx_power}
\end{equation}
This makes explicit that the strong-coupling gluon GTMD grows at small
\(x\) almost as \(1/x\), with the intercept softened by the finite
\(1/\sqrt{\lambda}\) correction.

The transverse ultraviolet endpoint is the fixed-spin seed of the
small-\(b_T\) Reggeized result.  After continuation to the BPST cut,
one may parameterize the spectral curve by
\begin{equation}
\Delta=2\pm i\varpi,
\qquad
j(\varpi)
=
j_0^g
-
\frac{\varpi^2}{2\sqrt{\lambda}},
\qquad
\varpi\ge0 .
\label{eq:secIV_bpst_cut_param}
\end{equation}
For \(0<\bar b\ll1\), the analytically continued transverse impact factor
contains the branch-cut factor
\begin{equation}
\bar b^{\,\Delta-2}
=
e^{\mp i\varpi \ell_b},
\qquad
\ell_b\equiv \ln\frac{1}{\bar b},
\qquad
\bar b\equiv \sqrt{2}\kappa_c b .
\label{eq:secIV_smallb_oscillatory_factor}
\end{equation}
This form is consistent with the fixed-spin ultraviolet kernel
\(\mathcal K_j^{\rm UV}\sim \bar b^{\Delta_c(j)-(j-2)}\).  Near the
Regge cut one rewrites the same power as
\begin{equation}
\bar b^{\Delta(j)-(j-2)}
=
\bar b^{4-j}\,\bar b^{\Delta(j)-2}.
\label{eq:secIV_regge_power_split}
\end{equation}
The first factor is analytic and slowly varying near the rightmost
singularity, so at leading Regge accuracy it is evaluated at
\(j=j_0^g\).  The second factor carries the nonanalytic branch-cut
phase and is responsible for diffusion in \(\ell_b\).  Thus the fixed-spin worldsheet prefactor has not been dropped; it is contained in
the explicit factor \(\bar b^{4-j_0^g}\) in the final small-\(x\) result.
The discontinuity is fixed by the combination
\(\bar b^{\Delta-2}/\Gamma(\Delta-2)\), which gives
\(2i\varpi\cos(\varpi\ell_b)\) on the BPST cut in the present soft-wall
normalization.  With
\begin{equation}
Y\equiv \ln\frac{1}{x},
\label{eq:secIV_rapidity_variable}
\end{equation}
the explicit small-\(x\), small-\(b_T\) result derived in
section~\ref{SEC_VI} is
\begin{widetext}
\begin{equation}
F_g(x,t,b_T;\mu,\zeta)
\Big|_{\xi\simeq0,\;x\ll1,\;\bar b\ll1}
\simeq
\mathcal N_g^{\rm UV}(t)\,
x^{-\left(1-\frac{2}{\sqrt{\lambda}}\right)}\,
S(b_T;\mu,\zeta)\,
\bar b^{4-j_0^g}
\frac{\lambda^{1/4}}{\sqrt{2\pi}\,Y^{3/2}}
\left(1-\sqrt\lambda\frac{\ell_b^2}{Y}\right)
\exp\!\left[
-\frac{\sqrt{\lambda}}{2}
\frac{\ell_b^2}{Y}
\right].
\label{eq:secIV_smallx_smallb_result}
\end{equation}
\end{widetext}
The smooth coefficient \(\mathcal N_g^{\rm UV}(t)\) contains the
\(\Gamma(\Delta-2)\)-stripped boundary moment evaluated at
\(j=j_0^g\).  Equivalently, the stripped small-\(x\) result is obtained
by deleting the universal factor \(S(b_T;\mu,\zeta)\).  The
\(Y^{-3/2}\) factor, the polynomial in \(\ell_b^2/Y\), and the Gaussian
are the branch-cut diffusion factors.  The worldsheet soft factor is
analytic in \(j\) and therefore factors out of the Regge integral, just
as it factors out of the fixed-spin moment in
Eq.~\eqref{eq:secIV_full_fixed_spin_factorization}.  Thus the small-\(x\)
formula is not an additional dynamical assumption: it is the Reggeized
reorganization of the same fixed-spin holographic representation.

Before turning to endpoint limits we emphasize a notational point.  The
quantity denoted by \(b_T\) below is the transverse separation conjugate
to the average partonic transverse momentum \(k_T\).  It is not the
impact parameter conjugate to \(\Delta_T\); the latter information is
carried by the off-forward variables \(t\) and \(\Delta_T\).

\section{Leading-saddle transverse-separation representations}
\label{SEC_V}

In this section we derive the three endpoint reduction formulas stated in Eq.~\eqref{eq:secIV_endpoint_moment_results} within the leading semiclassical and planar approximation, using the bulk-to-bulk and bulk-to-boundary technology of AdS Witten diagrams \cite{Witten:1998qj,Freedman:1998tz,DHoker:1999jke,Costa:2011mg,Costa:2014kfa,Hijano:2015zsa,Dyer:2017zef}.  The
universal vacuum area of the staple worldsheet has already been stripped
off in section~\ref{SEC_III}.  Thus \(\widetilde F_{j}^{g,{\rm ws}}\) denotes the stripped finite-separation
worldsheet conformal moment throughout this section.  The finite-
separation contribution to the full GTMD moment is obtained by multiplying
the final expressions by \(S(b_T;\mu,\zeta)\), while the strict local
boundary sector is added separately at \(b_T=0\).  The infrared endpoint
formulas in this section should be read with the cusp-subtracted ordering
stated explicitly in Sec.~\ref{subsec:cusp_subtracted_ir_ordering}: the shallow
cusp regions are absorbed into the soft factor or matched to the ultraviolet
sector, while the central finite-depth worldsheet saddle defines the displayed
infrared transfer kernels.

We use
\begin{equation}
\begin{gathered}
b\equiv |\bm b_T|,
\qquad
\bar b\equiv \sqrt{2}\,\kappa_c b,\\
K^2=-t,
\qquad
a_{t,c}\equiv -\frac{t}{4\kappa_c^2}.
\end{gathered}
\label{eq:secV_defs}
\end{equation}
The exchanged closed spin-\(j\) field has dimension
\begin{equation}
\Delta_c(j)=2+j+\gamma_c(j),
\label{eq:secV_delta}
\end{equation}
and we define
\begin{equation}
\begin{gathered}
\alpha_j\equiv \Delta_c(j)-(j-2)=4+\gamma_c(j),\\
\nu_j\equiv \Delta_c(j)-2=j+\gamma_c(j).
\end{gathered}
\label{eq:secV_alpha_nu}
\end{equation}
The exponent \(\alpha_j\) is the physical transverse ultraviolet power
of the finite-separation worldsheet source.

The strict boundary conformal moment is
\begin{equation}
\widetilde F_{j}^{g,{\rm bdry}}(\xi,t)
\equiv
\widetilde F_j^{g,{\rm full}}(\xi,t,0),
\label{eq:secV_boundary_moment_def}
\end{equation}
and admits the soft-wall mode decomposition
\begin{equation}
\widetilde F_{j}^{g,{\rm bdry}}(\xi,t)
=
\sum_{n=0}^{\infty}
\widetilde F_{j}^{g,{\rm bdry}}(\xi,t;n).
\label{eq:secV_boundary_mode_sum}
\end{equation}

\subsection{Boundary worldsheet profile}

\begin{figure}[t]
\centering
\begin{tikzpicture}[scale=0.78]
  \draw[->,line width=0.5pt] (-3.35,0) -- (3.45,0) node[right] {$u$};
  \draw[->,line width=0.5pt] (0,-3.35) -- (0,3.45) node[above] {$v$};
  \foreach \x in {-3,-2,-1,1,2,3}{
    \draw[black!45,line width=0.3pt] (\x,0.07) -- (\x,-0.07);
    \draw[black!45,line width=0.3pt] (0.07,\x) -- (-0.07,\x);
  }
  \node[font=\scriptsize] at (-3.05,-0.28) {$-3$};
  \node[font=\scriptsize] at (3.05,-0.28) {$3$};
  \node[font=\scriptsize] at (-0.28,3.05) {$3$};
  \node[font=\scriptsize] at (-0.34,-3.05) {$-3$};
  \foreach \lev in {0.80,0.60,0.40,0.20,0.10}{
    \pgfmathsetmacro{\umax}{ln(1/\lev + sqrt(1/(\lev*\lev)-1))}
    \pgfmathsetmacro{\udraw}{0.995*\umax}
    \draw[line width=0.55pt]
      plot[domain=-\udraw:\udraw,samples=90,smooth,variable=\u]
      ({\u},{ln(1/(\lev*cosh(\u)) + sqrt(1/(\lev*\lev*cosh(\u)*cosh(\u))-1))});
    \draw[line width=0.55pt]
      plot[domain=-\udraw:\udraw,samples=90,smooth,variable=\u]
      ({\u},{-ln(1/(\lev*cosh(\u)) + sqrt(1/(\lev*\lev*cosh(\u)*cosh(\u))-1))});
  }
  \fill (0,0) circle (1.4pt);
  \node[font=\scriptsize,anchor=west,fill=white,draw=black!35,rounded corners=1pt,inner sep=2pt]
    at (1.78,2.35) {contours: $0.8,0.6,0.4,0.2,0.1$};
  \node[font=\scriptsize] at (0,-3.78)
    {$z_b(u,v)=1/(\cosh u\,\cosh v)$};
\end{tikzpicture}
\caption{
Contour representation of the universal near-boundary worldsheet profile
\(z_b(u,v)=1/(\cosh u\,\cosh v)\).  The contours give fixed values of
\(z_b\); the maximum lies at \(u=v=0\), and the profile falls exponentially
in the cusp regions.  The physical worldsheet is obtained by the scale
transformation \(z_{\rm ws}=b\,z_b\), so changing
\(b=|\bm b_T|\) rescales the holographic depth without changing the
near-boundary conformal shape.
}
\label{fig:zb_profile}
\end{figure}
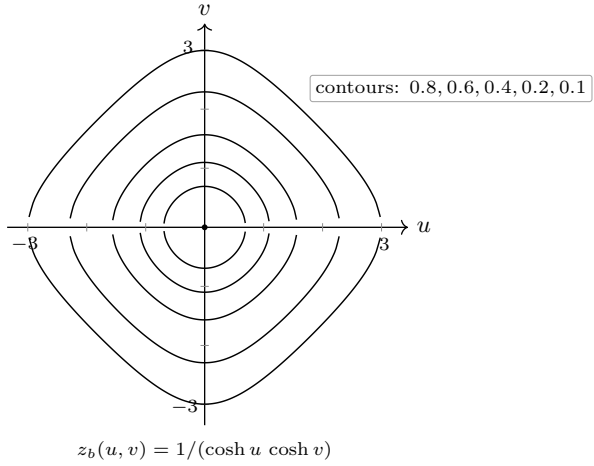

The conformal profile
\begin{equation}
z_b(u,v)=\frac{1}{\cosh u\,\cosh v}
\label{eq:zb_profile}
\end{equation}
is the universal near-boundary solution for the four-cusp
worldsheet associated with the staple Wilson-line contour.
It follows solely from the asymptotically AdS geometry and
is therefore independent of the infrared completion of the
background. The profile reaches its maximal value
\(z_b(0,0)=1\) at the center of the worldsheet and decays
exponentially for large worldsheet coordinates,
\begin{equation}
z_b(u,v)
\xrightarrow{|u|,|v|\gg 1}
4\,e^{-|u|-|v|},
\end{equation}
corresponding to the approach of the surface toward the
Wilson-line cusps on the AdS boundary, as illustrated in Fig.~\ref{fig:zb_profile}.

The physical worldsheet is obtained through the scaling
\begin{equation}
z_{\rm ws}(u,v;b)=b\,z_b(u,v),
\end{equation}
so that the transverse separation \(b=|\bm b_T|\)
controls the overall depth of the surface in the holographic
direction. As \(b_T\to0\), the entire worldsheet is pulled
toward the boundary and the exchanged spin-$j$ propagator
can be replaced by its ultraviolet factorized form. The
combination of the worldsheet insertion factor
\((\sqrt{2}\kappa_c z_{\rm ws})^{-(j-2)}\) with the
near-boundary behavior of the exchanged mode,
\(\psi_j(z_{\rm ws})\sim z_{\rm ws}^{\Delta_c(j)}\),
produces the characteristic power
\begin{equation}
z_{\rm ws}^{\Delta_c(j)-(j-2)}
=
b^{\,4+\gamma_c(j)}
z_b^{\,4+\gamma_c(j)},
\end{equation}
which is responsible for the ultraviolet endpoint behavior
\begin{equation}
{\cal K}^{\rm UV}_j(\lambda,b_T)
\propto
b^{\,4+\gamma_c(j)}.
\end{equation}
The conformal profile therefore acts as a universal geometric
weight governing the coupling of the bilocal gluon operator
to the classical worldsheet, while all hadronic information
remains encoded separately in the bulk wave functions and
exchange kernel.

\subsection{Starting Witten diagram}

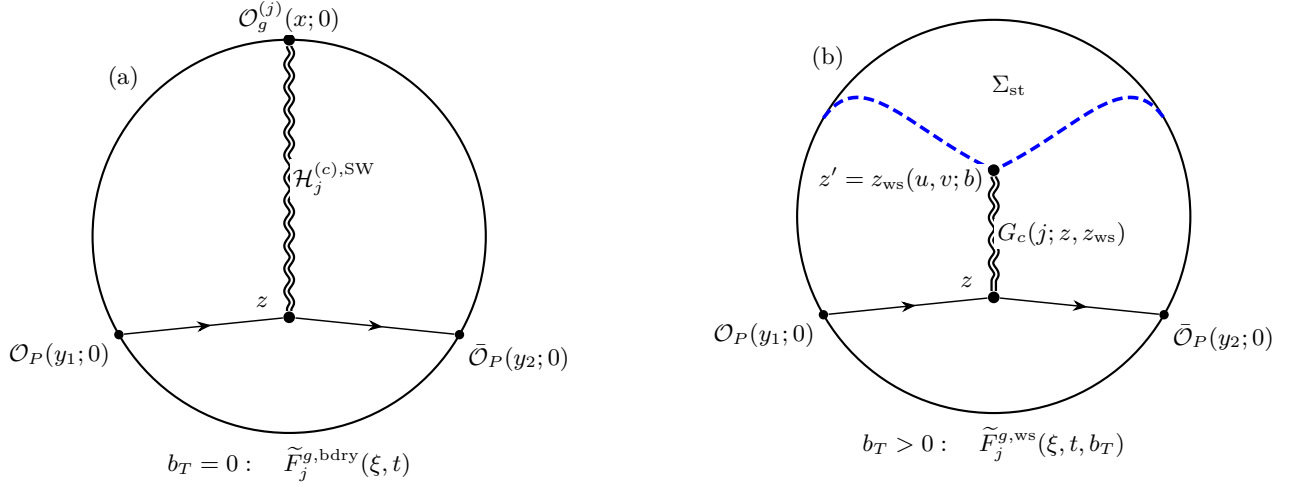
\begin{figure*}[t]
\centering
\tikzset{
mainboundary/.style={draw=black,line width=0.8pt},
mainstring/.style={draw=blue,line width=1.35pt,dash pattern=on 4pt off 2pt},
mainhadron/.style={
draw=black,
line width=0.55pt,
postaction={decorate},
decoration={markings,mark=at position .55 with {\arrow{Stealth}}}
},
maindoublewiggle/.style={
double,
decorate,
decoration={snake,amplitude=1.0pt,segment length=3.0mm},
double distance=0.8pt,
line width=0.85pt
},
mainvertex/.style={circle,fill=black,inner sep=1.55pt},
mainext/.style={circle,fill=black,inner sep=1.25pt}
}

\begin{minipage}{0.48\textwidth}
\centering
\begin{tikzpicture}[scale=1.02]
  \def\R{2.55}
  \draw[mainboundary] (0,0) circle (\R);

  \coordinate (Pin)  at (-150:\R);
  \coordinate (Pout) at ( -30:\R);
  \node[mainext] (ein)  at (Pin)  {};
  \node[mainext] (eout) at (Pout) {};

  \node[font=\small,below left]  at (ein)  {$\cO_P(y_1;0)$};
  \node[font=\small,below right] at (eout) {$\bar\cO_P(y_2;0)$};

  \node[mainvertex] (vb) at (0,-1.05) {};
  \node[font=\small,fill=white,inner sep=1pt] at (-0.35,-0.85) {$z$};

  \draw[mainhadron] (ein) -- (vb);
  \draw[mainhadron] (vb) -- (eout);

  \node[mainvertex] (bdry) at (90:\R) {};
  \node[font=\small,fill=white,inner sep=1pt] at (0,2.85)
    {$\cO_g^{(j)}(x;0)$};

  \draw[maindoublewiggle] (bdry) -- (vb)
    node[midway,right,font=\small,fill=white,inner sep=1pt]
    {$\mathcal H_j^{(c),{\rm SW}}$};

  \node[font=\small] at (-2.15,2.05) {(a)};
  \node[font=\small] at (0,-2.95)
    {$b_T=0:\quad \widetilde F_{j}^{g,{\rm bdry}}(\xi,t)$};
\end{tikzpicture}
\end{minipage}
\hfill
\begin{minipage}{0.48\textwidth}
\centering
\begin{tikzpicture}[scale=1.02]
  \def\R{2.55}
  \draw[mainboundary] (0,0) circle (\R);

  \coordinate (Pin)  at (-150:\R);
  \coordinate (Pout) at ( -30:\R);
  \node[mainext] (ein)  at (Pin)  {};
  \node[mainext] (eout) at (Pout) {};

  \node[font=\small,below left]  at (ein)  {$\cO_P(y_1;0)$};
  \node[font=\small,below right] at (eout) {$\bar\cO_P(y_2;0)$};

  \node[mainvertex] (vb) at (0,-1.05) {};
  \node[font=\small,fill=white,inner sep=1pt] at (-0.35,-0.85) {$z$};

  \draw[mainhadron] (ein) -- (vb);
  \draw[mainhadron] (vb) -- (eout);

  \coordinate (S1) at (150:\R);
  \coordinate (S2) at (30:\R);
  \coordinate (Sdip) at (0,0.60);

  \draw[mainstring]
    (S1)
    .. controls (-1.70,2.00) and (-1.05,1.08) .. (Sdip)
    .. controls ( 1.05,1.08) and ( 1.70,2.00) .. (S2);

  \node[font=\small,fill=white,inner sep=1pt] at (0.20,1.70)
    {$\Sigma_{\rm st}$};

  \node[mainvertex] (vt) at (Sdip) {};
  \node[font=\small,fill=white,inner sep=1pt] at (-1.20,0.48)
    {$z'=z_{\rm ws}(u,v;b)$};

  \draw[maindoublewiggle] (vt) -- (vb)
    node[midway,right,font=\small,fill=white,inner sep=1pt]
    {$G_c(j;z,z_{\rm ws})$};

  \node[font=\small] at (-2.15,2.05) {(b)};
  \node[font=\small] at (0,-2.95)
    {$b_T>0:\quad \widetilde F_{j}^{g,{\rm ws}}(\xi,t,b_T)$};
\end{tikzpicture}
\end{minipage}
\caption{
Fixed-spin Witten diagrams for the boundary and finite-separation GTMD
conformal moments.  In panel (a), the double-wiggly line is the
bulk-to-boundary transfer kernel.  In panel (b), the double-wiggly line is
the full bulk-to-bulk propagator \(G_c(j,z,z_{\rm ws};t)\) between the
hadron vertex and the worldsheet vertex. The thick-dashed-blue curve is the string defect in AdS.}
\label{fig:secV_fixed_j_witten_gtmd}
\end{figure*}

The fixed-spin boundary and finite-separation Witten diagrams are shown
in Fig.~\ref{fig:secV_fixed_j_witten_gtmd}; their structure is the standard factorized tree-level exchange familiar from holographic current and local-operator correlators \cite{Freedman:1998tz,DHoker:1999jke,Osborn:1993cr,Costa:2011mg,Costa:2014kfa}.  The finite-\(b_T\)
diagram is
\begin{widetext}
\begin{align}
\widetilde F_{j}^{g,{\rm ws}}(\xi,t,b_T;\epsilon)
=
\widetilde g_5^{\,2}g_j(\lambda)
\int_{-\infty}^{\infty}du
\int_{-\infty}^{\infty}dv\,
\left(\sqrt{2}\kappa_c z_{\rm ws}\right)^{-(j-2)}
\int_0^\infty dz\,
\rho_j(z;\xi)\,
G_c(j,z,z_{\rm ws};t).
\label{eq:secV_ws_witten}
\end{align}
\end{widetext}

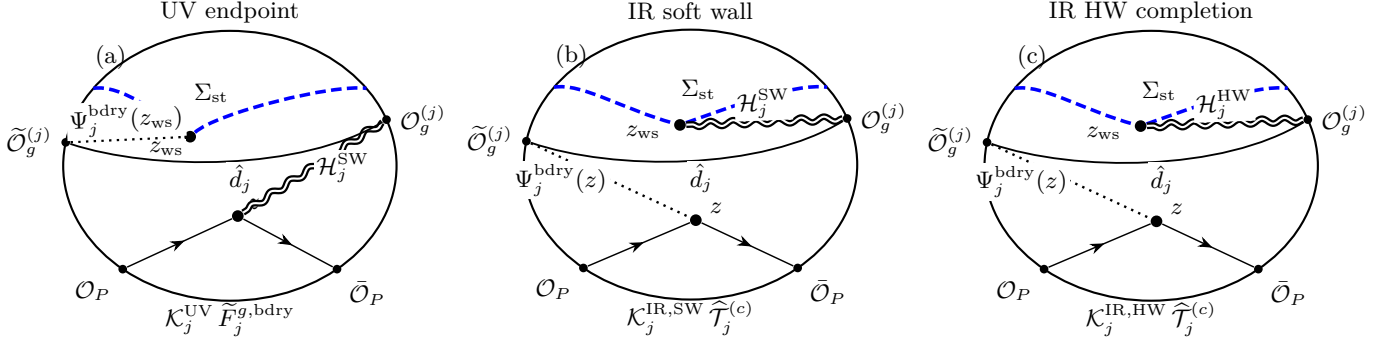
\begin{figure*}[t]
\centering
\tikzset{
facboundary/.style={draw=black,line width=0.8pt},
fachadron/.style={
draw=black,
line width=0.55pt,
postaction={decorate},
decoration={markings,mark=at position .55 with {\arrow{Stealth}}}
},
facstring/.style={draw=blue,line width=1.35pt,dash pattern=on 4pt off 2pt},
facbb/.style={
double,
decorate,
decoration={snake,amplitude=1.0pt,segment length=3.0mm},
double distance=0.8pt,
line width=0.85pt
},
facnorm/.style={dotted,line width=0.9pt},
facproj/.style={line width=0.7pt},
facvertex/.style={circle,fill=black,inner sep=1.55pt},
facext/.style={circle,fill=black,inner sep=1.15pt}
}
\begin{minipage}{0.32\textwidth}
\centering
\begin{tikzpicture}[scale=0.70]
  \def\Rx{3.15}
  \def\Ry{2.55}
  \draw[facboundary] (0,0) ellipse [x radius=\Rx cm,y radius=\Ry cm];

  \coordinate (OgL) at ({\Rx*cos(170)},{\Ry*sin(170)});
  \coordinate (OgR) at ({\Rx*cos(20)},{\Ry*sin(20)});
  \coordinate (P1)  at ({\Rx*cos(-130)},{\Ry*sin(-130)});
  \coordinate (P2)  at ({\Rx*cos(-50)},{\Ry*sin(-50)});

  \node[facext,label={[font=\small]left:$\widetilde{\cO}^{(j)}_g$}] (ol) at (OgL) {};
  \node[facext,label={[font=\small]right:$\cO^{(j)}_g$}] (or) at (OgR) {};
  \node[facext,label={[font=\small]below left:$\cO_P$}] (p1) at (P1) {};
  \node[facext,label={[font=\small]below right:$\bar\cO_P$}] (p2) at (P2) {};

  \node[facvertex] (vt) at (-0.75,0.55) {};
  \node[facvertex] (vb) at (0.15,-0.95) {};
  \node[font=\small,fill=white,inner sep=1pt] at (-1.20,0.33) {$z_{\rm ws}$};
  \node[font=\small,fill=white,inner sep=1pt] at (0.45,-0.72) {$z$};

  \draw[facstring]
    ({\Rx*cos(145)},{\Ry*sin(145)})
    .. controls (-2.05,1.55) and (-1.30,0.95) .. (vt)
    .. controls (-0.25,1.00) and (1.95,1.55) .. ({\Rx*cos(35)},{\Ry*sin(35)});
  \node[font=\small,fill=white,inner sep=1pt] at (-0.35,1.38) {$\Sigma_{\rm st}$};

  \draw[facnorm] (ol) -- (vt)
    node[midway,above,sloped,font=\small,fill=white,inner sep=1pt]
    {$\Psi_j^{\rm bdry}(z_{\rm ws})$};

  \draw[facbb] (or) -- (vb)
    node[midway,right,font=\small,fill=white,inner sep=1pt]
    {$\mathcal H_j^{\rm SW}$};

  \draw[facproj] (ol)
    .. controls (-1.45,-0.10) and (1.35,-0.10) ..
    node[pos=0.54,below,font=\small,fill=white,inner sep=1pt]
    {$\hat d_j$}
    (or);

  \draw[fachadron] (p1) -- (vb);
  \draw[fachadron] (vb) -- (p2);

  \node[font=\small] at (-2.25,2.05) {(a)};
  \node[font=\small] at (0,2.90) {UV endpoint};
  \node[font=\small] at (0,-2.90)
    {$\mathcal K_j^{\rm UV}\,\widetilde F_j^{g,{\rm bdry}}$};
\end{tikzpicture}
\end{minipage}
\hfill
\begin{minipage}{0.32\textwidth}
\centering
\begin{tikzpicture}[scale=0.70]
  \def\Rx{3.15}
  \def\Ry{2.55}
  \draw[facboundary] (0,0) ellipse [x radius=\Rx cm,y radius=\Ry cm];

  \coordinate (OgL) at ({\Rx*cos(170)},{\Ry*sin(170)});
  \coordinate (OgR) at ({\Rx*cos(20)},{\Ry*sin(20)});
  \coordinate (P1)  at ({\Rx*cos(-130)},{\Ry*sin(-130)});
  \coordinate (P2)  at ({\Rx*cos(-50)},{\Ry*sin(-50)});

  \node[facext,label={[font=\small]left:$\widetilde{\cO}^{(j)}_g$}] (ol) at (OgL) {};
  \node[facext,label={[font=\small]right:$\cO^{(j)}_g$}] (or) at (OgR) {};
  \node[facext,label={[font=\small]below left:$\cO_P$}] (p1) at (P1) {};
  \node[facext,label={[font=\small]below right:$\bar\cO_P$}] (p2) at (P2) {};

  \node[facvertex] (vt) at (-0.20,0.75) {};
  \node[facvertex] (vb) at (0.10,-1.05) {};
  \node[font=\small,fill=white,inner sep=1pt] at (-0.90,0.62) {$z_{\rm ws}$};
  \node[font=\small,fill=white,inner sep=1pt] at (0.48,-0.82) {$z$};

  \draw[facstring]
    ({\Rx*cos(145)},{\Ry*sin(145)})
    .. controls (-2.05,1.55) and (-1.30,1.05) .. (vt)
    .. controls (0.65,1.05) and (1.95,1.55) .. ({\Rx*cos(35)},{\Ry*sin(35)});
  \node[font=\small,fill=white,inner sep=1pt] at (0.15,1.42) {$\Sigma_{\rm st}$};

  \draw[facbb] (or) -- (vt)
    node[midway,above,sloped,font=\small,fill=white,inner sep=1pt]
    {$\mathcal H_j^{\rm SW}$};

  \draw[facnorm] (ol) -- (vb)
    node[midway,left,font=\small,fill=white,inner sep=1pt]
    {$\Psi_j^{\rm bdry}(z)$};

  \draw[facproj] (ol)
    .. controls (-1.45,-0.12) and (1.35,-0.12) ..
    node[pos=0.54,below,font=\small,fill=white,inner sep=1pt]
    {$\hat d_j$}
    (or);

  \draw[fachadron] (p1) -- (vb);
  \draw[fachadron] (vb) -- (p2);

  \node[font=\small] at (-2.25,2.05) {(b)};
  \node[font=\small] at (0,2.90) {IR soft wall};
  \node[font=\small] at (0,-2.90)
    {$\mathcal K_j^{\rm IR,SW}\,\widehat{\mathcal T}^{(c)}_j$};
\end{tikzpicture}
\end{minipage}
\hfill
\begin{minipage}{0.32\textwidth}
\centering
\begin{tikzpicture}[scale=0.70]
  \def\Rx{3.15}
  \def\Ry{2.55}
  \draw[facboundary] (0,0) ellipse [x radius=\Rx cm,y radius=\Ry cm];

  \coordinate (OgL) at ({\Rx*cos(170)},{\Ry*sin(170)});
  \coordinate (OgR) at ({\Rx*cos(20)},{\Ry*sin(20)});
  \coordinate (P1)  at ({\Rx*cos(-130)},{\Ry*sin(-130)});
  \coordinate (P2)  at ({\Rx*cos(-50)},{\Ry*sin(-50)});

  \node[facext,label={[font=\small]left:$\widetilde{\cO}^{(j)}_g$}] (ol) at (OgL) {};
  \node[facext,label={[font=\small]right:$\cO^{(j)}_g$}] (or) at (OgR) {};
  \node[facext,label={[font=\small]below left:$\cO_P$}] (p1) at (P1) {};
  \node[facext,label={[font=\small]below right:$\bar\cO_P$}] (p2) at (P2) {};

  \node[facvertex] (vt) at (-0.20,0.75) {};
  \node[facvertex] (vb) at (0.10,-1.05) {};
  \node[font=\small,fill=white,inner sep=1pt] at (-0.90,0.62) {$z_{\rm ws}$};
  \node[font=\small,fill=white,inner sep=1pt] at (0.48,-0.82) {$z$};

  \draw[facstring]
    ({\Rx*cos(145)},{\Ry*sin(145)})
    .. controls (-2.05,1.55) and (-1.30,1.05) .. (vt)
    .. controls (0.65,1.05) and (1.95,1.55) .. ({\Rx*cos(35)},{\Ry*sin(35)});
  \node[font=\small,fill=white,inner sep=1pt] at (0.15,1.42) {$\Sigma_{\rm st}$};

  \draw[facbb] (or) -- (vt)
    node[midway,above,sloped,font=\small,fill=white,inner sep=1pt]
    {$\mathcal H_j^{\rm HW}$};

  \draw[facnorm] (ol) -- (vb)
    node[midway,left,font=\small,fill=white,inner sep=1pt]
    {$\Psi_j^{\rm bdry}(z)$};

  \draw[facproj] (ol)
    .. controls (-1.45,-0.12) and (1.35,-0.12) ..
    node[pos=0.54,below,font=\small,fill=white,inner sep=1pt]
    {$\hat d_j$}
    (or);

  \draw[fachadron] (p1) -- (vb);
  \draw[fachadron] (vb) -- (p2);

  \node[font=\small] at (-2.25,2.05) {(c)};
  \node[font=\small] at (0,2.90) {IR HW completion};
  \node[font=\small] at (0,-2.90)
    {$\mathcal K_j^{\rm IR,HW}\,\widehat{\mathcal T}^{(c)}_j$};
\end{tikzpicture}
\end{minipage}
\caption{
Endpoint-factorized fixed-\(j\) Witten diagrams for GTMD conformal
moments.  Dotted lines denote normalized boundary modes
\(\Psi_j^{(c),{\rm bdry}}\).  Double-wiggly lines denote
bulk-to-boundary transfer kernels \(\mathcal H_j^{(c)}\).
The symbol $\hat d_j$ denotes the fixed-spin conformal projector $\tilde{\cal O}_g^{(j)}{\hat d_j}{\cal O}_g^{(j)}$ that
selects the spin-$j$ component of the boundary gluonic bilocal operator.
Since it carries no radial dynamics and can be absorbed in $g_j(\lambda)$, it is suppressed in the subsequent
endpoint-factorized representations.}
\label{fig:secV_endpoint_factorized_witten_gtmd}
\end{figure*}

\subsection{Endpoint factorization of the propagator}

The endpoint reductions follow from a simple property of the
bulk-to-bulk propagator.  When one endpoint of the propagator approaches
the AdS boundary, the propagator factorizes into a normalized
near-boundary mode attached to the shallow endpoint and a
bulk-to-boundary transfer kernel from the boundary to the deeper endpoint.
For the spin-\(j\) exchange considered here, the normalized boundary mode
is
\begin{equation}
\Psi_j^{(c),{\rm bdry}}(z;\epsilon)
=
-
\bigg[
\frac{
\left(\sqrt{2}\kappa_c\epsilon\right)^{4-\Delta_c(j)}
}{
\Delta_c(j)
}
\bigg]
\left(\sqrt{2}\kappa_c z\right)^{\Delta_c(j)} .
\label{eq:secV_boundary_mode}
\end{equation}
The soft-wall bulk-to-boundary kernel scales as
\begin{equation}
\mathcal H_j^{(c),{\rm SW}}(K,z;\epsilon)
=
\left(\sqrt{2}\kappa_c\epsilon\right)^{\Delta_c(j)-4}
\widehat{\mathcal H}_j^{(c),{\rm SW}}(K,z),
\label{eq:secV_HSW_scaling}
\end{equation}
where
\begin{widetext}
\begin{equation}
\widehat{\mathcal H}_j^{(c),{\rm SW}}(K,z)
=
\frac{
\Gamma(A_{j,t})
}{
\Gamma(\Delta_c(j)-2)
}
\left(\sqrt{2}\kappa_c z\right)^{\Delta_c(j)}
U\!\left(
A_{j,t},B_j;2\kappa_c^2 z^2
\right),
\label{eq:secV_HSW_finite}
\end{equation}
\end{widetext}
with
\begin{align}
A_{j,t}
&=
\frac{a_{t,c}}{2}
+
\frac{\Delta_c(j)}{2}
=
1+\frac{j+\gamma_c(j)+a_{t,c}}{2},
\nonumber\\
B_j&=\Delta_c(j)-1=1+j+\gamma_c(j).
\label{eq:secV_A_B}
\end{align}
The radial dependence is governed by the Tricomi confluent
hypergeometric function \(U(A,B;x)\), which is the solution of the
confluent hypergeometric equation that decays for large positive
argument.  This choice ensures the correct infrared behavior of the
transfer kernel in the soft-wall background.  Near the boundary one should
not keep only the regular branch of \(U\).  For
\(B_j=\Delta_c(j)-1>1\), the Tricomi function contains the
non-normalizable small-\(x\) branch
\begin{equation}
U(A,B;x)
\sim
\frac{\Gamma(B-1)}{\Gamma(A)}x^{1-B}+\cdots,
\qquad x\to0,
\label{eq:secV_tricomi_small_x_branch}
\end{equation}
so \[(\sqrt2\kappa_c z)^{\Delta_c(j)}U(A,\Delta_c(j)-1;2\kappa_c^2z^2)\]
contains the cutoff-normalized boundary-source behavior before
multiplication by the prefactor in Eq.~\eqref{eq:secV_HSW_scaling}.  The
normalizable near-boundary power that enters the endpoint splitting is the
separate factor \(\Psi_j^{(c),{\rm bdry}}\sim z^{\Delta_c(j)}\).
The cutoff power in the boundary mode is \(4-\Delta_c(j)\), the inverse of
the transfer-kernel cutoff power \(\Delta_c(j)-4\) in
Eq.~\eqref{eq:secV_HSW_scaling}.  The factors therefore cancel in the product
\(\Psi_j^{(c),{\rm bdry}}\mathcal H_j^{(c)}\), leaving finite endpoint
kernels.  The overall sign in Eq.~\eqref{eq:secV_boundary_mode} follows the
decay-constant convention and is absorbed into the endpoint-kernel
normalization.

There are two possible endpoint orderings.  In the ultraviolet endpoint,
\(b_T\to0^+\), the worldsheet profile satisfies
\(z_{\rm ws}(u,v;b)\ll z\).  The point closest to the boundary is then the
worldsheet insertion, and the propagator reduces to
\begin{equation}
G_c(j,z,z_{\rm ws};t)
\simeq
\Psi_j^{(c),{\rm bdry}}(z_{\rm ws};\epsilon)\,
\mathcal H_j^{(c),{\rm SW}}(K,z;\epsilon).
\label{eq:secV_UV_factorized_propagator}
\end{equation}
This is the graphical decomposition shown in
Fig.~\ref{fig:secV_endpoint_factorized_witten_gtmd}(a).  Since the
bulk-to-boundary kernel remains on the hadron side, the \(z\)-integral is
the same one that appears in the strict boundary moment
\(\widetilde F_{j}^{g,{\rm bdry}}(\xi,t)\).  The endpoint dependence on \(b_T\) is
therefore entirely carried by the worldsheet factor multiplying that
boundary moment.

In the infrared endpoint the ordering is reversed
\begin{equation}
z\ll z_{\rm ws}(u,v;b).
\end{equation}
The hadron vertex is now the shallower point.  The normalized
near-boundary mode attaches to the target side, while the transfer kernel
is evaluated at the worldsheet point:
\begin{equation}
G_c(j,z,z_{\rm ws};t)
\simeq
\Psi_j^{(c),{\rm bdry}}(z;\epsilon)\,
\mathcal H_j^{(c)}(K,z_{\rm ws};\epsilon).
\label{eq:secV_IR_factorized_propagator}
\end{equation}
This is the decomposition shown in
Fig.~\ref{fig:secV_endpoint_factorized_witten_gtmd}(b,c).  Because the
near-boundary mode is now attached to the target, the target integral no
longer gives the full conformal moment.  It gives instead the target
projection
\begin{equation}
\widehat{\mathcal T}^{(c)}_j(\xi)
=
\int_0^\infty dz\,
\rho_j(z;\xi)
\left(\sqrt{2}\kappa_c z\right)^{\Delta_c(j)} .
\label{eq:secV_target_projection}
\end{equation}
The remaining \(b_T\)-dependence is determined by the transfer kernel
on the worldsheet side.  Choosing the soft-wall transfer gives the
algebraic infrared tail, while replacing the deep-infrared transfer by
the gap-matched hard-wall kernel gives the confining exponential tail.

Thus the splitting of the bulk-to-bulk propagator is not an additional
dynamical assumption.  It is the standard endpoint expansion of a Witten
diagram.  The side closer to the boundary supplies the normalized
boundary mode; the opposite side is probed by the corresponding
bulk-to-boundary transfer kernel.  This is why the UV endpoint reduces to
\(\mathcal K_j^{\rm UV}\widetilde F_{j}^{g,{\rm bdry}}(\xi,t)\), whereas the two IR
endpoints reduce to
\(\mathcal K_j^{\rm IR}\widehat{\mathcal T}^{(c)}_j(\xi)\).

The three endpoint factorizations are therefore
\begin{widetext}
\begin{align}
{\rm UV:}\qquad
G_c(j,z,z_{\rm ws};t)\big|_{z_{\rm ws}\ll z}
&\simeq
\Psi_j^{(c),{\rm bdry}}(z_{\rm ws};\epsilon)\,
\mathcal H_j^{(c),{\rm SW}}(K,z;\epsilon),
\nonumber\\
{\rm IR\;SW:}\qquad
G_c(j,z,z_{\rm ws};t)\big|_{z\ll z_{\rm ws}}
&\simeq
\Psi_j^{(c),{\rm bdry}}(z;\epsilon)\,
\mathcal H_j^{(c),{\rm SW}}(K,z_{\rm ws};\epsilon),
\nonumber\\
{\rm IR\;HW:}\qquad
G_c(j,z,z_{\rm ws};t)\big|_{z\ll z_{\rm ws}}
&\simeq
\Psi_j^{(c),{\rm bdry}}(z;\epsilon)\,
\mathcal H_j^{(c),{\rm HW}}(M_{\rm gap},z_{\rm ws};\epsilon).
\label{eq:secV_endpoint_factorizations}
\end{align}
\end{widetext}

\subsection{Ultraviolet endpoint}

We now apply the first endpoint ordering in
Eq.~\eqref{eq:secV_endpoint_factorizations}.  In the limit
\(b_T\to0^+\), the near-boundary worldsheet lies entirely in the region
\begin{equation}
z_{\rm ws}(u,v;b)=b\,z_b(u,v)\ll z,
%\qquad
%z_b(u,v)=\frac{1}{\cosh u\,\cosh v}.
\end{equation}
The bulk-to-bulk propagator then factorizes with the normalized boundary
mode attached to the worldsheet side:
\begin{equation}
G_c(j,z,z_{\rm ws};t)
\simeq
\Psi_j^{(c),{\rm bdry}}(z_{\rm ws};\epsilon)\,
\mathcal H_j^{(c),{\rm SW}}(K,z;\epsilon).
\end{equation}
Substituting this into Eq.~\eqref{eq:secV_ws_witten}, the \(z\)-integral
is precisely the one defining the boundary conformal moment.  The only
new factor is the worldsheet integral.  Its power follows from combining
the fixed-spin worldsheet insertion with the near-boundary behavior of
the exchanged mode:
\begin{equation}
\left(\sqrt{2}\kappa_c z_{\rm ws}\right)^{-(j-2)}
\left(\sqrt{2}\kappa_c z_{\rm ws}\right)^{\Delta_c(j)}
\propto
z_{\rm ws}^{\,\Delta_c(j)-(j-2)} .
\end{equation}
Using
\begin{equation}
\Delta_c(j)=2+j+\gamma_c(j),
\qquad
\alpha_j\equiv \Delta_c(j)-(j-2)=4+\gamma_c(j),
\end{equation}
one obtains the ultraviolet power
\begin{equation}
z_{\rm ws}^{\,\alpha_j}
=
b^{\alpha_j} z_b(u,v)^{\alpha_j}.
\label{eq:secV_UV_power_counting}
\end{equation}

Equivalently, using the soft-wall spectral decomposition of the boundary
moment,
\begin{equation}
\widetilde F_{j}^{g,{\rm bdry}}(\xi,t)
=
\sum_{n=0}^{\infty}
\widetilde F_{j}^{g,{\rm bdry}}(\xi,t;n),
\end{equation}
the finite-\(b_T\) moment becomes
\begin{equation}
\widetilde F_{j}^{g,{\rm ws}}(\xi,t,b_T)
=
\sum_{n=0}^{\infty}
\mathcal K^{\rm UV}_j(\lambda,b_T;n)\,
\widetilde F_{j}^{g,{\rm bdry}}(\xi,t;n),
\label{eq:secV_UV_mode_sum}
\end{equation}
where
\begin{align}
\mathcal K^{\rm UV}_j(\lambda,b_T;n)
&=
-
\frac{
\widetilde g_5^{\,2}g_j(\lambda)
}{
\Delta_c(j)
}
\bar b^{\,\alpha_j}
\mathcal I^{(4),{\rm UV}}_{j,n}(\bar b),
\nonumber\\
\bar b&=\sqrt{2}\kappa_c b .
\label{eq:secV_UV_mode_kernel}
\end{align}
and
\begin{align}
\mathcal I^{(4),{\rm UV}}_{j,n}(\bar b)
&=
\int_{-\infty}^{\infty}du
\int_{-\infty}^{\infty}dv\,
z_b(u,v)^{\alpha_j}
\nonumber\\
&\hspace{0.7cm}\times
\frac{
L_n^{\Delta_c(j)-2}\!\left(\bar b^{\,2}z_b(u,v)^2\right)
}{
L_n^{\Delta_c(j)-2}(0)
}.
\label{eq:secV_UV_Ijn}
\end{align}
For \(\bar b\ll1\), the Laguerre ratio is
\(1+\mathcal O(\bar b^2)\), so the leading kernel is independent of the
radial mode \(n\).  The remaining integral is purely geometric
\begin{align}
\mathcal I^{(4)}_j
&=
\int_{-\infty}^{\infty}du
\int_{-\infty}^{\infty}dv\,
z_b(u,v)^{\alpha_j}
\nonumber\\
&=
\pi
\left[
\frac{
\Gamma\!\left(\frac{\alpha_j}{2}\right)
}{
\Gamma\!\left(\frac{\alpha_j+1}{2}\right)
}
\right]^2,
\qquad
\Re\alpha_j>0 .
\label{eq:secV_UV_geometric_integral}
\end{align}
Therefore all radial modes acquire the same leading worldsheet factor, and the
mode sum reconstructs the boundary moment
\begin{equation}
\widetilde F_{j}^{g,{\rm ws}}(\xi,t,b_T)
\xrightarrow[b_T\to0^+]{}
\mathcal K^{\rm UV}_j(\lambda,b_T)\,
\widetilde F_{j}^{g,{\rm bdry}}(\xi,t),
\label{eq:secV_UV_reduction}
\end{equation}
with
\begin{widetext}
\begin{equation}
\mathcal K^{\rm UV}_j(\lambda,b_T)
=
-
\widetilde g_5^{\,2}g_j(\lambda)
\frac{\pi}{\Delta_c(j)}
\left[
\frac{
\Gamma\!\left(\frac{\Delta_c(j)-(j-2)}{2}\right)
}{
\Gamma\!\left(\frac{\Delta_c(j)-(j-2)+1}{2}\right)
}
\right]^2
\bar b^{\,\Delta_c(j)-(j-2)} .
\label{eq:secV_UV_kernel_exact}
\end{equation}
\end{widetext}
In the compact form,
\begin{equation}
\mathcal K^{\rm UV}_j(\lambda,b_T)
=
\phi^{\rm Wst}_0(j,\lambda)\,
\bar b^{\,4+\gamma_c(j)},
\label{eq:secV_UV_kernel_phi}
\end{equation}
where
\begin{equation}
\phi^{\rm Wst}_0(j,\lambda)
=
-
\widetilde g_5^{\,2}g_j(\lambda)
\frac{\pi}{\Delta_c(j)}
\left[
\frac{
\Gamma\!\left(2+\frac{\gamma_c(j)}{2}\right)
}{
\Gamma\!\left(\frac{5+\gamma_c(j)}{2}\right)
}
\right]^2 .
\label{eq:secV_phi_Wst}
\end{equation}

As a useful normalization check, consider the graviton channel \(j=2\).
For the protected stress-tensor trajectory one has \(\gamma_c(2)=0\), so
\(\alpha_2=4\).  The worldsheet source then carries no longitudinal-derivative
factor, \(z_{\rm ws}^{-(j-2)}=1\), while the normalizable spin-two wave
contributes \(z_{\rm ws}^{\Delta_c(2)}=z_{\rm ws}^4\).  Hence the finite-
separation saddle gives \(\mathcal K_2^{\rm UV}\propto \bar b^4\), exactly as
Eq.~\eqref{eq:secV_UV_kernel_phi} states.  This check also makes explicit why
adding an extra \(z_{\rm ws}^{2(j-2)}\) factor to the present current
normalization would spoil the endpoint power counting.

This result also clarifies why the strict point \(b_T=0\) and the
limit \(b_T\to0^+\) should not be identified.  Equation
\eqref{eq:secV_UV_reduction} is the contribution of the
finite-separation classical worldsheet saddle only; it is not the full
operator-product limit of the physical, rapidity-renormalized GTMD.  The
usual local OPE is represented holographically by a separate boundary or
contact Witten diagram, which defines the strictly local moment,
\begin{equation}
\widetilde F_j^{g,{\rm full}}(\xi,t,0)=\widetilde F_j^{g,{\rm bdry}}(\xi,t).
\end{equation}
For \(\Re[4+\gamma_c(j)]>0\), the finite-separation saddle alone obeys
\begin{equation}
\lim_{b_T\to0^+}
\widetilde F_{j}^{g,{\rm ws}}(\xi,t,b_T)=0 .
\end{equation}
This vanishing is therefore a statement about one holographic saddle
sector and not a claim that the physical GTMD or its collinear OPE
vanishes at \(b_T=0\).

\subsection{Cusp-subtracted infrared ordering}
\label{subsec:cusp_subtracted_ir_ordering}

The infrared endpoint requires one additional qualification.  The
four-cusp profile satisfies \(z_{\rm ws}=b\,z_b(u,v)\) with
\(z_b(u,v)\to0\) in the cusp regions \(|u|,|v|\to\infty\).  Hence the
ordering \(z\ll z_{\rm ws}\) is not uniform over the full
\((u,v)\)-plane, even when \(b\to\infty\).  The large-argument
asymptotics of \(U(A,B;x)\), \(K_\nu(x)\), or the repulsive-wall transfer
kernel should therefore not be inserted under the unsubtracted full
worldsheet integral.

In the following infrared formulas the perimeter and cusp regions are
understood to be subtracted into the soft factor \(S\), and the remaining
stripped worldsheet source is restricted to the central region in which
\(z_{\rm ws}\) is of order \(b\).  Equivalently, one may introduce a
factorization scale \(z_\star\) and split the worldsheet into
\(z_{\rm ws}>z_\star\) and \(z_{\rm ws}<z_\star\).  The shallow cusp sector
renormalizes the Wilson-line soft factor or matches onto the ultraviolet
boundary sector, while the central sector gives the infrared transfer
kernels displayed below.  The large-\(b_T\) endpoint statement is
therefore a cusp-subtracted, finite-depth saddle statement.  The constants
that appear in the infrared kernels, such as \(\mathcal C^{\rm IR,SW}_{j,t}\),
\(\widehat z_{\rm IR}\), \(\kappa_{\rm IR}\), \(\delta_j\), and
\(\sigma_j^{\rm RW}\), are model- and saddle-dependent data.  The infrared
formulas below are conditional endpoint asymptotics of the chosen confining
completion, not universal predictions of strong coupling alone.

\subsection{Soft-wall infrared transfer endpoint}

We next take the opposite ordering,
\begin{equation}
z\ll z_{\rm ws}(u,v;b).
\end{equation}
Now the hadron vertex lies closer to the boundary than the worldsheet
insertion.  The normalized boundary mode is therefore attached to the
target side:
\begin{equation}
G_c(j,z,z_{\rm ws};t)
\simeq
\Psi_j^{(c),{\rm bdry}}(z;\epsilon)\,
\mathcal H_j^{(c),{\rm SW}}(K,z_{\rm ws};\epsilon).
\label{eq:secV_IR_SW_endpoint}
\end{equation}
Substituting this into Eq.~\eqref{eq:secV_ws_witten}, the target integral
is no longer the boundary conformal moment.  It is the projection of the
target density onto the near-boundary spin-\(j\) mode,
\begin{equation}
\widehat{\mathcal T}^{(c)}_j(\xi)
=
\int_0^\infty dz\,
\rho_j(z;\xi)
\left(\sqrt{2}\kappa_c z\right)^{\Delta_c(j)} .
\label{eq:secV_target_projection_ir_sw}
\end{equation}
The remaining worldsheet integral defines the soft-wall infrared kernel:
\begin{equation}
\widetilde F_{j,{\rm IR},{\rm SW}}^{g,{\rm ws}}(\xi,t,b_T)
=
\mathcal K^{\rm IR,SW}_j(\lambda,b_T,t)\,
\widehat{\mathcal T}^{(c)}_j(\xi).
\label{eq:secV_IR_SW_factorization}
\end{equation}
Explicitly,
\begin{align}
\mathcal K^{\rm IR,SW}_j(\lambda,b_T,t)
&=
-
\frac{
\widetilde g_5^{\,2}g_j(\lambda)
}{
\Delta_c(j)
}
\int du\,dv\,
\left(\sqrt{2}\kappa_c z_{\rm ws}\right)^{-(j-2)}
\nonumber\\
&\quad\times
\widehat{\mathcal H}_j^{(c),{\rm SW}}(K,z_{\rm ws}).
\label{eq:secV_IR_SW_kernel_master}
\end{align}
Using
\(z_{\rm ws}=b\,\widehat z(u,v;b)\) and the soft-wall transfer function
\eqref{eq:secV_HSW_finite}, one obtains
\begin{align}
\mathcal K^{\rm IR,SW}_j(\lambda,b_T,t)
&=
-
\widetilde g_5^{\,2}g_j(\lambda)
\frac{
\Gamma(A_{j,t})
}{
\Delta_c(j)\,
\Gamma(\Delta_c(j)-2)
}
\bar b^{\,\alpha_j}
\nonumber\\
&\hspace{0.7cm}\times
\mathcal I^{(4),{\rm IR,SW}}_j(\bar b,t),
\label{eq:secV_IR_SW_kernel_exact}
\end{align}
with
\begin{align}
\mathcal I^{(4),{\rm IR,SW}}_j(\bar b,t)
&=
\int_{-\infty}^{\infty}du
\int_{-\infty}^{\infty}dv\,
\widehat z(u,v;b)^{\alpha_j}
\nonumber\\
&\hspace{0.7cm}\times
U\!\left(
A_{j,t},B_j;
\bar b^{\,2}\widehat z(u,v;b)^2
\right).
\label{eq:secV_IR_SW_integral}
\end{align}

The large-\(b_T\) behavior follows from the large-argument asymptotics
of the Tricomi function,
\begin{equation}
U(A,B;x)
\sim
x^{-A}
\left[
1+\mathcal O(x^{-1})
\right],
\qquad x\to\infty .
\end{equation}
With \(x=\bar b^2\widehat z^2\), the combination appearing in the
worldsheet integral behaves as
\begin{align}
\bar b^{\,\alpha_j}\widehat z^{\,\alpha_j}
U\!\left(A_{j,t},B_j;\bar b^2\widehat z^2\right)
&\sim
\bar b^{\,\alpha_j-2A_{j,t}}
\widehat z^{\,\alpha_j-2A_{j,t}}
\nonumber\\
&=
\bar b^{\,2-j-a_{t,c}}
\widehat z^{\,2-j-a_{t,c}} .
\label{eq:secV_IR_SW_asymptotic_power}
\end{align}
The anomalous dimension cancels from the leading infrared power.  It
remains only in the overall spin-dependent coefficient.  After the cusp/perimeter regions have been subtracted into the soft factor and the remaining deep
worldsheet integral is dominated by a finite infrared saddle, define
\begin{equation}
\mathcal C^{\rm IR,SW}_{j,t}
=
\int du\,dv\,
\widehat z(u,v;b)^{2-j-a_{t,c}},
\label{eq:secV_IR_SW_Cjt}
\end{equation}
evaluated in that saddle approximation.  Then
\begin{align}
\mathcal K^{\rm IR,SW}_j(\lambda,b_T,t)
&\sim
-
\widetilde g_5^{\,2}g_j(\lambda)
\frac{
\Gamma\!\left(1+\frac{j+\gamma_c(j)+a_{t,c}}{2}\right)
}{
\left(2+j+\gamma_c(j)\right)
\Gamma\!\left(j+\gamma_c(j)\right)
}
\nonumber\\
&\quad\times
\mathcal C^{\rm IR,SW}_{j,t}\,
\bar b^{\,2-j-a_{t,c}}
\left[
1+\mathcal O(\bar b^{-2})
\right].
\label{eq:secV_IR_SW_large_b}
\end{align}
and we obtain
\begin{align}
\widetilde F_{j,{\rm IR},{\rm SW}}^{g,{\rm ws}}(\xi,t,b_T)
&\xrightarrow[b_T\to\infty]{}
\mathcal K^{\rm IR,SW}_j(\lambda,b_T,t)\,
\widehat{\mathcal T}^{(c)}_j(\xi),
\nonumber\\
\mathcal K^{\rm IR,SW}_j
&\sim
C_j(t)\,\bar b^{\,2-j-a_{t,c}} .
\label{eq:secV_IR_SW_reduction}
\end{align}
The soft-wall infrared endpoint is therefore algebraic.  This reflects
the fact that the soft-wall transfer function suppresses the deep bulk
only through the power-law tail of \(U(A,B;x)\).  This algebraic tail is a
model artifact of using the soft-wall transfer deep in the infrared; it is
not advertised here as a universal confining prediction.  The hard-wall and
repulsive-wall infrared completions below are the confining alternatives used to model
a finite transverse correlation length.

\subsection{Gap-matched hard-wall infrared completion}

The preceding endpoint uses the soft-wall transfer all the way into the
deep worldsheet region and therefore gives an algebraic tail.  To
represent a confining transverse mass gap, we keep the same endpoint
ordering and the same target projection
\(\widehat{\mathcal T}^{(c)}_j(\xi)\), but replace only the
worldsheet-side transfer kernel by a hard-wall, gap-matched transfer:
\begin{equation}
G_c(j,z,z_{\rm ws};t)
\simeq
\Psi_j^{(c),{\rm bdry}}(z;\epsilon)\,
\mathcal H_j^{(c),{\rm HW}}(M_{\rm gap},z_{\rm ws};\epsilon).
\label{eq:secV_IR_HW_endpoint}
\end{equation}
The finite hard-wall transfer is
\begin{align}
\widehat{\mathcal H}^{(c),{\rm HW}}_j(K,z)
&=
\frac{
2^{1-\nu_j}
}{
\Gamma(\nu_j)
}
\left(
\frac{K}{\sqrt{2}\kappa_c}
\right)^{\nu_j}
\left(\sqrt{2}\kappa_c z\right)^2
\nonumber\\
&\hspace{0.7cm}\times
\left[
\mathcal R_j(K,z_0)I_{\nu_j}(Kz)
+
K_{\nu_j}(Kz)
\right],
\label{eq:secV_HW_transfer}
\end{align}
where
\begin{equation}
\nu_j=\Delta_c(j)-2=j+\gamma_c(j),
\end{equation}
and
\begin{equation}
\mathcal R_j(K,z_0)
=
-
\frac{
\left.
\partial_z
\left[
z^2K_{\nu_j}(Kz)
\right]
\right|_{z=z_0}
}{
\left.
\partial_z
\left[
z^2I_{\nu_j}(Kz)
\right]
\right|_{z=z_0}
}.
\label{eq:secV_HW_reflection}
\end{equation}
For a large wall position, \(Kz_0\gg1\),
\begin{equation}
\mathcal R_j(K,z_0)
\sim
\frac{K_{\nu_j}(Kz_0)}{I_{\nu_j}(Kz_0)}
\sim
\pi e^{-2Kz_0}.
\end{equation}
Hence the reflected \(I_{\nu_j}\) branch is exponentially suppressed at
the worldsheet point whenever
\begin{equation}
Kz_0\gg1,
\qquad
K\left(z_0-z_{\rm ws}^{\rm IR}\right)\gg1 .
\end{equation}
The infrared transfer is then dominated by the decaying branch.  We first
write the deep-infrared transfer with a general transverse mass gap
\(M_{\rm gap}\),
\begin{equation}
\eta_{\rm gap}\equiv\frac{M_{\rm gap}}{\sqrt{2}\kappa_c},
\end{equation}
so that
\begin{align}
\widehat{\mathcal H}^{(c),{\rm HW}}_j(M_{\rm gap},z_{\rm ws})
&\longrightarrow
\widehat{\mathcal H}^{(c),{\rm HW,gap}}_j(M_{\rm gap};z_{\rm ws})
\nonumber\\
&=
\frac{2^{1-\nu_j}}{\Gamma(\nu_j)}
\eta_{\rm gap}^{\nu_j}
\left(\sqrt{2}\kappa_c z_{\rm ws}\right)^2
\nonumber\\
&\quad\times
K_{\nu_j}\!\left(M_{\rm gap} z_{\rm ws}\right).
\label{eq:secV_HW_gap_transfer}
\end{align}
The frequently used choice \(M_{\rm gap}=\sqrt{2}\kappa_c\) is the
single-parameter specialization \(\eta_{\rm gap}=1\); it is a model
matching choice, not an independent derivation of the mass gap.  The
gap-matched hard-wall completion should therefore be read as a phenomenological
infrared completion of the transfer kernel, not as a consequence of the soft-wall
saddle itself.

The gap-matched hard-wall completion therefore factorizes as
\begin{equation}
\widetilde F_{j,{\rm IR},{\rm HW}}^{g,{\rm ws}}(\xi,t,b_T)
=
\mathcal K^{\rm IR,HW,gap}_j(\lambda,b_T)\,
\widehat{\mathcal T}^{(c)}_j(\xi),
\label{eq:secV_IR_HW_gap_factorization}
\end{equation}
with
\begin{align}
\mathcal K^{\rm IR,HW,gap}_j(\lambda,b_T)
&=
-
\frac{
\widetilde g_5^{\,2}g_j(\lambda)
}{
\Delta_c(j)
}
\frac{
2^{1-\nu_j}
}{
\Gamma(\nu_j)
}
\eta_{\rm gap}^{\nu_j}
\bar b^{\,4-j}
\nonumber\\
&\hspace{0.7cm}\times
\mathcal I^{(4),{\rm HW,gap}}_j(\bar b),
\label{eq:secV_IR_HW_gap_kernel}
\end{align}
where
\begin{align}
\mathcal I^{(4),{\rm HW,gap}}_j(\bar b)
&=
\int du\,dv\,
\widehat z(u,v;b)^{4-j}
\nonumber\\
&\quad\times
K_{\nu_j}\!\left(\eta_{\rm gap}\bar b\,\widehat z(u,v;b)\right).
\label{eq:secV_IR_HW_gap_integral}
\end{align}
Equivalently, relative to the soft-wall transfer, the deep-infrared
replacement is
\begin{align}
&\left(\sqrt{2}\kappa_c z_{\rm ws}\right)^{\Delta_c(j)}
U\!\left(A_{j,t},B_j;2\kappa_c^2z_{\rm ws}^2\right)
\nonumber\\
&\hspace{1.0cm}\longrightarrow
\left(\sqrt{2}\kappa_c z_{\rm ws}\right)^2
K_{\nu_j}\!\left(M_{\rm gap} z_{\rm ws}\right).
\label{eq:secV_SW_to_HW_replacement}
\end{align}

Using
\begin{equation}
K_{\nu_j}(x)
=
\sqrt{\frac{\pi}{2x}}\,
e^{-x}
\left[
1+\mathcal O(x^{-1})
\right],
\qquad x\to\infty ,
\end{equation}
and writing the infrared worldsheet saddle as
\begin{equation}
z_{\rm ws}^{\rm IR}(b)
=
b\,\widehat z_{\rm IR},
\qquad
\widehat z_{\rm IR}=\mathcal O(1),
\end{equation}
one finds
\begin{align}
\mathcal I^{(4),{\rm HW,gap}}_j(\bar b)
&\sim
(\eta_{\rm gap}\bar b)^{-1/2}
\widehat z_{\rm IR}^{\,\frac{7}{2}-j}
 e^{-\eta_{\rm gap}\bar b\,\widehat z_{\rm IR}}
\nonumber\\
&\quad\times
\left[
1+\mathcal O(\bar b^{-1})
\right].
\end{align}
Consequently
\begin{align}
\mathcal K^{\rm IR,HW,gap}_j(\lambda,b_T)
&\sim
\mathcal N^{\rm IR,HW}_j(\lambda)\,
\eta_{\rm gap}^{\nu_j-1/2}
\bar b^{\,\frac{7}{2}-j}
 e^{-\eta_{\rm gap}\bar b\,\widehat z_{\rm IR}}
\nonumber\\
&\quad\times
\left[
1+\mathcal O(\bar b^{-1})
\right].
\label{eq:secV_HW_large_b_bar}
\end{align}
The displayed normalization keeps the explicit gap factor.  In compact
phenomenological forms this factor may equivalently be absorbed into
\(\mathcal N^{\rm IR,HW}_j\) or into the fitted infrared scale.
Defining
\begin{equation}
\kappa_{\rm IR}
=
M_{\rm gap}\,\widehat z_{\rm IR}
=
\eta_{\rm gap}\sqrt{2}\kappa_c\,\widehat z_{\rm IR},
\end{equation}
this becomes
\begin{align}
\mathcal K^{\rm IR,HW,gap}_j(\lambda,b_T)
&\sim
\widetilde{\mathcal N}^{\rm IR,HW}_j(\lambda)\,
(\kappa_{\rm IR}b)^{\frac{7}{2}-j}
e^{-\kappa_{\rm IR}b}
\nonumber\\
&\quad\times
\left[
1+\mathcal O((\kappa_{\rm IR}b)^{-1})
\right].
\label{eq:secV_HW_large_b_kappa}
\end{align}
For the specialization \(M_{\rm gap}=\sqrt{2}\kappa_c\) and \(\widehat z_{\rm IR}=1\), or if the product \(\eta_{\rm gap}\widehat z_{\rm IR}\) is absorbed into the fitted infrared scale, this is
\begin{equation}
\mathcal K^{\rm IR,HW,gap}_j(\lambda,b_T)
\sim
\bar b^{\,\frac{7}{2}-j}e^{-\bar b}.
\label{eq:secV_HW_Eq32_form}
\end{equation}
The gap-matched hard-wall completion therefore replaces the soft-wall algebraic
tail by the gap-induced confining exponential falloff.

\subsection{Leading-saddle endpoint reductions}

The endpoint limits may be summarized compactly as follows, with common
arguments suppressed only in the kernel labels:
\begin{align}
{\rm UV:}\quad
\widetilde F_{j}^{g,{\rm ws}}
&\xrightarrow[b_T\to0^+]{}
\mathcal K^{\rm UV}_j\,
\widetilde F_{j}^{g,{\rm bdry}},
\nonumber\\
\mathcal K^{\rm UV}_j
&=
\phi^{\rm Wst}_0(j,\lambda)\,
\bar b^{\,4+\gamma_c(j)},
\nonumber\\[2pt]
{\rm IR\;SW:}\quad
\widetilde F_{j,{\rm IR},{\rm SW}}^{g,{\rm ws}}
&\xrightarrow[b_T\to\infty]{}
\mathcal K^{\rm IR,SW}_j\,
\widehat{\mathcal T}^{(c)}_j,
\nonumber\\
\mathcal K^{\rm IR,SW}_j
&\sim
C_j(t)\,\bar b^{\,2-j-a_{t,c}},
\nonumber\\[2pt]
{\rm IR\;HW:}\quad
\widetilde F_{j,{\rm IR},{\rm HW}}^{g,{\rm ws}}
&\xrightarrow[b_T\to\infty]{}
\mathcal K^{\rm IR,HW,gap}_j\,
\widehat{\mathcal T}^{(c)}_j,
\nonumber\\
\mathcal K^{\rm IR,HW,gap}_j
&\sim
\widetilde C_j^{\rm HW}
(\kappa_{\rm IR}b)^{\frac{7}{2}-j}
e^{-\kappa_{\rm IR}b},
\nonumber\\[2pt]
{\rm IR\;RW:}\quad
\widetilde F_{j,{\rm IR},{\rm RW}}^{g,{\rm ws}}
&\xrightarrow[b_T\to\infty]{}
\mathcal K^{\rm IR,RW}_j\,
\widehat{\mathcal T}^{(c)}_j,
\nonumber\\
\mathcal K^{\rm IR,RW}_j
&\sim
\bar b^{\,\delta_j}\,
e^{-\sigma^{\rm RW}_j{\bar b}^2}.
\label{eq:secV_endpoint_reduction_theorem}
\end{align}
The distinction between the four cases is entirely controlled by
which endpoint of the bulk-to-bulk propagator lies closest to the
boundary and by the infrared realization of the exchanged channel.
In the ultraviolet endpoint, the worldsheet insertion is the
near-boundary endpoint, so the endpoint kernel multiplies the full
boundary conformal moment. In the infrared endpoints, the target side
is the near-boundary endpoint, so the target collapses to
$\widehat{\mathcal T}^{(c)}_j(\xi)$, while the worldsheet-side transfer
determines the large-$b_T$ behavior. The soft-wall transfer yields
an algebraic tail. The gap-matched hard-wall completion yields a
confining exponential falloff. The repulsive-wall completion yields
a Gaussian suppression generated by the repulsive-wall Schr\"odinger
potential, with the detailed coefficients $\sigma^{\rm RW}_j$ and $\delta_j$
derived in Appendix~\ref{app:RW_GTMD_endpoint}. Unlike the hard-wall
result, the Gaussian behavior does not arise from a lowest-mass pole
approximation, but from the large-distance asymptotics of the
repulsive-wall transfer kernel itself.  All three infrared expressions
are conditional endpoint asymptotics: the constants
\(C^{\rm IR}_{j,t}\), \(\widehat z_{\rm IR}\), \(\kappa_{\rm IR}\),
\(\sigma_j^{\rm RW}\), and \(\delta_j\) are saddle and model data, not
universal strong-coupling predictions.

\section{Analytic Continuation and the Low-\texorpdfstring{$x$}{x} Regime}
\label{SEC_VI}

The small-\(x\) regime is obtained by analytically continuing the
fixed-spin conformal moments from even integer spin \(j\ge2\) to complex
\(j\), followed by inverse Mellin transformation in Bjorken \(x\)
\cite{Mueller:2005ed,Brower:2006ea,Costa:2012cb,Brower:2010wf,
Brower:2013}.  No new nonperturbative input is introduced at this stage;
Reggeization is a reorganization of the fixed-spin amplitudes derived
above.

This strong-coupling construction is complementary to perturbative
small-\(x\) and Color Glass Condensate approaches to gluon GTMDs and
Wigner distributions
\cite{Balitsky:1995ub,Kovchegov:1999yj,Iancu:2003xm,Weigert:2005us,
Dominguez:2011wm,Hatta:2016dxp,Hagiwara:2016kam,Zhou:2016rnt,
Benic:2026gtmd}.  Here the high-energy behavior is determined by the
analytic structure of the holographic conformal moments in the complex
\(j\)-plane.

At finite transverse separation the worldsheet sector factorizes as
\begin{equation}
F_j^{g,{\rm ws}}(\xi,t,b_T;\mu,\zeta)
=
S(b_T;\mu,\zeta)\,
\widetilde F_j^{g,{\rm ws}}(\xi,t,b_T;\mu),
\label{eq:secVI_soft_factorization}
\end{equation}
where \(S\) is analytic in \(j\).  The nontrivial analytic continuation
is therefore carried by the stripped amplitude \(\widetilde F_j^{g,{\rm ws}}\).
The rightmost singularity is the BPST branch point
\cite{Brower:2006ea,Brower:2010wf,Costa:2012cb,Brower:2013},
\begin{equation}
j_0^g=2-\frac{2}{\sqrt\lambda},
\label{eq:secVI_intercept}
\end{equation}
with local parameterization
\begin{equation}
\Delta_c=2\pm i\varpi,
\qquad
j(\varpi)=j_0^g-\frac{\varpi^2}{2\sqrt\lambda},
\qquad
\varpi\ge0 .
\label{eq:secVI_BPST_parameterization}
\end{equation}

In the DGLAP region \(\xi\simeq0\), the inverse Mellin transform of the
worldsheet sector is
\begin{equation}
F_g^{\rm ws}(x,t,b_T)
=
\frac{1}{2\pi i}
\int_{c-i\infty}^{c+i\infty}
dj\,x^{-(j-1)}F_j^{g,{\rm ws}}(0,t,b_T),
\label{eq:secVI_Mellin}
\end{equation}
where the contour lies to the right of all singularities.  Deforming the
contour onto the BPST cut gives
\begin{align}
F_g(x,t,b_T)
&=
x^{1-j_0^g}
\frac{1}{2\pi i}
\int_0^\infty d\varpi\,
\frac{\varpi}{\sqrt\lambda}
\nonumber\\
&\quad\times
\exp\!\left[-\frac{\varpi^2}{2\sqrt\lambda}\,Y\right]
\operatorname{Disc}_{\varpi}
\Big[F^{g,{\rm ws}}_{j(\varpi)}(0,t,b_T)\Big],
\label{eq:secVI_cut_representation}
\end{align}
with \(Y\equiv\ln\frac1x .\)
The diffusion approximation is the regime
\begin{equation}
Y\gg1,
\qquad
\frac{\ell_b^2}{Y}=\text{fixed},
\qquad
\ell_b\equiv \ln\frac1{\bar b},
\qquad
\bar b=\sqrt2\kappa_c b_T .
\label{eq:secVI_diffusion_regime}
\end{equation}

For \(0<\bar b\ll1\), the fixed-spin ultraviolet endpoint gives
\begin{align}
\widetilde F_j^{g,{\rm ws}}(0,t,b_T)
&\simeq
\mathcal K_j^{\rm UV}(\lambda,b_T)\,
\widetilde F_j^{g,{\rm bdry}}(0,t),
\nonumber\\
\mathcal K_j^{\rm UV}
&\sim
\bar b^{\Delta_c(j)-(j-2)} .
\label{eq:secVI_UV_kernel}
\end{align}
Near the cut we split
\begin{align}
\bar b^{\Delta_c(j)-(j-2)}
&=
\bar b^{4-j}\,\bar b^{\Delta_c(j)-2}
\nonumber\\
&=
\bar b^{4-j_0^g}
\left[1+\mathcal O\!\left(\frac{\varpi^2}{\sqrt\lambda}\ln\bar b\right)\right]
\bar b^{\Delta_c-2} .
\label{eq:secVI_power_split}
\end{align}
The soft-wall transfer kernel also contains the normalization
\(1/\Gamma(\Delta_c-2)\).  We therefore write, on the two sides of the
cut,
\begin{equation}
\widetilde F_{j(\varpi)}^{g,{\rm bdry}}(0,t)
\Big|_{\Delta_c=2\pm i\varpi}
=
\frac{1}{\Gamma(\pm i\varpi)}\,
\widehat{\widetilde F}_{j(\varpi)}^{g,{\rm bdry}}(0,t),
\label{eq:secVI_hatted_moment}
\end{equation}
where \(\widehat{\widetilde F}_{j}^{g,{\rm bdry}}\equiv
\Gamma(\Delta_c(j)-2)\widetilde F_j^{g,{\rm bdry}}\) is smooth at
\(j=j_0^g\).  The branch-cut dependence is then
\begin{align}
\operatorname{Disc}_{\varpi}
\left[
\frac{\bar b^{\Delta_c-2}}{\Gamma(\Delta_c-2)}
\right]
&=
\frac{e^{-i\varpi\ell_b}}{\Gamma(i\varpi)}
-
\frac{e^{+i\varpi\ell_b}}{\Gamma(-i\varpi)}
\nonumber\\
&\simeq
2i\varpi\cos(\varpi\ell_b)
+\mathcal O(\varpi^2),
\label{eq:secVI_disc_cos}
\end{align}
where \(1/\Gamma(\pm i\varpi)=\pm i\varpi+\mathcal O(\varpi^2)\) has
been used, while the phase \(\varpi\ell_b\) is kept unexpanded.

Substitution into Eq.~\eqref{eq:secVI_cut_representation} leaves the
explicit Gaussian integral
\begin{align}
I(Y,\ell_b)
&\equiv
\frac{1}{\pi\sqrt\lambda}
\int_0^\infty d\varpi\,\varpi^2
\exp\!\left[-\frac{Y}{2\sqrt\lambda}\varpi^2\right]
\cos(\varpi\ell_b)
\nonumber\\
&=
\frac{\lambda^{1/4}}{\sqrt{2\pi}\,Y^{3/2}}
\left(1-\sqrt\lambda\frac{\ell_b^2}{Y}\right)
\exp\!\left[-\frac{\sqrt\lambda}{2}\frac{\ell_b^2}{Y}\right].
\label{eq:secVI_gaussian_integral}
\end{align}
Thus, for \(\xi\simeq0\), \(x\ll1\), and \(\bar b\ll1\), the
finite-separation ultraviolet contribution to the small-\(x\) GTMD is
\begin{widetext}
\begin{align}
F_g(x,t,b_T)
\simeq
\mathcal N_g^{\rm UV}(t)
\,x^{-(j_0^g-1)}
\,S(b_T;\mu,\zeta)
\,\bar b^{4-j_0^g}
\times
\frac{\lambda^{1/4}}{\sqrt{2\pi}\,Y^{3/2}}
\left(1-\sqrt\lambda\frac{\ell_b^2}{Y}\right)
\times
\exp\!\left[-\frac{\sqrt\lambda}{2}\frac{\ell_b^2}{Y}\right].
\label{eq:secVI_final_smallx}
\end{align}
\end{widetext}
The polynomial prefactor in Eq.~\eqref{eq:secVI_final_smallx} is part of the
leading BPST branch-cut diffusion approximation.  It should not be
interpreted as a positivity statement or as a sign prediction outside the
intended hierarchy
\begin{align}
Y&\gg1,
\qquad
0<\bar b\ll1,
\nonumber\\
\varpi&\sim\left(\frac{\sqrt\lambda}{Y}\right)^{1/2}\ll1,
\qquad
\left|\frac{\varpi^2}{\sqrt\lambda}\ln\bar b\right|\ll1 .
\label{eq:secVI_regge_window}
\end{align}
with \(\ell_b^2/Y\) in the BPST diffusion domain and not so large that
subleading Regge-cut terms or nonleading impact-factor corrections compete
with the displayed leading approximation.

The smooth coefficient is
\begin{equation}
\mathcal N_g^{\rm UV}(t)
\propto
\phi_0^{\rm Wst}(j_0^g,\lambda)
\widehat{\widetilde F}_{j_0^g}^{\;g,{\rm bdry}}(0,t),
\label{eq:secVI_Nuv}
\end{equation}
up to a smooth convention-dependent normalization.  Equation
\eqref{eq:secVI_final_smallx} displays the BPST power
\(x^{-(j_0^g-1)}\), diffusion in \(\ell_b\), and the analytic worldsheet
soft factor.  At fixed \(Y\), \(\bar b\to0^+\) sends
\(\ell_b\to\infty\), so the Gaussian suppresses the finite-separation
worldsheet saddle.  This is not the collinear point; the strict
\(b_T=0\) value is supplied by the separate boundary Witten diagram.

We now compare this strong-coupling construction with perturbative GTMD
factorization.

\section{Comparison with perturbative GTMD factorization}
\label{SEC_VII}

It is useful to compare the leading-saddle holographic construction with
the perturbative GTMD factorization and matching program.  In perturbative
QCD, staple-linked GTMDs require rapidity renormalization and soft-factor
subtraction.  The subtracted object is scheme dependent: one may absorb a
square root of the soft factor into each matrix element, divide an
unsubtracted correlator by a full soft factor, or use an equivalent
rapidity-renormalized definition.  Recent one-loop matching calculations
express GTMDs in the perturbative small-\(b_T\), or large-\(k_T\), region in
terms of GPDs \cite{Bertone:2022nxa,Bertone:2025gtmd}.  Schematically,
\begin{widetext}
\begin{equation}
F^{[Y]}_{i/H}(x,\xi,b_T,t;\mu,\zeta)
=
\sum_{j,\Gamma}
C^{[Y/\Gamma]}_{i/j}(x,\xi,b_T;\mu,\zeta)
\otimes
F^{[\Gamma]}_{j/H}(x,\xi,t;\mu)
+
\mathcal O(b_T\Lambda_{\rm QCD}),
\label{eq:secVII_pqcd_matching_bspace}
\end{equation}
\end{widetext}
where \(Y\) labels the GTMD polarization channel and \(\Gamma\) labels the
GPD channel.  Beyond the simplest unpolarized channel, GTMD and GPD
polarizations can mix under matching and evolution \cite{Bertone:2025gtmd}.

In momentum space the one-loop matching coefficients generate the standard
perturbative radiative tail.  For the unpolarized channel this has the
schematic large-\(k_T\) form
\begin{widetext}
\begin{equation}
F^g(x,\xi,k_T,t;\mu,\zeta)
\sim
\frac{\alpha_s}{\pi}\frac{1}{k_T^2}
\left[
C_{g/g}\otimes H^g+C_{g/q}\otimes H^q+\cdots
\right](x,\xi,t;\mu),
\qquad
k_T\gg\Lambda_{\rm QCD},
\label{eq:secVII_pqcd_tail}
\end{equation}
\end{widetext}
up to logarithms and tensor structures.  Thus perturbative small-\(b_T\)
matching is a weak-coupling region factorization, established order by
order in \(\alpha_s\).

The holographic result has a different status and a different
normalization convention.  It uses the piecewise saddle decomposition of
Eq.~\eqref{eq:intro_piecewise_factorization}; the shorthand
\[F_j^g=F_{j,\bdry}^g+F_{j,\ws}^g+\cdots\] means only a decomposition into
distinct saddle sectors.  Only the finite-separation sector obeys the
worldsheet soft factorization
\begin{equation}
F_{j,\ws}^g
=
S\,\widetilde F_j^{g,{\rm ws}},
\qquad b_T>0 .
\label{eq:secVII_holographic_factorization}
\end{equation}
A standard rapidity-renormalized convention is obtained by the explicit
redistribution
\begin{equation}
F_{j,\ws}^{g,{\rm sub}[\alpha]}
=
S^{-\alpha}F_{j,\ws}^{g,{\rm unsub}}
=
S^{1-\alpha}\widetilde F_j^{g,{\rm ws}},
\label{eq:secVII_subtracted_map}
\end{equation}
where \(\alpha\) specifies the subtraction convention.  In a full-soft
convention \(\alpha=1\), the subtracted finite-separation worldsheet object
coincides with \(\widetilde F_j^{g,{\rm ws}}\); in a square-root convention
\(\alpha=1/2\), a residual \(S^{1/2}\) remains in a single matrix element.
The factorization follows from the classical worldsheet saddle and bulk
locality, with corrections of order \(1/\sqrt\lambda\) and \(1/N_c^2\), not
from a perturbative separation of soft, collinear, and hard momentum
regions.

The small-transverse-separation limits should therefore be compared with
care.  In perturbative QCD, the fully subtracted small-\(b_T\) GTMD has an
operator-product expansion onto GPDs with perturbative coefficient
functions.  In the present strong-coupling calculation, the
finite-separation worldsheet saddle instead has the ultraviolet overlap
\begin{widetext}
\begin{equation}
\widetilde F_j^{g,{\rm ws}}(\xi,t,b_T)
\xrightarrow[b_T\to0^+]{}
\mathcal K_j^{\rm UV}(b_T)\,
\widetilde F_j^{g,{\rm bdry}}(\xi,t),
\qquad
\mathcal K_j^{\rm UV}\sim\bar b^{4+\gamma_c(j)} .
\label{eq:secVII_holo_uv}
\end{equation}
\end{widetext}
For \(\Re[4+\gamma_c(j)]>0\), this finite-separation worldsheet
contribution vanishes as \(b_T\to0^+\).  The strict point is instead
\begin{equation}
F_j^g(\xi,t,0)=F_{j,\bdry}^g(\xi,t)
=\widetilde F_j^{g,{\rm bdry}}(\xi,t),
\label{eq:secVII_strict_point}
\end{equation}
up to subleading saddle sectors.  Thus there is no contradiction with the
perturbative OPE or with the collinear identity: the limit \(b_T\to0^+\) of
the worldsheet saddle and the value at \(b_T=0\) are different holographic
saddle sectors.

The comparison can be summarized as follows.  Perturbative GTMD
factorization describes the weak-coupling short-distance expansion of a
subtracted GTMD and gives coefficient functions multiplying GPDs.  The
holographic construction describes the leading strong-coupling
finite-separation worldsheet contribution and separates the universal
Wilson-line area from a target Witten diagram.  Both frameworks identify
GPD/conformal moments as the collinear objects controlling the
short-distance behavior, but they organize the soft factor and the
\(b_T\to0\) limit differently.

\section{Conclusions}
\label{SEC_VIII}

We have developed a fixed-spin strong-coupling formulation of unpolarized
gluon GTMD conformal moments using gauge/string duality.  For each even spin
$j$, the Wilson-line geometry is represented by a classical staple
worldsheet, while the target structure is carried by bulk spin-$j$ exchange.
The resulting leading-saddle factorization is
\begin{equation}
\begin{aligned}
F_j^g(\xi,t,0;\mu,\zeta)
&=
\widetilde F_j^{g,{\rm bdry}}(\xi,t;\mu),
\\[3pt]
F_j^g(\xi,t,b_T;\mu,\zeta)
&=
S(b_T;\mu,\zeta)\,
\widetilde F_j^{g,{\rm ws}}(\xi,t,b_T;\mu)
\\[-1pt]
&\quad+
\mathcal O(N_c^{-2},\lambda^{-1/2}),
\qquad b_T>0 .
\end{aligned}
\label{eq:conclusion_piecewise_factorization}
\end{equation}
The first line is the strict local boundary sector, and the second line is
the finite-separation worldsheet sector.  The soft factor is universal and
contains the Wilson-line vacuum area; the stripped amplitude contains the
target-dependent Witten diagram.

The transverse endpoint behavior follows from the radial ordering of the
bulk-to-bulk propagator.  In the ultraviolet endpoint, the worldsheet
insertion is the shallow endpoint and the stripped amplitude reduces to a
universal overlap kernel multiplying the boundary conformal moment.  In the
infrared endpoint, the target side is the shallow endpoint and the result
reduces to a target projection multiplied by a transfer kernel.  The
soft-wall transfer gives an algebraic tail, the gap-matched hard-wall
completion gives an exponential tail with scale \(\kappa_{\rm IR}\), and the
repulsive-wall completion gives a Gaussian tail.  The ultraviolet kernel is
the universal saddle prediction; the infrared tails are data of the chosen
confining completion.

Analytic continuation in $j$ gives the low-$x$ Regge regime governed by the
BPST spectral curve.  This continuation reorganizes the same fixed-spin
amplitudes and produces the strong-coupling diffusion in rapidity and
logarithmic transverse separation.  No extra nonperturbative input is needed
beyond the fixed-spin Witten-diagram data.

For phenomenology, the construction provides a conformal-moment-space
architecture for holographic-string GTMD parametrizations.  Existing
string-exchange GPD conformal moments \cite{Mamo:2024vjh,Mamo:2024jwp} can
be dressed by the transverse kernels derived here and then transformed back
to $x$-space through an inverse Mellin-Barnes integral.  This supplies the
analytic transverse dressing and Regge organization needed for applications
to gluon-GTMD observables, including exclusive double quarkonium, exclusive
$\pi^0$, and exclusive heavy vector or axial-vector meson production
\cite{Bhattacharya:2018lgm,Bhattacharya:2024pi0,Bhattacharya:2026qnd}.

The framework is ready for extensions to polarized distributions, quark
GTMDs, generalized Wigner distributions, and alternative confining
holographic backgrounds, as well as for comparisons with perturbative QCD in
kinematic regimes where both descriptions can be meaningfully contrasted.

\section*{Acknowledgments}
K.M. thanks Jefferson Lab for hospitality during the completion of this work.  K.M. is  supported by the DOE Grant  DE-FG02-04ER41309 and  NSF Grant  2412625.  I.Z. is supported by the DOE grant DE-FG-88ER40388. Both are partially supported by the DOE Quark-Gluon Tomography (QGT) Topical Collaboration, Award DE-SC0023646.

\section*{Data Availability}
No data were created or analyzed in this article.  The results are analytic and are fully specified by the equations and model assumptions stated in the text.

\appendix

\section{Conventions and Kinematics}

Light-front coordinates are defined as
\begin{equation}
x^\pm=\frac{x^0\pm x^3}{\sqrt{2}},
\qquad
x_\perp=(x^1,x^2).
\end{equation}
The average hadron momentum is $\bar P=(P+P')/2$ and the momentum transfer is $\Delta=P'-P$, with invariant momentum transfer $t=\Delta^2$.
The skewness parameter is defined by
\begin{equation}
\xi=-\frac{\Delta^+}{2\bar P^+}.
\end{equation}
Fourier transforms in transverse space follow the convention
\begin{equation}
\int\frac{d^2k_\perp}{(2\pi)^2}
e^{i\bm k_\perp\cdot\bm b_T}
=
\int_0^\infty\frac{k_\perp dk_\perp}{2\pi}
J_0(k_\perp b_T).
\end{equation}

%\begin{widetext}

\section{Finite-Skewness Conformal Moments and the Mellin Limit}
\label{app:conformal_mellin_limit}

This appendix records the finite-skewness moment convention used in the
main text.  At finite skewness, the diagonal conformal moments of gluon
GPDs and GTMDs are Gegenbauer moments, not ordinary Mellin moments.  We
use the gluon Gegenbauer basis with index \(5/2\), as in
Ref.~\cite{Mamo:2024vjh}.  The orthogonality relation is
\begin{equation}
\int_{-1}^{1}dy\,(1-y^2)^2
C_n^{5/2}(y)C_m^{5/2}(y)
=
h_n^{(5/2)}\delta_{nm},
\end{equation}
where
\begin{equation}
h_n^{(5/2)}
=
\frac{\pi\,2^{-4}\Gamma(n+5)}
{n!\,(n+5/2)\Gamma(5/2)^2} .
\end{equation}
For the spin convention of the present paper the gluon twist-two operator
contains \(j-2\) light-cone derivatives, so the Gegenbauer index is
\(n=j-2\).  The corresponding normalized conformal moment is
\begin{widetext}
\begin{equation}
\mathbb F_j^g(\xi,t,b_T)
=
\frac{\Gamma(5/2)\Gamma(j-1)}
{2^{j-2}\Gamma(j+1/2)}
\int_{-1}^{1}dx\,
\xi^{\,j-2}
C_{j-2}^{5/2}\!\left(\frac{x}{\xi}\right)
F_g(x,\xi,t,b_T),
\qquad j=2,4,\ldots .
\label{eq:app_exact_gluon_conformal_moment}
\end{equation}
\end{widetext}
The prefactor is chosen so that the highest-power term of the
Gegenbauer polynomial has unit Mellin normalization.  Indeed,
\begin{equation}
C_{j-2}^{5/2}(y)
=
\frac{2^{j-2}\Gamma(j+1/2)}
{\Gamma(5/2)\Gamma(j-1)}y^{j-2}
+\text{lower powers},
\end{equation}
and therefore
\begin{equation}
\lim_{\xi\to0}
\frac{\Gamma(5/2)\Gamma(j-1)}
{2^{j-2}\Gamma(j+1/2)}
\xi^{\,j-2}
C_{j-2}^{5/2}\!\left(\frac{x}{\xi}\right)
=
x^{j-2}.
\end{equation}
Consequently,
\begin{equation}
\mathbb F_j^g(0,t,b_T)
=
\int_{-1}^{1}dx\,x^{j-2}F_g(x,0,t,b_T).
\label{eq:app_conformal_to_mellin}
\end{equation}
This is the precise sense in which the paper uses the phrase
conformal/Mellin moment.  The holographic fixed-spin exchange computes
the conformal moment at finite \(\xi\), and its \(\xi\to0\) limit is the
Mellin moment used for the inverse Mellin transform in
Sec.~\ref{SEC_VI}.

\section{Spin-\texorpdfstring{\(j\)}{j} Worldsheet Vertex, Coupling Normalization, and Radial Power Counting}
\label{app:spinj_vertex_power}

This appendix fixes the radial power convention for the spin-\(j\) source
on the open string worldsheet and records the normalization of the
spin-\(j\) coupling used in the main text.  The two points are logically
separate.  The factor \(g_j(\lambda)\) is a dimensionless string-tension
coupling.  It carries powers of \(\alpha'^{-1}\) and of the reference AdS
or soft-wall scale, but it carries no dependence on \(z_{\rm ws}\) or on
\(b_T\).  The entire small-\(b_T\) power comes instead from the explicit
radial factor in the worldsheet source and from the near-boundary wave of
the exchanged spin-\(j\) field.

The starting point is the Polyakov coupling of a fundamental string to a
metric perturbation,
\begin{equation}
\delta S_{\rm P}
=
\frac{1}{4\pi\alpha'}
\int d^2\sigma\sqrt h\,h^{ab}
\partial_aX^M\partial_bX^N\,\delta G_{MN}(X).
\label{eq:app_graviton_polyakov}
\end{equation}
This is the \(j=2\) graviton vertex.  A higher-spin field on the leading
closed-string trajectory is represented by a symmetric spin-\(j\) bulk
field contracted with a worldsheet current built from tangent vectors,
\begin{widetext}
\begin{equation}
V_j[C]
=
\sqrt{2\kappa_5^2}\,g_j(\lambda)(\sqrt2\kappa_c)^j
\int_{\Sigma_C} d^2\sigma\,
\mathcal J_{\rm ws}^{M_1\cdots M_j}(\sigma)
\,z^{-(j-2)}
K^{(j)}_{M_1\cdots M_j;\nu_1\cdots\nu_j}(X(\sigma);a).
\label{eq:app_current_vertex}
\end{equation}
\end{widetext}
Here \(\mathcal J_{\rm ws}^{M_1\cdots M_j}\) is the renormalized
worldsheet current, including the induced measure and local vielbein or
Jacobian factors appropriate to the closed-channel BPST normalization.
In conformal coordinates on the near-boundary four-cusp solution,
\begin{equation}
z_{\rm ws}(u,v;b)=b\,z_b(u,v),
\qquad
z_b(u,v)=\frac{1}{\cosh u\cosh v},
\end{equation}
the induced worldsheet metric is related to the unit-size solution by a
scale transformation.  Therefore the measure-current combination in
Eq.~\eqref{eq:app_current_vertex} is a dimensionless function of \(u,v\)
after renormalization.  The change of variables \(z_{\rm ws}=b z_b\) does
not produce an additional power of \(b\).  This is why the source in
Eq.~\eqref{eq:secIV_operator_source} may be written with \(du\,dv\) and
with the explicit radial factor only.

The dimensionless coupling \(g_j(\lambda)\) follows from the powers of the
string tension in the spin-\(j\) vertex.  In an unrescaled AdS convention,
\begin{equation}
\frac{R^2}{\alpha'}=\sqrt\lambda,
\end{equation}
and a dimensionless spin-\(j\) closed-string coupling can be parameterized
as
\begin{equation}
g_j(\lambda)
=
c_j\left(\frac{R^2}{\alpha'}\right)^{j/2}
=
c_j\lambda^{j/4},
\label{eq:app_gj_unrescaled}
\end{equation}
where \(c_j\) is a convention-dependent numerical normalization fixed by
matching the two-point function of the spin-\(j\) operator.  In the
rescaled soft-wall coordinates used,
\begin{equation}
R_{\rm eff}=\frac{1}{\sqrt2\kappa_c},
\qquad
\sqrt\lambda=\frac{R_{\rm eff}^2}{\alpha'}=
\frac{1}{2\kappa_c^2\alpha'},
\end{equation}
so the same statement is equivalently written as
\begin{equation}
g_j(\lambda)
=
c_j\left(\frac{1}{\alpha'\kappa_c^2}\right)^{j/2}
=
c_j\left(2\sqrt\lambda\right)^{j/2}.
\label{eq:app_gj_rescaled}
\end{equation}
The factor \(2^{j/2}\) and the constant \(c_j\) are convention dependent
and may be absorbed into the definition of the renormalized current or of
\(\widetilde g_5^{\,2}g_j\).  Crucially, Eqs.~\eqref{eq:app_gj_unrescaled}
and \eqref{eq:app_gj_rescaled} contain only \(\lambda\) and fixed model
scales; they contain no \(z_{\rm ws}\) and no \(b_T\).  Hence \(g_j\) cannot
alter the transverse endpoint power.

The near-boundary radial power is now fixed unambiguously.  Contracting
the vertex with a normalizable spin-\(j\) wave gives
\begin{equation}
z_{\rm ws}^{-(j-2)}\psi_j(z_{\rm ws})
\sim
z_{\rm ws}^{\Delta_c(j)-(j-2)}
=
z_{\rm ws}^{4+\gamma_c(j)}.
\label{eq:app_current_power}
\end{equation}
Using \(z_{\rm ws}=b z_b(u,v)\), the worldsheet integral therefore gives
\begin{equation}
\int du\,dv\,z_b(u,v)^{4+\gamma_c(j)}\,b^{4+\gamma_c(j)},
\end{equation}
up to a \(b\)-independent constant.  This proves that the ultraviolet
kernel is governed by
\begin{equation}
\mathcal K_j^{\rm UV}
\sim
\bar b^{\Delta_c(j)-(j-2)}
=
\bar b^{4+\gamma_c(j)}.
\end{equation}
A formula with an extra \(z_{\rm ws}^{2(j-2)}\) corresponds to a different
boundary-frame normalization of the worldsheet current.  Combining that
factor with the present BPST propagator convention would double count the
vielbein conversion.

\section{Fixed-spin Witten diagrams and soft-wall endpoint kernels}
\label{APP_WITTEN_UV_KERNEL}

This appendix collects the technical ingredients underlying the
fixed-spin Witten-diagram representation derived in
Sec.~\ref{SEC_V}.  The leading-saddle endpoint reduction formulas established in the
main text follow from a simple property of holographic propagation:
when one endpoint of the exchanged bulk propagator approaches the AdS
boundary, the propagator factorizes into a universal near-boundary
wavefunction and a bulk transfer kernel.  This factorization isolates
the universal Wilson-line geometry from the target-dependent radial
structure and makes the endpoint limits analytically tractable.

Throughout this appendix the universal worldsheet soft factor
associated with the vacuum area of the staple contour has been stripped
off.  If the complete GTMD matrix element is desired, every expression
below should be multiplied by
\begin{equation}
S(b_T;\mu,\zeta),
\end{equation}
which contains the full rapidity dependence but no information on the
hadronic target.

We use
\begin{equation}
b\equiv |\bm b_T|,
\qquad
\bar b\equiv \sqrt{2}\,\kappa_c b,
\qquad
\bar\epsilon_c\equiv \kappa_c\epsilon ,
\end{equation}
where \(\kappa_c\) denotes the soft-wall confinement scale and
\(\epsilon\) is the ultraviolet radial cutoff.  Momentum transfer is
parameterized through
\begin{equation}
K^2=-t,
\qquad
a_{t,c}\equiv -\frac{t}{4\kappa_c^2}.
\label{eq:app_unified_defs}
\end{equation}

The exchanged object is a closed-channel spin-\(j\) bulk field dual to
the twist-two gluon operator of conformal spin \(j\).  Its scaling
dimension is
\begin{equation}
\Delta_c(j)
=
2+j+\gamma_c(j),
\end{equation}
where \(\gamma_c(j)\) is the corresponding strong-coupling anomalous
dimension.
Two combinations appear repeatedly
\begin{equation}
\nu_j
= \Delta_c(j)-2
= j+\gamma_c(j),
\end{equation}
and
\begin{equation}
\alpha_j
=\Delta_c(j)-(j-2)
=4+\gamma_c(j).
\label{eq:app_unified_alpha_nu_def}
\end{equation}
The parameter \(\nu_j\) controls the radial behavior of the exchanged
spin-\(j\) mode.  The combination \(\alpha_j\) governs the ultraviolet
endpoint behavior of the finite-separation GTMD.  Physically,
\(\alpha_j\) arises because the Wilson-line insertion contributes the
projection factor
\((z_{\rm ws})^{-(j-2)}\),
whereas the near-boundary spin-(j) wavefunction contributes
\((z_{\rm ws})^{\Delta_c(j)}\).  Their product yields the universal
endpoint power
\begin{equation}
(z_{\rm ws})^{\alpha_j}
=
(z_{\rm ws})^{4+\gamma_c(j)}.
\end{equation}
The endpoint reductions derived in Sec.~\ref{SEC_V} therefore depend
only on the universal near-boundary structure of the exchanged
spin-(j) field.  Once this structure is extracted, all dependence on
the hadronic target enters through radial overlap integrals.

\subsection{Closed spin-\texorpdfstring{\(j\)}{j} soft-wall channel}

The soft-wall model provides a convenient realization of fixed-spin
exchange in a confining holographic background.  The exchanged
spin-(j) field propagates through a discrete tower of normalizable
modes whose masses increase linearly with the radial excitation number.
These modes furnish a complete basis for the bulk propagator and allow
both bulk-to-bulk and bulk-to-boundary propagation to be represented
spectrally.

The normalizable closed-channel soft-wall modes are
\begin{equation}
\psi_n^{(c)}(j,z)
=
c_n^{(c)}(j)\,
z^{\Delta_c(j)}
L_n^{\Delta_c(j)-2}
\!\left(2\kappa_c^2z^2\right),
\label{eq:app_unified_sw_modes}
\end{equation}
with normalization
\begin{equation}
c_n^{(c)}(j)
=
\left[
\frac{
2^{\Delta_c(j)+1}
\kappa_c^{2\Delta_c(j)}
\Gamma(n+1)
}{
\Gamma\!\left(
n+\Delta_c(j)-1
\right)
}
\right]^{1/2}.
\label{eq:app_unified_sw_mode_norm}
\end{equation}
These modes satisfy
\begin{equation}
\frac{1}{2\kappa_c^2}
\int_0^\infty dz\,
\sqrt g\,
e^{-2\kappa_c^2z^2}
|g^{xx}|
\psi_n^{(c)}
\psi_m^{(c)}
=
\delta_{nm},
\label{eq:app_unified_sw_orthonormality}
\end{equation}
and possess masses
\begin{equation}
M_{c,n}^2(j)
=
8\kappa_c^2
\left(
n+\frac{\Delta_c(j)}{2}
\right).
\label{eq:app_unified_sw_masses}
\end{equation}

The fixed-spin bulk-to-bulk propagator therefore admits the spectral
representation
\begin{equation}
G_c(j,z,z';t)
=
2\kappa_c^2
\sum_{n=0}^{\infty}
\frac{
\psi_n^{(c)}(j,z)
\psi_n^{(c)}(j,z')
}{
-t+M_{c,n}^2(j)
}.
\label{eq:app_unified_sw_bulk_bulk}
\end{equation}
The corresponding decay constants are defined through the ultraviolet
behavior of the normalizable modes,
\begin{equation}
\mathcal F_n^{(c)}(j,\epsilon)
=
\frac{1}{2\kappa_c^2}
\frac{1}{\widetilde g_5}
\left[
-\sqrt g\,
e^{-2\kappa_c^2z^2}
|g^{xx}|
\partial_z
\psi_n^{(c)}(j,z)
\right]_{z=\epsilon}.
\label{eq:app_unified_sw_decay_constants}
\end{equation}
Using these decay constants, the bulk-to-boundary transfer kernel may
be written in spectral form as
\begin{widetext}
\begin{align}
\mathcal H_j^{(c),{\rm SW}}
(K,z;\epsilon)
=
\sum_{n=0}^{\infty}
\frac{
\widetilde g_5
\mathcal F_n^{(c)}(j,\epsilon)
\psi_n^{(c)}(j,z)
}{
-t+M_{c,n}^2(j)
}
=
\mathcal N_c
(j,\bar K_c,\bar\epsilon_c)
\left(
\sqrt2\kappa_c z
\right)^{\Delta_c(j)}
U\!\left(
A_{j,t},
B_j;
2\kappa_c^2z^2
\right),
\label{eq:app_unified_sw_bulk_boundary}
\end{align}
\end{widetext}
where \(U(A,B;x)\) is Tricomi's confluent hypergeometric function,
\begin{equation}
\bar K_c
=
\frac{K}{\kappa_c},
\qquad
A_{j,t}
=
\frac{a_{t,c}}{2}
+
\frac{\Delta_c(j)}{2},
\qquad
B_j
=
\Delta_c(j)-1 .
\label{eq:app_unified_A_B_def}
\end{equation}
Near the AdS boundary the relevant Tricomi solution is controlled by its
non-normalizable branch, not by the regular term alone.  For
\(B=\Delta_c(j)-1>1\),
\begin{equation}
U(A,B;x)
\sim
\frac{\Gamma(B-1)}{\Gamma(A)}x^{1-B}+\cdots,
\qquad x\to0 .
\label{eq:app_tricomi_small_x_branch}
\end{equation}
Consequently
\begin{align}
z^{\Delta_c(j)}
U(A,\Delta_c(j)-1;2\kappa_c^2z^2)
&\sim
z^{4-\Delta_c(j)}
\nonumber\\
&\quad+
\text{normalizable branch}.
\label{eq:app_tricomi_boundary_scaling}
\end{align}
before the cutoff normalization in \(\mathcal N_c\) is applied.  This
singular branch is what allows
\(\mathcal H_j^{(c),{\rm SW}}(K,\epsilon;\epsilon)=1\).  The normalized
boundary mode used in endpoint factorization is the separate normalizable
factor \(\Psi_j^{(c),{\rm bdry}}\sim z^{\Delta_c(j)}\).  Deep in the bulk,
\(
U(A,B;x)\sim x^{-A}
\),
which generates the algebraic infrared behavior characteristic of the
soft-wall model.
The normalization is fixed by
\begin{equation}
\mathcal H_j^{(c),{\rm SW}}
(K,\epsilon;\epsilon)
=
1,
\end{equation}
giving
\begin{equation}
\mathcal N_c
(j,\bar K_c,\bar\epsilon_c)
=
(\sqrt2\bar\epsilon_c)^{\Delta_c(j)-4}
\frac{
\Gamma(A_{j,t})
}{
\Gamma(\Delta_c(j)-2)
}.
\label{eq:app_unified_sw_Nc}
\end{equation}

When one endpoint of the propagator approaches the boundary, the
relevant normalized boundary wavefunction is
\begin{widetext}
\begin{equation}
\Psi_j^{(c),{\rm bdry}}
(z;\epsilon)
=
\frac{
\psi_0^{(c)}(j,z\rightarrow0)
}{
\widetilde g_5
\mathcal F_0^{(c)}(j,\epsilon)
}
=
-
\frac{
(\sqrt2\kappa_c z)^{\Delta_c(j)}
(\sqrt2\bar\epsilon_c)^{4-\Delta_c(j)}
}{
\Delta_c(j)
}.
\label{eq:app_unified_boundary_mode}
\end{equation}
\end{widetext}
This mode carries the complementary cutoff power
\((\sqrt2\bar\epsilon_c)^{4-\Delta_c(j)}\), which cancels the transfer-kernel
factor \((\sqrt2\bar\epsilon_c)^{\Delta_c(j)-4}\).  The minus sign matches the
decay-constant convention used in Eq.~\eqref{eq:secV_boundary_mode}.

\subsection{Finite-Separation Diagram and Cutoff Cancellation}

The stripped fixed-spin worldsheet Witten diagram takes the form
\begin{widetext}
\begin{align}
\widetilde F_j^{g,{\rm ws}}(\xi,t,b_T;\epsilon)
=
\widetilde g_5^{\,2}
g_j(\lambda)
\int du\,dv\,
(\sqrt2\kappa_c z_{\rm ws})^{-(j-2)}
\int_0^\infty dz\,
\rho_j(z;\xi)\,
G_c(j,z,z_{\rm ws};t).
\label{eq:app_unified_ws_witten}
\end{align}
\end{widetext}
The worldsheet coordinate \(z_{\rm ws}\) specifies the radial position
at which the exchanged spin-(j) field couples to the Wilson-line
surface.  Near the ultraviolet boundary the classical four-cusp
worldsheet assumes the universal form
\begin{equation}
z_{\rm ws}(u,v)
=
b\,z_b(u,v),
\qquad
z_b(u,v)
=
\frac{1}{\cosh u\,\cosh v}.
\label{eq:app_unified_zb_profile}
\end{equation}

In the deep infrared the confining geometry may deform the saddle.
It is therefore convenient to write
\begin{equation}
\begin{gathered}
z_{\rm ws}(u,v;b)=b\,\widehat z(u,v;b),\\
\widehat z(u,v;b)\rightarrow z_b(u,v)
\quad (b\rightarrow0).
\end{gathered}
\label{eq:app_unified_zhat_profile}
\end{equation}

The corresponding boundary conformal moment is
\begin{equation}
\widetilde F_j^{g,{\rm bdry}}(\xi,t;\epsilon)
=
\int_0^\infty dz\,
\rho_j(z;\xi)\,
\mathcal H_j^{(c),{\rm SW}}
(K,z;\epsilon).
\label{eq:app_unified_boundary_moment_cutoff}
\end{equation}

Using the spectral decomposition of the transfer kernel,
\begin{equation}
\widetilde F_j^{g,{\rm bdry}}(\xi,t;\epsilon)
=
\sum_{n=0}^{\infty}
\widetilde F_j^{g,{\rm bdry}}(\xi,t;n;\epsilon),
\end{equation}
where
\begin{equation}
\widetilde F_j^{g,{\rm bdry}}(\xi,t;n;\epsilon)
=
\int_0^\infty dz\,
\rho_j(z;\xi)
\frac{
\widetilde g_5
\mathcal F_n^{(c)}(j,\epsilon)
\psi_n^{(c)}(j,z)
}{
-t+M_{c,n}^2(j)
}.
\end{equation}

The entire cutoff dependence factorizes:
\begin{equation}
\widetilde F_j^{g,{\rm bdry}}(\xi,t;n;\epsilon)
=
(\sqrt2\bar\epsilon_c)^{\Delta_c(j)-4}
\widetilde F_j^{g,{\rm bdry}}(\xi,t;n),
\end{equation}
and therefore
\begin{equation}
\widetilde F_j^{g,{\rm bdry}}(\xi,t;\epsilon)
=
(\sqrt2\bar\epsilon_c)^{\Delta_c(j)-4}
\widetilde F_j^{g,{\rm bdry}}(\xi,t).
\label{eq:app_unified_lower_cutoff_scaling}
\end{equation}
This cutoff dependence cancels exactly against the complementary cutoff
factor carried by the normalized boundary mode
\eqref{eq:app_unified_boundary_mode}.  Consequently, all endpoint
kernels derived below are finite and regulator independent.

\subsection{Ultraviolet Endpoint Reduction}

We first consider the ultraviolet transverse limit
\begin{equation}
b_T\rightarrow0^+,
\qquad
\xi,\;t
\;\;{\rm fixed}.
\end{equation}
In this regime the worldsheet remains parametrically close to the AdS
boundary,
\begin{equation}
z_{\rm ws}
=
b\,z_b(u,v)
\ll z ,
\label{eq:app_uv_ordering}
\end{equation}
for the values of \(z\) that dominate the hadronic radial integral.
Physically, the exchanged spin-\(j\) field first couples to the
near-boundary worldsheet insertion and subsequently propagates into the
bulk toward the hadronic vertex.  The bulk propagator therefore
factorizes according to
\begin{equation}
G_c(j,z,z_{\rm ws};t)
\simeq
\Psi_j^{(c),{\rm bdry}}
(z_{\rm ws};\epsilon)\,
\mathcal H_j^{(c),{\rm SW}}
(K,z;\epsilon).
\label{eq:app_uv_factorization}
\end{equation}
The first factor represents the universal boundary wavefunction of the
exchanged spin-\(j\) mode evaluated at the worldsheet endpoint,
whereas the second factor transfers the excitation from the boundary
into the bulk target.

Substituting Eq.~\eqref{eq:app_uv_factorization} into the Witten
diagram gives
\begin{widetext}
\begin{align}
\widetilde F_j^{g,{\rm ws}}(\xi,t,b_T)
=
-
\frac{
\widetilde g_5^{\,2}g_j(\lambda)
}{
\Delta_c(j)
}
\int du\,dv
\;
(\sqrt2\kappa_c z_{\rm ws})^{-(j-2)}
(\sqrt2\kappa_c z_{\rm ws})^{\Delta_c(j)}
\widetilde F_j^{g,{\rm bdry}}(\xi,t),
\end{align}
\end{widetext}
where the cutoff dependence has cancelled exactly between the boundary
mode and the bulk-to-boundary kernel.
The entire worldsheet dependence therefore enters through the power
\begin{equation}
(\sqrt2\kappa_c z_{\rm ws})^{\alpha_j},
\qquad
\alpha_j
=
\Delta_c(j)-(j-2)
=
4+\gamma_c(j).
\label{eq:app_uv_alpha}
\end{equation}
Using
\begin{equation}
z_{\rm ws}
=
b\,z_b(u,v),
\end{equation}
one obtains
\begin{align}
\widetilde F_j^{g,{\rm ws}}(\xi,t,b_T)
=
\mathcal K_j^{\rm UV}
(\lambda,b_T)
\,
\widetilde F_j^{g,{\rm bdry}}(\xi,t),
\label{eq:app_uv_reduction}
\end{align}
with
\begin{equation}
\mathcal K_j^{\rm UV}
=
-
\frac{
\widetilde g_5^{\,2}g_j(\lambda)
}{
\Delta_c(j)
}
\bar b^{\,\alpha_j}
\mathcal I_j^{\rm UV}(\bar b).
\label{eq:app_uv_kernel_master}
\end{equation}

The dimensionless worldsheet integral is
\begin{widetext}
\begin{align}
\mathcal I_j^{\rm UV}(\bar b)
=
\int_{-\infty}^{\infty}du
\int_{-\infty}^{\infty}dv\,
z_b(u,v)^{\alpha_j}
\frac{
L_n^{\Delta_c(j)-2}
\!\left(
\bar b^2 z_b(u,v)^2
\right)
}{
L_n^{\Delta_c(j)-2}(0)
}.
\label{eq:app_uv_integral}
\end{align}
\end{widetext}
Since
\(
\bar b\ll1,
\)
the Laguerre polynomial ratio approaches unity,
\begin{equation}
\frac{
L_n^{\Delta_c(j)-2}
(\bar b^2 z_b^2)
}{
L_n^{\Delta_c(j)-2}(0)
}
=
1+\mathcal O(\bar b^2).
\end{equation}
To leading order the result becomes independent of the radial mode
number \(n\), and the remaining integral depends only on the geometry
of the near-boundary worldsheet,
\begin{align}
\mathcal I_j^{\rm UV}
&=
\int du\,dv\,
z_b(u,v)^{\alpha_j}
\nonumber\\
&=
\pi
\left[
\frac{
\Gamma(\alpha_j/2)
}{
\Gamma((\alpha_j+1)/2)
}
\right]^2 ,
\qquad
{\rm Re}\,\alpha_j>0 .
\label{eq:app_uv_geometric}
\end{align}

The ultraviolet kernel therefore takes the closed form
\begin{equation}
\mathcal K_j^{\rm UV}
=
-
\widetilde g_5^{\,2}
g_j(\lambda)
\frac{\pi}{\Delta_c(j)}
\left[
\frac{
\Gamma\!\left(
2+\frac{\gamma_c(j)}{2}
\right)
}{
\Gamma\!\left(
\frac{5+\gamma_c(j)}{2}
\right)
}
\right]^2
\bar b^{\,4+\gamma_c(j)}.
\label{eq:app_uv_kernel_final}
\end{equation}
The important point is that the dependence on the hadronic target has
completely factorized.  The ultraviolet limit is controlled entirely
by the near-boundary geometry of the worldsheet and by the conformal
dimension of the exchanged spin-\(j\) field.  The target enters only
through the boundary conformal moment
\(\widetilde F_j^{g,{\rm bdry}}(\xi,t)\), hence
\begin{equation}
\widetilde F_j^{g,{\rm ws}}(\xi,t,b_T)
\xrightarrow[b_T\to0^+]{}
\mathcal K_j^{\rm UV}
(\lambda,b_T)
\,
\widetilde F_j^{g,{\rm bdry}}(\xi,t).
\label{eq:app_uv_theorem}
\end{equation}

The strict point \(b_T=0\) is supplied instead by the ordinary
boundary or contact Witten diagram,
\begin{equation}
F_j^g(\xi,t,0)
=
\widetilde F_j^{g,{\rm bdry}}(\xi,t),
\end{equation}
and is therefore not identical to the worldsheet limit
\(b_T\to0^+\).  For
\({\rm Re}[4+\gamma_c(j)]>0\), the finite-separation worldsheet saddle
alone satisfies
\begin{equation}
\lim_{b_T\to0^+}
\widetilde F_j^{g,{\rm ws}}(\xi,t,b_T)
=
0.
\end{equation}
This is not the full short-distance OPE of the physical GTMD.  The
local OPE is represented by the separate boundary saddle, while the
finite-separation worldsheet correlator describes a different saddle
sector controlled by the same fixed-spin exchange channel.

\subsection{Soft-wall Infrared Endpoint Reduction}

We now consider the opposite radial ordering,
\begin{equation}
z\ll z_{\rm ws},
\label{eq:app_ir_sw_ordering}
\end{equation}
which corresponds to a worldsheet insertion located deeper in the bulk
than the region dominating the hadronic radial overlap.
Physically, the exchanged spin-\(j\) excitation is first created at the
hadronic vertex near the boundary and subsequently propagates toward the
deep worldsheet endpoint.  The endpoint factorization is therefore the
reverse of the ultraviolet case.  The universal boundary wavefunction
attaches to the target side of the diagram, while the transfer kernel
is evaluated at the worldsheet endpoint.

The bulk propagator factorizes according to
\begin{equation}
G_c(j,z,z_{\rm ws};t)
\simeq
\Psi_j^{(c),{\rm bdry}}
(z;\epsilon)\,
\mathcal H_j^{(c),{\rm SW}}
(K,z_{\rm ws};\epsilon).
\label{eq:app_ir_sw_factorization}
\end{equation}
The normalized boundary mode now depends on the target coordinate
\(z\).  Consequently the target radial integral no longer produces the
boundary conformal moment
\(\widetilde F_j^{g,{\rm bdry}}\).  Instead it generates the projection
\begin{equation}
\widehat{\mathcal T}^{(c)}_j(\xi)
=
\int_0^\infty dz\,
\rho_j(z;\xi)
\left(
\sqrt2\kappa_c z
\right)^{\Delta_c(j)}.
\label{eq:app_ir_sw_target_projection}
\end{equation}

This quantity measures the overlap of the target with the
near-boundary spin-\(j\) wavefunction and contains all target
dependence of the infrared endpoint.

Substituting
Eq.~\eqref{eq:app_ir_sw_factorization}
into the finite-separation Witten diagram yields
\begin{equation}
\widetilde F_{j,{\rm IR},{\rm SW}}^{g,{\rm ws}}
(\xi,t,b_T)
=
\mathcal K_j^{\rm IR,SW}
(\lambda,b_T,t)
\,
\widehat{\mathcal T}^{(c)}_j(\xi),
\label{eq:app_ir_sw_reduction}
\end{equation}
where
\begin{widetext}
\begin{align}
\mathcal K_j^{\rm IR,SW}
(\lambda,b_T,t)
=
-
\frac{
\widetilde g_5^{\,2}
g_j(\lambda)
}{
\Delta_c(j)
}
\int du\,dv\,
(\sqrt2\kappa_c z_{\rm ws})^{-(j-2)}
\widehat{\mathcal H}_j^{(c),{\rm SW}}
(K,z_{\rm ws}).
\label{eq:app_ir_sw_kernel_master}
\end{align}
\end{widetext}
Using
\begin{equation}
z_{\rm ws}
=
b\,\widehat z(u,v;b),
\end{equation}
together with the finite transfer function
\begin{equation}
\widehat{\mathcal H}_j^{(c),{\rm SW}}
=
\frac{
\Gamma(A_{j,t})
}{
\Gamma(\Delta_c(j)-2)
}
(\sqrt2\kappa_c z)^{\Delta_c(j)}
U(A_{j,t},B_j;2\kappa_c^2z^2),
\end{equation}
one obtains
\begin{align}
\mathcal K_j^{\rm IR,SW}
&=
-
\widetilde g_5^{\,2}
g_j(\lambda)
\frac{
\Gamma(A_{j,t})
}{
\Delta_c(j)\,
\Gamma(\Delta_c(j)-2)
}
\bar b^{\,\alpha_j}
\nonumber\\
&\hspace{0.5cm}\times
\mathcal I_j^{\rm IR,SW}
(\bar b,t),
\label{eq:app_ir_sw_kernel_exact}
\end{align}
where
\begin{align}
\mathcal I_j^{\rm IR,SW}
(\bar b,t)
&=
\int_{-\infty}^{\infty}du
\int_{-\infty}^{\infty}dv
\,
\widehat z(u,v;b)^{\alpha_j}
\nonumber\\
&\hspace{0.5cm}\times
U\!\left(
A_{j,t},
B_j;
\bar b^2\widehat z(u,v;b)^2
\right).
\label{eq:app_ir_sw_integral}
\end{align}

The physical origin of the infrared behavior is now transparent.
Unlike the ultraviolet endpoint, the transfer function is evaluated at
large radial distance.  The asymptotic behavior of the endpoint is
therefore controlled entirely by the large-argument behavior of the
Tricomi function. For
\(
x\rightarrow\infty,
\)
one has
\begin{equation}
U(A,B;x)
=
x^{-A}
\left[
1+\mathcal O(x^{-1})
\right].
\label{eq:app_ir_sw_tricomi_large}
\end{equation}
Substituting
\(
x=\bar b^2\widehat z^2
\)
gives
\begin{align}
&
\bar b^{\,\alpha_j}
\widehat z^{\,\alpha_j}
U(A_{j,t},B_j;\bar b^2\widehat z^2)
\nonumber\\
&\sim
\bar b^{\,\alpha_j-2A_{j,t}}
\widehat z^{\,\alpha_j-2A_{j,t}}
\nonumber\\
&=
\bar b^{\,2-j-a_{t,c}}
\widehat z^{\,2-j-a_{t,c}} .
\label{eq:app_ir_sw_power}
\end{align}

A noteworthy feature is the complete cancellation of the anomalous
dimension \(\gamma_c(j)\) from the leading infrared power.  The
anomalous dimension affects only the overall normalization through the
prefactor multiplying the integral.  The infrared power itself is
determined solely by the spin \(j\) and the momentum-transfer
parameter \(a_{t,c}\).

If the infrared worldsheet geometry is dominated by a finite confining
saddle, we define
\begin{equation}
\mathcal C^{\rm IR,SW}_{j,t}
=
\int du\,dv\,
\widehat z(u,v;b)^{2-j-a_{t,c}}.
\label{eq:app_ir_sw_Cjt}
\end{equation}
The large-\(b_T\) kernel then becomes
\begin{widetext}
\begin{align}
\mathcal K_j^{\rm IR,SW}
(\lambda,b_T,t)
\sim
-
\widetilde g_5^{\,2}
g_j(\lambda)
\frac{
\Gamma\!\left(
1+\frac{j+\gamma_c(j)+a_{t,c}}{2}
\right)
}{
(2+j+\gamma_c(j))
\Gamma(j+\gamma_c(j))
}
\mathcal C^{\rm IR,SW}_{j,t}
\,
\bar b^{\,2-j-a_{t,c}}
\left[
1+\mathcal O(\bar b^{-2})
\right].
\label{eq:app_ir_sw_kernel_asymptotic}
\end{align}
\end{widetext}
so that
\begin{equation}
\widetilde F_{j,{\rm IR},{\rm SW}}^{g,{\rm ws}}
(\xi,t,b_T)
\xrightarrow[b_T\to\infty]{}
\mathcal K_j^{\rm IR,SW}
(\lambda,b_T,t)
\,
\widehat{\mathcal T}^{(c)}_j(\xi),
\end{equation}
with
\begin{equation}
\mathcal K_j^{\rm IR,SW}
\sim
C_j(t)
\,
\bar b^{\,2-j-a_{t,c}}.
\label{eq:app_ir_sw_final}
\end{equation}

The soft-wall infrared endpoint therefore exhibits an algebraic
large-distance tail.  This behavior is a direct consequence of the
power-law decay of the Tricomi transfer function in the deep bulk.
Although confinement is encoded through the soft-wall background, the
infrared propagation remains algebraic rather than exponential.  This tail
should be regarded as a model artifact of the soft-wall transfer, not as a
universal confining prediction.  The latter requires an additional
infrared scale in the transfer kernel, which motivates the hard-wall and
repulsive-wall infrared completions discussed below.

\section{Gap-matched hard-wall infrared completion and exponential tail}
\label{APP_WITTEN_IR_HW_KERNEL}

The soft-wall infrared endpoint discussed in
Appendix~\ref{APP_WITTEN_UV_KERNEL}
produces an algebraic large-\(b_T\) behavior.  This is a direct
consequence of the large-distance behavior of the soft-wall transfer
function,
\(
U(A,B;x)\sim x^{-A}
\),
which decays only as a power.

A confining gauge theory is expected to exhibit instead an infrared
correlation length and therefore an exponential suppression at large
transverse separation.  To model this behavior analytically we replace
only the deep-infrared transfer between the worldsheet and the target by
a hard-wall transfer function.  The near-boundary normalization and the
target-side projection remain those of the soft-wall construction.  The
resulting setup should therefore be viewed as a gap-matched infrared
completion of the soft-wall endpoint whose purpose is to isolate the confining infrared tail.

The physical picture is simple.  The target creates a spin-\(j\)
excitation near the boundary.  That excitation propagates through the
bulk toward a worldsheet insertion located deep in the infrared.  The
soft-wall description yields algebraic propagation.  The hard-wall
description introduces a mass gap and converts the long-distance
propagation into exponential attenuation.

\subsection{Hard-wall spin-\texorpdfstring{\(j\)}{j} transfer channel}

The hard-wall model occupies the AdS interval
\(
0<z<z_0,
\)
where \(z_0\) denotes the infrared wall.  Normalizable modes satisfy a
Neumann condition at \(z=z_0\), corresponding to vanishing flux through
the confining boundary.
The normalizable spin-\(j\) modes are
\begin{equation}
\psi_n^{(c),{\rm HW}}(j,z)
=
c_n^{(c),{\rm HW}}(j)\,
z^2
J_{\nu_j}
\!\left(
M^{\rm HW}_{c,n}(j)\,z
\right),
\label{eq:hw_modes_reorg}
\end{equation}
with
\begin{equation}
\nu_j
=
\Delta_c(j)-2
=
j+\gamma_c(j).
\end{equation}
The discrete masses are determined by the Neumann condition
\begin{equation}
\left.
\partial_z
\Big[
z^2
J_{\nu_j}(Mz)
\Big]
\right|_{z=z_0}
=
0,
\label{eq:hw_neumann_reorg}
\end{equation}
so that
\begin{equation}
M^{\rm HW}_{c,n}(j)
=
\frac{\chi_{\nu_j,n}}{z_0},
\label{eq:hw_mass_reorg}
\end{equation}
where \(\chi_{\nu_j,n}\) is the corresponding root.
For the graviton trajectory,
\(j=2\),
\(\nu_j=2\),
Eq.~\eqref{eq:hw_neumann_reorg}
reduces to the familiar tensor-glueball condition
\begin{equation}
J_1(Mz_0)=0.
\end{equation}
The modes obey
\begin{equation}
\frac{1}{2\kappa_c^2}
\int_0^{z_0}
dz\,
\sqrt g\,|g^{xx}|
\psi_n^{(c),{\rm HW}}
\psi_m^{(c),{\rm HW}}
=
\delta_{nm},
\label{eq:hw_orthonormality_reorg}
\end{equation}
which fixes the normalization
\begin{eqnarray}
c_n^{(c),{\rm HW}}(j)
=
\frac{\sqrt2\,\kappa_c}
{
\left[
\displaystyle
\int_0^{z_0}
dz\,z\,
J_{\nu_j}^2
\!\left(
\frac{\chi_{\nu_j,n}z}{z_0}
\right)
\right]^{1/2}
}.\nonumber\\
\end{eqnarray}

The resulting bulk propagator is
\begin{equation}
G_c^{\rm HW}(j,z,z';t)
=
2\kappa_c^2
\sum_{n=0}^{\infty}
\frac{
\psi_n^{(c),{\rm HW}}(j,z)
\psi_n^{(c),{\rm HW}}(j,z')
}{
-t+\big(M^{\rm HW}_{c,n}(j)\big)^2
}.
\label{eq:hw_bulk_bulk_reorg}
\end{equation}
Exactly as in the soft-wall channel, one may construct a
bulk-to-boundary transfer kernel through the decay constants
\begin{equation}
\mathcal F_n^{(c),{\rm HW}}
=
\frac{1}{2\kappa_c^2}
\frac{1}{\widetilde g_5}
\left[
-\sqrt g\,|g^{xx}|
\partial_z
\psi_n^{(c),{\rm HW}}
\right]_{z=\epsilon}.
\end{equation}
Summing over the normalizable tower yields
\begin{widetext}
\begin{align}
\mathcal H_j^{(c),{\rm HW}}
(K,z;\epsilon)
=
\mathcal N_j^{\rm HW}(K,\epsilon)
\,z^2
\Big[
\mathcal R_j(K,z_0)
I_{\nu_j}(Kz)
+
K_{\nu_j}(Kz)
\Big].
\label{eq:hw_transfer_reorg}
\end{align}
\end{widetext}
The modified Bessel functions have a simple physical interpretation.
The \(K_{\nu_j}\) branch represents a mode that decays into the
infrared,
\[
K_{\nu_j}(Kz)
\sim
e^{-Kz},
\]
while the \(I_{\nu_j}\) branch grows exponentially,
\[
I_{\nu_j}(Kz)
\sim
e^{Kz}.
\]
The coefficient
\begin{equation}
\mathcal R_j(K,z_0)
=
-
\frac{
\partial_z
\left[
z^2K_{\nu_j}(Kz)
\right]_{z=z_0}
}{
\partial_z
\left[
z^2I_{\nu_j}(Kz)
\right]_{z=z_0}
}
\label{eq:hw_reflection_reorg}
\end{equation}
is a reflection coefficient generated by the infrared wall.  It ensures
that the linear combination in
Eq.~\eqref{eq:hw_transfer_reorg}
satisfies the Neumann boundary condition.

Near the ultraviolet boundary the hard-wall kernel exhibits exactly the
same cutoff dependence as the soft-wall kernel,
\begin{equation}
\mathcal H_j^{(c),{\rm HW}}
=
(\sqrt2\bar\epsilon_c)^{\Delta_c(j)-4}
\widehat{\mathcal H}_j^{(c),{\rm HW}}
+
\mathcal O
\!\left(
\epsilon^{\Delta_c(j)-2}
\right),
\end{equation}
with finite transfer function
\begin{align}
\widehat{\mathcal H}_j^{(c),{\rm HW}}
(K,z)
=
\frac{
2^{1-\nu_j}
}{
\Gamma(\nu_j)
}
\left(
\frac{K}{\sqrt2\kappa_c}
\right)^{\nu_j}
(\sqrt2\kappa_c z)^2
\nonumber\\
\times
\Big[
\mathcal R_j(K,z_0)I_{\nu_j}(Kz)
+
K_{\nu_j}(Kz)
\Big].
\label{eq:hw_finite_transfer_reorg}
\end{align}

The matching of the cutoff dependence to the soft-wall channel is
essential.  It allows the same normalized boundary mode
\(\Psi_j^{(c),{\rm bdry}}\) to be used in the endpoint factorization,
ensuring that the ultraviolet structure of the theory remains unchanged.

\subsection{Infrared Endpoint Factorization}

We now return to the infrared radial ordering
\(z\ll z_{\rm ws},
\)
for which the exchanged spin-\(j\) mode is created near the target and
propagates toward a worldsheet insertion located deep in the bulk.
Exactly as in the soft-wall infrared analysis, the propagator
factorizes when one endpoint remains close to the boundary,
\begin{equation}
G_c(j,z,z_{\rm ws};t)
\simeq
\Psi_j^{(c),{\rm bdry}}(z;\epsilon)\,
\mathcal H_j^{(c),{\rm HW}}(K,z_{\rm ws};\epsilon).
\label{eq:hw_endpoint_factorization_reorg}
\end{equation}
The normalized boundary mode therefore attaches to the target side of
the diagram, while the hard-wall transfer function controls the
propagation from the target toward the infrared worldsheet vertex.

As in the soft-wall case, the target dependence factorizes into the
projection
\begin{equation}
\widehat{\mathcal T}^{(c)}_j(\xi)
=
\int_0^\infty dz\,
\rho_j(z;\xi)
\left(
\sqrt2\kappa_c z
\right)^{\Delta_c(j)},
\label{eq:hw_target_projection_reorg}
\end{equation}
which measures the overlap of the target wavefunction with the
near-boundary spin-\(j\) mode.
After cancellation of the ultraviolet cutoff factors, the finite
infrared contribution assumes the form
\begin{equation}
\widetilde F_{j,{\rm IR},{\rm HW}}^{g,{\rm ws}}
(\xi,t,b_T)
=
\mathcal K_j^{\rm IR,HW}
(\lambda,b_T,t)
\,
\widehat{\mathcal T}^{(c)}_j(\xi),
\label{eq:hw_ir_factorized_reorg}
\end{equation}
with
\begin{align}
\mathcal K_j^{\rm IR,HW}
&=
-
\frac{
\widetilde g_5^{\,2}g_j(\lambda)
}{
\Delta_c(j)
}
\int du\,dv\,
(\sqrt2\kappa_c z_{\rm ws})^{-(j-2)}
\nonumber\\
&\quad\times
\widehat{\mathcal H}^{(c),{\rm HW}}_j
(K,z_{\rm ws}).
\label{eq:hw_kernel_master_reorg}
\end{align}
Using
\begin{equation}
z_{\rm ws}
=
b\,\widehat z(u,v;b),
\end{equation}
the kernel may be written as
\begin{equation}
\mathcal K_j^{\rm IR,HW}
=
-
\frac{
\widetilde g_5^{\,2}g_j(\lambda)
}{
\Delta_c(j)
}
\frac{
2^{1-\nu_j}
}{
\Gamma(\nu_j)
}
\left(
\frac{K}{\sqrt2\kappa_c}
\right)^{\nu_j}
\bar b^{\,4-j}
\,
\mathcal I_j^{\rm IR,HW},
\label{eq:hw_kernel_exact_reorg}
\end{equation}
where
\begin{widetext}
\begin{align}
\mathcal I_j^{\rm IR,HW}
=
\int du\,dv\,
\widehat z(u,v;b)^{4-j}
\Big[
\mathcal R_j(K,z_0)
I_{\nu_j}(Kb\widehat z)
+
K_{\nu_j}(Kb\widehat z)
\Big].
\label{eq:hw_integral_exact_reorg}
\end{align}
\end{widetext}
At this stage no approximation has been made.  The expression above is
the exact gap-matched soft-wall/hard-wall infrared-completion kernel.

\subsection{Decaying-Branch Approximation}

The physical origin of confinement becomes apparent in the large-\(z\)
behavior of the modified Bessel functions.
For large argument,
\begin{equation}
I_\nu(x)
\sim
\frac{e^x}{\sqrt{2\pi x}},
\qquad
K_\nu(x)
\sim
\sqrt{\frac{\pi}{2x}}
e^{-x},
\qquad
x\rightarrow\infty .
\label{eq:hw_bessel_large_reorg}
\end{equation}
The \(I_\nu\) branch grows exponentially toward the infrared,
whereas the \(K_\nu\) branch decays exponentially.  Confining
propagation is therefore associated with the \(K_\nu\) branch.

Whether the growing branch can be neglected depends on the reflection
coefficient generated by the wall.  Using the asymptotic behavior of
the Bessel functions one finds
\begin{equation}
\mathcal R_j(K,z_0)
\sim
\pi
e^{-2Kz_0}
\left[
1+\mathcal O((Kz_0)^{-1})
\right].
\label{eq:hw_reflection_large_reorg}
\end{equation}
At a worldsheet point \(z_{\rm ws}\),
\begin{equation}
\frac{
\mathcal R_j(K,z_0)
I_{\nu_j}(Kz_{\rm ws})
}{
K_{\nu_j}(Kz_{\rm ws})
}
\sim
e^{-2K(z_0-z_{\rm ws})}.
\label{eq:hw_ratio_reorg}
\end{equation}
The reflected branch is therefore exponentially suppressed provided
\begin{equation}
Kz_0\gg1,
\qquad
K(z_0-z_{\rm ws}^{\rm IR})\gg1,
\label{eq:hw_dec_conditions_reorg}
\end{equation}
where \(z_{\rm ws}^{\rm IR}\) denotes the infrared worldsheet saddle.

Under these conditions the transfer function simplifies to the purely
decaying form
\begin{equation}
\widehat{\mathcal H}^{(c),{\rm HW}}_j
(K,z)
\longrightarrow
\frac{
2^{1-\nu_j}
}{
\Gamma(\nu_j)
}
\left(
\frac{K}{\sqrt2\kappa_c}
\right)^{\nu_j}
(\sqrt2\kappa_c z)^2
K_{\nu_j}(Kz).
\label{eq:hw_decaying_transfer_reorg}
\end{equation}
The hard wall therefore converts the infrared transfer into the
propagation of a massive mode with inverse correlation length \(K\).

\subsection{Gap Matching and Confining Exponential Tail}

The mixed construction is intended to model the deep infrared region.
The natural scale controlling that propagation is therefore not the
external momentum transfer but the confining mass gap.  We first write
\begin{equation}
K_{\rm IR}=M_{\rm gap},
\qquad
\eta_{\rm gap}\equiv\frac{M_{\rm gap}}{\sqrt2\kappa_c},
\label{eq:hw_gap_matching_general_reorg}
\end{equation}
and apply this replacement only inside the infrared transfer function.
The target projection and ultraviolet normalization remain unchanged.  This
hard-wall replacement is a phenomenological infrared completion of the
transfer kernel; it is not derived from the soft-wall saddle.  In the
soft-wall normalization used for the compact endpoint formulas, the minimal
gap-matched choice is
\begin{equation}
M_{\rm gap}=\sqrt2\,\kappa_c,
\qquad \eta_{\rm gap}=1 .
\label{eq:hw_gap_matching_reorg}
\end{equation}

The resulting gap-matched transfer is
\begin{align}
\widehat{\mathcal H}^{(c),{\rm HW,gap}}_j(z;M_{\rm gap})
&=
\frac{2^{1-\nu_j}}{\Gamma(\nu_j)}
\left(\frac{M_{\rm gap}}{\sqrt2\kappa_c}\right)^{\nu_j}
\nonumber\\
&\quad\times
(\sqrt2\kappa_c z)^2
K_{\nu_j}\!\left(M_{\rm gap}z\right).
\label{eq:hw_gap_transfer_reorg}
\end{align}
The infrared factorization becomes
\begin{equation}
\widetilde F_{j,{\rm IR},{\rm HW}}^{g,{\rm ws}}
(\xi,t,b_T)
=
\mathcal K_j^{\rm IR,HW,gap}
(\lambda,b_T)
\,
\widehat{\mathcal T}^{(c)}_j(\xi),
\label{eq:hw_gap_factorization_reorg}
\end{equation}
where
\begin{equation}
\mathcal K_j^{\rm IR,HW,gap}
=
-
\frac{
\widetilde g_5^{\,2}g_j(\lambda)
}{
\Delta_c(j)
}
\frac{
2^{1-\nu_j}
}{
\Gamma(\nu_j)
}
\eta_{\rm gap}^{\nu_j}
\bar b^{\,4-j}
\mathcal I_j^{\rm HW,gap}(\bar b),
\label{eq:hw_gap_kernel_reorg}
\end{equation}
with
\begin{equation}
\mathcal I_j^{\rm HW,gap}
=
\int du\,dv\,
\widehat z(u,v;b)^{4-j}
K_{\nu_j}
\!\left(
\eta_{\rm gap}\bar b\,\widehat z(u,v;b)
\right).
\label{eq:hw_gap_integral_reorg}
\end{equation}
The physical difference from the soft-wall endpoint is now explicit.
The Tricomi function has been replaced by a modified Bessel function,
\begin{equation}
U(A,B;x)
\longrightarrow
K_{\nu_j}(x),
\end{equation}
thereby converting algebraic propagation into exponential propagation.

For large argument,
\begin{equation}
K_{\nu_j}(x)
=
\sqrt{\frac{\pi}{2x}}
e^{-x}
\left[
1+\mathcal O(x^{-1})
\right].
\label{eq:hw_asymptotic_K_reorg}
\end{equation}
Let the infrared worldsheet saddle satisfy
\begin{equation}
z_{\rm ws}^{\rm IR}
=
b\,\widehat z_{\rm IR},
\qquad
\widehat z_{\rm IR}
=
\mathcal O(1).
\label{eq:hw_ir_saddle_reorg}
\end{equation}
The dominant infrared contribution then behaves as
\begin{equation}
\mathcal K_j^{\rm IR,HW,gap}
\sim
\mathcal N_j^{\rm IR,HW}
(\lambda)
\,
\eta_{\rm gap}^{\nu_j-1/2}
\bar b^{\frac72-j}
\exp
\!\left[
-\eta_{\rm gap}\bar b\,\widehat z_{\rm IR}
\right].
\label{eq:hw_large_b_bar_reorg}
\end{equation}

Introducing the physical infrared correlation scale
\begin{equation}
\kappa_{\rm IR}
=
M_{\rm gap}\widehat z_{\rm IR}
=
\eta_{\rm gap}\sqrt2\kappa_c
\widehat z_{\rm IR},
\label{eq:hw_kappa_ir_reorg}
\end{equation}
one finally obtains
\begin{align}
\mathcal K_j^{\rm IR,HW,gap}
&\sim
\widetilde{\mathcal N}_j^{\rm IR,HW}
(\lambda)
(\kappa_{\rm IR}b)^{\frac72-j}
e^{-\kappa_{\rm IR}b}
\nonumber\\
&\quad\times
\left[
1+\mathcal O((\kappa_{\rm IR}b)^{-1})
\right].
\label{eq:hw_final_tail_reorg}
\end{align}

This result is the hard-wall counterpart of the soft-wall infrared
endpoint statement.  The soft-wall transfer produces the algebraic tail

\[
\mathcal K_j^{\rm IR,SW}
\sim
b^{\,2-j-a_{t,c}},
\]
whereas the gap-matched hard-wall transfer produces the confining
behavior

\[
\mathcal K_j^{\rm IR,HW}
\sim
e^{-\kappa_{\rm IR}b}.
\]
The exponential decay reflects the existence of a finite infrared
correlation length and provides a simple holographic realization of
confinement in transverse-coordinate space.

\section{Repulsive-wall infrared completion for finite-separation GTMDs}
\label{app:RW_GTMD_endpoint}

This appendix gives a self-contained definition of the repulsive-wall (RW)
endpoint used in the main text.  The RW construction is a model completion for the
deep infrared transfer kernel.  Its ultraviolet behavior is asymptotically
AdS and therefore gives the same small-\(b_T\) power as the soft-wall and
hard-wall channels, while its infrared Whittaker tail gives a Gaussian
suppression.  All constants introduced below are model and scheme data,
not universal predictions.

\subsection{Definitions and scheme parameters}

The repulsive-wall background is
\begin{widetext}
\begin{equation}
ds^2=e^{2A(z)}(dz^2+\eta_{\mu\nu}dx^\mu dx^\nu),
\qquad
e^{2A(z)}=\left(\frac{R}{z}\right)^2e^{a\kappa_c^2z^2},
\qquad a>0 .
\label{eq:RW_metric_GTMD}
\end{equation}
\end{widetext}
The dimensionless parameter \(a\) fixes the strength of the repulsive wall.
The soft-wall scale is related to a phenomenological QCD scale by a scheme
choice,
\begin{equation}
\kappa_c=\delta_\kappa\Lambda_{\rm QCD},
\qquad
R^{-1}=\delta_R\Lambda_{\rm QCD},
\end{equation}
and the constants \(\delta_\kappa\), \(\delta_R\), and \(a\) are part of the
infrared model definition.  We also define
\begin{equation}
\widetilde t\equiv \frac{t}{\kappa_c^2},
\qquad
\widetilde K^2\equiv\frac{K^2}{\kappa_c^2}=-\widetilde t,
\qquad
x\equiv a\kappa_c^2z^2 .
\label{eq:RW_dimensionless_defs}
\end{equation}
The quantity \(S_j\) denotes the dimensionless spin-dependent constant
term in the RW Schr\"odinger potential after the universal near-boundary
\(1/z^2\) singularity has been extracted.  Its explicit value depends on
the chosen five-dimensional spin-\(j\) action and on the curvature-coupling
convention.  Once that convention is fixed, \(S_j\) is fixed; it is not an
additional function of \(b_T\).

With these definitions the Kummer index used below is
\begin{align}
A^{\rm RW}_{j,t}
&=
\frac{S_j+3\Delta_c(j)-6-3\widetilde t/a}{6}
\nonumber\\
&=
\frac{S_j+3\Delta_c(j)-6+3\widetilde K^2/a}{6}.
\label{eq:RW_A_index}
\end{align}
and
\begin{equation}
B_j=\Delta_c(j)-1.
\label{eq:RW_B_index}
\end{equation}
Changing the sign convention for \(\widetilde t\) simply changes the first
form of Eq.~\eqref{eq:RW_A_index}; the transfer kernel is unchanged if the
same convention is used throughout.

\subsection{RW Green function}

The two independent radial branches are
\begin{equation}
{\cal K}^{\rm RW}_{M,j}(z)
=
e^{-x/2}x^{\Delta_c(j)/2}
M\!\left(A^{\rm RW}_{j,t},B_j;x\right),
\label{eq:RW_M_branch}
\end{equation}
and
\begin{equation}
{\cal K}^{\rm RW}_{U,j}(z)
=
e^{-x/2}x^{\Delta_c(j)/2}
U\!\left(A^{\rm RW}_{j,t},B_j;x\right).
\label{eq:RW_U_branch}
\end{equation}
The ordered Green function is
\begin{align}
G^{\rm RW}_c(j,z,z';t)
&=
{\cal N}^{\rm RW}_j(t)\,
{\cal K}^{\rm RW}_{M,j}(z_<)\,
{\cal K}^{\rm RW}_{U,j}(z_>),
\nonumber\\
z_<&\equiv\min(z,z'),
\qquad
z_>\equiv\max(z,z').
\label{eq:RW_G_ordered}
\end{align}
with normalization
\begin{equation}
{\cal N}^{\rm RW}_j(t)
=
-\frac12
\frac{\Gamma(A^{\rm RW}_{j,t})}{\Gamma(\Delta_c(j)-1)}
\left(\frac{3a\kappa_c^2}{2}\right)^{\Delta_c(j)-2} .
\label{eq:RW_norm}
\end{equation}
Overall convention-dependent factors are absorbed into
\(\widetilde g_5^{\,2}g_j(\lambda)\).  The limiting behaviors are
\begin{align}
{\cal K}^{\rm RW}_{M,j}(z)
&\sim x^{\Delta_c(j)/2}
\quad (x\to0),
\nonumber\\
{\cal K}^{\rm RW}_{U,j}(z)
&\sim e^{-x/2}x^{\Delta_c(j)/2-A^{\rm RW}_{j,t}}
\quad (x\to\infty).
\label{eq:RW_limits}
\end{align}
The second relation is the origin of the Gaussian infrared suppression.

\begin{widetext}

\subsection{Endpoint reductions}

The stripped finite-separation RW moment is
\begin{equation}
\widetilde F^{g,{\rm RW}}_{j,\ws}(\xi,t,b_T)
=
\widetilde g_5^{\,2}g_j(\lambda)
\int du\,dv\,
(\sqrt2\kappa_c z_{\rm ws})^{-(j-2)}
\int_0^\infty dz\,\rho_j(z;\xi)G^{\rm RW}_c(j,z,z_{\rm ws};t).
\label{eq:RW_witten}
\end{equation}
For \(b_T\to0^+\), \(z_{\rm ws}=b z_b(u,v)\ll z\), and the RW channel is
asymptotically AdS.  Therefore
\begin{equation}
\widetilde F^{g,{\rm RW}}_{j,\ws}(\xi,t,b_T)
\xrightarrow[b_T\to0^+]{}
\mathcal K_j^{\rm UV,RW}(\lambda,b_T)\,
\widetilde F^{g,{\rm RW}}_{j,\bdry}(\xi,t),
\qquad
\mathcal K_j^{\rm UV,RW}\propto\bar b^{4+\gamma_c(j)} .
\label{eq:RW_UV_endpoint}
\end{equation}
\end{widetext}
Thus the ultraviolet power is independent of the infrared completion.

For the infrared ordering \(z\ll z_{\rm ws}\), the target side supplies the
near-boundary mode and the worldsheet side probes the decaying RW branch.
The target dependence becomes
\begin{equation}
\widehat{\mathcal T}^{(c),{\rm RW}}_j(\xi)
=
\int_0^\infty dz\,\rho_j(z;\xi)(\sqrt2\kappa_c z)^{\Delta_c(j)} .
\label{eq:RW_target_projection}
\end{equation}
The remaining kernel is
\begin{widetext}
\begin{equation}
\mathcal K_j^{\rm IR,RW}(\lambda,b_T,t)
=
{\cal N}^{\rm IR,RW}_{j,t}(\lambda)\,
\bar b^{\alpha_j}
\int du\,dv\,\widehat z(u,v;b)^{\alpha_j}
 e^{-x_{\rm ws}/2}
U\!\left(A^{\rm RW}_{j,t},B_j;x_{\rm ws}\right),
\label{eq:RW_IR_kernel_master}
\end{equation}
where
\begin{equation}
\alpha_j\equiv\Delta_c(j)-(j-2),
\qquad
x_{\rm ws}=a\kappa_c^2z_{\rm ws}^2
=\frac{a}{2}\bar b^2\widehat z(u,v;b)^2 .
\label{eq:RW_xws}
\end{equation}
Using \(U(A,B;x)\sim x^{-A}\) at large \(x\), the integrand behaves as
\begin{equation}
\bar b^{\alpha_j}\widehat z^{\alpha_j}e^{-x_{\rm ws}/2}
U(A^{\rm RW}_{j,t},B_j;x_{\rm ws})
\sim
\left(\frac{a}{2}\right)^{-A^{\rm RW}_{j,t}}
\bar b^{\alpha_j-2A^{\rm RW}_{j,t}}
\widehat z^{\alpha_j-2A^{\rm RW}_{j,t}}
\exp\!\left[-\frac{a}{4}\bar b^2\widehat z^2\right].
\label{eq:RW_integrand_asymptotic}
\end{equation}
\end{widetext}
If the cusp-subtracted central worldsheet is dominated by a finite saddle
\(\widehat z_{\rm IR}=O(1)\), then
\begin{align}
\mathcal K_j^{\rm IR,RW}(\lambda,b_T,t)
&\sim
C^{\rm IR,RW}_{j,t}\,
\bar b^{\delta^{\rm RW}_{j,t}}
\exp\!\left[-\sigma^{\rm RW}_{j,t}\bar b^2\right]
\nonumber\\
&\quad\times
\left[1+\mathcal O(\bar b^{-2})\right].
\label{eq:RW_IR_endpoint}
\end{align}
with
\begin{equation}
\delta^{\rm RW}_{j,t}\equiv\alpha_j-2A^{\rm RW}_{j,t},
\qquad
\sigma^{\rm RW}_{j,t}\equiv\frac{a}{4}\widehat z_{\rm IR}^{\,2}.
\label{eq:RW_delta_sigma}
\end{equation}
At fixed \(t\) we abbreviate \(\delta^{\rm RW}_{j,t}\) and \(\sigma^{\rm RW}_{j,t}\) as \(\delta_j\) and \(\sigma^{\rm RW}_j\) in the main text.  Equivalently, defining the physical RW infrared scale
\begin{equation}
\kappa_{\rm RW}=\sqrt a\,\kappa_c\widehat z_{\rm IR},
\end{equation}
Eq.~\eqref{eq:RW_IR_endpoint} may be written as
\begin{equation}
\mathcal K_j^{\rm IR,RW}
\sim
\widetilde C^{\rm IR,RW}_{j,t}
(\kappa_{\rm RW}b)^{\delta^{\rm RW}_{j,t}}
\exp\!\left[-\frac12\kappa_{\rm RW}^2b^2\right].
\label{eq:RW_IR_physical_scale}
\end{equation}
The power \(\delta^{\rm RW}_{j,t}\), the coefficient
\(C^{\rm IR,RW}_{j,t}\), and the Gaussian width are fixed only after the RW
scheme \((a,S_j,\delta_\kappa,\delta_R,\widehat z_{\rm IR})\) is chosen.
They are therefore model data, not universal strong-coupling numbers.

\section{Worldsheet soft factor and rapidity evolution}
\label{app:soft_factor}

This appendix gives the self-contained strong-coupling soft-factor
construction used in the main text. The inputs are the standard
string representation of Wilson loops at large \(N_c\) and large
\(\lambda\), the near-boundary cusp minimal surface, and the usual
perimeter subtraction for Wilson-line self energies.

\subsection{Definition and geometric decomposition}

Let \(C_b(n_A,n_B)\) be the regulated staple contour entering the gluon
GTMD operator, with transverse separation \(b_T\) and two non-lightlike
eikonal directions \(n_A,n_B\).  In Minkowski signature we write
\begin{equation}
 n_A^2=n_B^2=1,
 \qquad
 n_A\!\cdot n_B=\cosh\chi,
\end{equation}
where \(\chi\) is the relative rapidity.  The lightlike limit is obtained
only after the soft factor is renormalized.  In Euclidean signature the
same cusp is described by an opening angle \(\theta\), followed by the
analytic continuation \(\theta\to i\chi\).

The vacuum soft factor is represented holographically by the minimal
worldsheet ending on \(C_b\),
\begin{align}
 S(b_T;\mu,\zeta)
 &=
 \exp\!\left[-S_{\rm soft}^{\rm ren}[C_b]\right],
 \nonumber\\
 S_{\rm soft}^{\rm ren}[C_b]
 &=
 \frac{\sqrt\lambda}{2\pi}\,A_{\rm ren}[C_b].
 \label{eq:appG_soft_def}
\end{align}
Here \(A_{\rm ren}\) is the classical area after subtracting the
perimeter divergences of the isolated Wilson-line segments.  The
normalization in Eq.~\eqref{eq:appG_soft_def} is the unsubtracted
worldsheet convention used in the main text.  
A Collins-subtracted TMD
scheme redistributes the same vacuum factor between the matrix element
and the subtraction; it does not change the geometric separation between
\(S\) and the target Witten diagram.

At leading saddle the renormalized area is local in geometric regions of
the contour.  For one staple,
\begin{widetext}
\begin{equation}
A_{\rm ren}^{\rm staple}(b_T;\chi)
=
A_{\rm cusp}^{(1)}(\chi)+A_{\rm cusp}^{(2)}(\chi)
+A_{\rm strip}^{\rm ren}(b_T;z_{\rm IR})
+A_{\rm match}(b_T,\chi)
+A_{\rm IR}(b_T;z_{\rm IR}) .
\label{eq:appG_area_decomposition}
\end{equation}
\end{widetext}
The two cusp terms are universal and controlled by the asymptotically
AdS region.  The strip and infrared terms are finite after perimeter
subtraction and depend on the infrared completion of the background.  The
factorization formula in the main text uses the full vacuum soft factor,
which contains the appropriate product of the two staple factors in the
chosen GTMD convention.

\subsection{Cusp contribution and Collins-Soper evolution}

Near a cusp the bulk geometry is asymptotically \(\mathrm{AdS}_5\),
\begin{equation}
 ds^2=\frac{R^2}{z^2}\left(dz^2+d\rho^2+\rho^2d\varphi^2+d x_\perp^2\right),
\end{equation}
and the leading cusp surface is scale invariant,
\begin{equation}
 z(\rho,\varphi)=\rho f(\varphi),
 \qquad
 f(\pm\theta/2)=0 .
\end{equation}
The Nambu-Goto density therefore contains \(d\rho/\rho\), and the
renormalized cusp area takes the general form
\begin{equation}
 A_{\rm cusp}^{(E)}(\theta)
 =
 \mathfrak c_E(\theta)\ln\frac{L}{\epsilon}+\text{finite},
\label{eq:appG_euclidean_cusp_general}
\end{equation}
where \(\mathfrak c_E(\theta)\) is determined by the finite-angle
minimal-surface problem.  After analytic continuation to the regulated
Minkowski cusp, we write
\begin{equation}
 A_{\rm cusp}^{(M)}(\chi)
 =
 \mathfrak c_M(\chi)\ln\frac{L}{\epsilon}+\text{finite}.
\label{eq:appG_minkowski_cusp_general}
\end{equation}
For the present GTMD application only the large-rapidity behavior is
used:
\begin{equation}
 \mathfrak c_M(\chi)=\chi+\mathcal O(\chi^0),
 \qquad
 \partial_\chi\mathfrak c_M(\chi)\to1
 \quad (\chi\to\infty).
\label{eq:appG_large_chi_cusp}
\end{equation}
We intentionally leave \(\mathfrak c_M(\chi)\) unevaluated at finite
\(\chi\).  The familiar perturbative/eikonal finite-angle expression should not be read as the exact strong-coupling
AdS minimal-surface result.

With the Nambu-Goto normalization used in this paper, the coefficient
multiplying the single-cusp logarithm in \(-\ln S^{1/2}\) is
\begin{equation}
 \Gamma_{\rm cusp}^{\rm ws}(\lambda)
 =
 \frac{\sqrt\lambda}{2\pi} .
\label{eq:appG_cusp_coeff}
\end{equation}
This is a convention for the cusp logarithm assigned to the single-staple
factor \(S^{1/2}\).  Equivalently, the finite-angle function
\(\mathfrak c_M\) in this appendix includes the cusp geometry assigned to one
staple in the scheme used in the main text.  A per-geometric-cusp convention,
a full Wilson loop convention, a two-staple soft factor, or a square-root
subtracted collinear matrix element convention changes the quoted coefficient
by the corresponding factor of two.  All such factors are bookkeeping
conventions and do not affect the separation of the vacuum worldsheet area
from the target Witten diagram.

The Collins-Soper kernel is defined by
\begin{equation}
K(b_T,\mu)
=
-\frac{\partial}{\partial\ln\sqrt\zeta}
\ln S^{1/2}(b_T;\mu,\zeta).
\label{eq:appG_CS_def}
\end{equation}
In the geometric regulator, \(\chi\) is the rapidity variable and
\(\partial/\partial\ln\sqrt\zeta=\eta_\zeta\,
\partial/\partial\chi\), with \(\eta_\zeta\) fixed by the chosen
rapidity scheme.  Hence
\begin{equation}
\frac{\partial}{\partial\chi}
\ln S^{1/2}(b_T;\chi)
=
-\Gamma_{\rm cusp}^{\rm ws}(\lambda)
\partial_\chi\mathfrak c_M(\chi)
\ln\frac{L}{\epsilon}
+\cdots .
\label{eq:appG_dchi_soft_general}
\end{equation}
For \(\chi\gg1\), Eq.~\eqref{eq:appG_large_chi_cusp} gives the standard
linear cusp growth.  Identifying \(\mu\sim1/\epsilon\), splitting
\[\ln(L/\epsilon)=\ln(\mu b_T)+\ln(L/b_T)+\text{const.}\,,\] and
absorbing the \(\mu\)-independent piece into the finite nonperturbative
term gives the structural Collins-Soper form
\begin{widetext}
\begin{align}
 K_{\rm SC}^{(\eta_\zeta)}(b_T,\mu)
 =
 -\eta_\zeta\,\Gamma_{\rm cusp}^{\rm ws}(\lambda)\ln(\mu^2 b_T^2)
 +K_{\rm NP}^{\rm SC}(b_T)
 +C_{\eta_\zeta}^{\rm SC},
 \label{eq:appG_K_SC_split}
\end{align}
\end{widetext}
where \(C_{\eta_\zeta}^{\rm SC}\) is an additive scheme constant.  The main
text uses the canonical geometric identification \(\eta_\zeta=1\).  In a
different rapidity scheme, the replacement
\(\Gamma_{\rm cusp}^{\rm ws}\to\eta_\zeta\Gamma_{\rm cusp}^{\rm ws}\), together
with a change of \(C_{\eta_\zeta}^{\rm SC}\), gives the same logarithmic
Collins-Soper structure.

\subsection{Quadratic transverse mismatch}

The leading transverse correction near the cusp follows by expanding the
Nambu-Goto action to quadratic order in a displacement \(Y\) orthogonal
to the cusp plane.  With boundary data
\(Y(-\theta/2)=0\), \(Y(\theta/2)=b_T\), the minimizing solution is
\(\rho\)-independent by scale invariance.  The quadratic action is
\begin{equation}
\Delta S^{(2)}
=
\frac{\sqrt\lambda}{4\pi}
\int d\rho\,d\varphi\,
\sqrt g\,g^{ab}\partial_aY\partial_bY .
\label{eq:appG_quad_action}
\end{equation}
The conserved flux equation gives
\begin{equation}
 y'(\varphi)=\frac{j_0 C(\theta)}{f(\varphi)^2},
 \qquad
 j_0=
 \frac{b_T}
 {C(\theta)\displaystyle\int_{-\theta/2}^{\theta/2}d\varphi\,f(\varphi)^{-2}},
\end{equation}
where \(C(\theta)\) is the first-integral constant of the cusp profile.
The on-shell result is
\begin{equation}
\Delta S_{\rm mismatch}^{(E)}
=
\frac{\sqrt\lambda}{4\pi}
\,b_T^2\,\mathcal F(\theta)
\ln\frac{\rho_c}{\epsilon},
\label{eq:appG_mismatch_E}
\end{equation}
with
\begin{equation}
 \mathcal F(\theta)
 =
 \frac{1}
 {C(\theta)\displaystyle\int_{-\theta/2}^{\theta/2}d\varphi\,f(\varphi)^{-2}}
\end{equation}
in units where the AdS radius is one.  Its Minkowski continuation is
\(\mathcal F^{(M)}(\chi)=\mathcal F(\theta\to i\chi)\).  This term
contributes to the finite transverse-distance dependence of the soft
factor.  It is subleading relative to the universal cusp logarithm in the
leading rapidity evolution.

\subsection{Strip contribution and infrared sensitivity}

The smooth long segments of the staple define a separate minimal-surface
problem: two long parallel boundary lines of length \(L\) separated by
\(b_T\).  In pure AdS or in the short-distance region of a confining
background, the connected saddle gives
\begin{equation}
\frac{A_{\rm strip}^{\rm ren}}{L}
=
-\frac{4\pi^2}{\Gamma(1/4)^4}\,\frac{1}{b_T},
\qquad
b_T\ll z_{\rm IR}.
\label{eq:appG_strip_small}
\end{equation}
In a hard-wall background with wall position \(z_0\), the connected saddle
reaches the wall at
\begin{equation}
 b_c=2z_0\frac{\sqrt\pi\,\Gamma(3/4)}{\Gamma(1/4)},
\end{equation}
and for \(b_T\gtrsim b_c\) the renormalized area is wall pinned,
\begin{equation}
\frac{A_{\rm strip}^{\rm ren}}{L}
=
\frac{b_T}{z_0^2}.
\label{eq:appG_strip_large}
\end{equation}
Soft-wall backgrounds replace this sharp crossover by a smooth one: the
surface stabilizes at \(z\sim z_{\rm IR}\sim1/\kappa\), giving a linear
large-\(b_T\) area with an effective string tension set by the
infrared scale.  The strip term is rapidity independent at leading saddle
and therefore contributes to \(K_{\rm NP}^{\rm SC}(b_T)\), not to the
universal cusp coefficient.

\subsection{Final form used in the GTMD factorization}

Combining the pieces, the soft factor used in the fixed-spin GTMD moment
can be written schematically as
\begin{widetext}
\begin{equation}
\ln S(b_T;\mu,\zeta)
=
-2\left\{
\Gamma_{\rm cusp}^{\rm ws}(\lambda)
\mathfrak c_M(\chi)\ln\frac{L}{\epsilon}
+
\frac{\sqrt\lambda}{2\pi}
\left[
\frac{b_T^2}{2}\mathcal F_{\rm ws}^{(M)}(\chi)
\ln\frac{\rho_c}{\epsilon}
+A_{\rm strip}^{\rm ren}(b_T;z_{\rm IR})
+A_{\rm IR}
+\cdots
\right]
\right\},
\label{eq:appG_final_soft}
\end{equation}
\end{widetext}
where the overall factor of two corresponds to the two staple factors in
the vacuum soft factor; it may be redistributed by changing the TMD
subtraction convention.  The finite-angle functions
\(\mathfrak c_M\) and \(\mathcal F_{\rm ws}^{(M)}\) are regulator and
scheme dependent.  Their large-rapidity cusp limit gives the universal
Collins-Soper coefficient, while all target structure remains in the
stripped fixed-spin worldsheet Witten diagram \(\widetilde F_{j}^{g,{\rm ws}}\).

\bibliography{GTMD-ref}

@article{Collins:1981uk,
    author = "Collins, John C. and Soper, Davison E.",
    title = "{Back-To-Back Jets in QCD}",
    reportNumber = "OITS-166",
    doi = "10.1016/0550-3213(81)90339-4",
    journal = "Nucl. Phys. B",
    volume = "193",
    pages = "381",
    year = "1981",
    note = "[Erratum: Nucl. Phys. B 213, 545 (1983)]"
}

@article{Collins:1984kg,
    author = "Collins, John C. and Soper, Davison E. and Sterman, George F.",
    title = "{Transverse Momentum Distribution in Drell-Yan Pair and W and Z Boson Production}",
    reportNumber = "ITP-SB-84-22",
    doi = "10.1016/0550-3213(85)90479-1",
    journal = "Nucl. Phys. B",
    volume = "250",
    pages = "199--224",
    year = "1985"
}

@article{Ji:2004wu,
    author = "Ji, Xiang-dong and Ma, Jian-Ping and Yuan, Feng",
    title = "{QCD factorization for semi-inclusive deep-inelastic scattering at low transverse momentum}",
    eprint = "hep-ph/0404183",
    archivePrefix = "arXiv",
    doi = "10.1103/PhysRevD.71.034005",
    journal = "Phys. Rev. D",
    volume = "71",
    pages = "034005",
    year = "2005"
}

@book{Collins:2011zzd,
    author = "Collins, John",
    title = "{Foundations of Perturbative QCD}",
    publisher = "Cambridge University Press",
    address = "Cambridge, UK",
    series = "Cambridge Monographs on Particle Physics, Nuclear Physics and Cosmology",
    doi = "10.1017/CBO9780511975592",
    year = "2011"
}

@article{Bacchetta:2006tn,
    author = "Bacchetta, Alessandro and Diehl, Markus and Goeke, Klaus and Metz, Andreas and Mulders, Piet J. and Schlegel, Marc",
    title = "{Semi-inclusive deep inelastic scattering at small transverse momentum}",
    eprint = "hep-ph/0611265",
    archivePrefix = "arXiv",
    doi = "10.1088/1126-6708/2007/02/093",
    journal = "JHEP",
    volume = "02",
    pages = "093",
    year = "2007"
}

@article{Diehl:2003ny,
    author = "Diehl, Markus",
    title = "{Generalized parton distributions}",
    eprint = "hep-ph/0307382",
    archivePrefix = "arXiv",
    reportNumber = "DESY-THESIS-2003-018",
    doi = "10.1016/j.physrep.2003.08.002",
    journal = "Phys. Rept.",
    volume = "388",
    pages = "41--277",
    year = "2003"
}

@article{Belitsky:2005qn,
    author = "Belitsky, A. V. and Radyushkin, A. V.",
    title = "{Unraveling hadron structure with generalized parton distributions}",
    eprint = "hep-ph/0504030",
    archivePrefix = "arXiv",
    doi = "10.1016/j.physrep.2005.06.002",
    journal = "Phys. Rept.",
    volume = "418",
    pages = "1--387",
    year = "2005"
}

@article{Meissner:2007rx,
    author = "Meissner, S. and Metz, A. and Goeke, K.",
    title = "{Relations between generalized and transverse momentum dependent parton distributions}",
    eprint = "hep-ph/0703176",
    archivePrefix = "arXiv",
    doi = "10.1103/PhysRevD.76.034002",
    journal = "Phys. Rev. D",
    volume = "76",
    pages = "034002",
    year = "2007"
}

@article{Meissner:2009ww,
  author        = {Meissner, Stefan and Metz, Andreas and Schlegel, Marc},
  title         = {Generalized parton correlation functions for a spin-1/2 hadron},
  journal       = {JHEP},
  volume        = {08},
  year          = {2009},
  pages         = {056},
  doi           = {10.1088/1126-6708/2009/08/056},
  eprint        = {0906.5323},
  archivePrefix = {arXiv},
  primaryClass  = {hep-ph}
}

@article{Belitsky:2003nz,
    author = "Belitsky, A. V. and Ji, X. and Yuan, F.",
    title = "{Quark imaging in the proton via quantum phase-space distributions}",
    eprint = "hep-ph/0307383",
    archivePrefix = "arXiv",
    reportNumber = "UMD-PP-03-093, DOE-ER-40762-292, JLAB-THY-03-93",
    doi = "10.1103/PhysRevD.69.074014",
    journal = "Phys. Rev. D",
    volume = "69",
    pages = "074014",
    year = "2004"
}

@article{Lorce:2011dv,
    author = "Lorc{\'e}, C{\'e}dric and Pasquini, Barbara and Vanderhaeghen, Marc",
    title = "{Unified framework for generalized and transverse-momentum dependent parton distributions within a 3Q light-cone picture of the nucleon}",
    eprint = "1102.4704",
    archivePrefix = "arXiv",
    primaryClass = "hep-ph",
    doi = "10.1007/JHEP05(2011)041",
    journal = "JHEP",
    volume = "05",
    pages = "041",
    year = "2011"
}

@article{Lorce:2011kd,
    author = "Lorc{\'e}, C{\'e}dric and Pasquini, Barbara",
    title = "{Quark Wigner Distributions and Orbital Angular Momentum}",
    eprint = "1106.0139",
    archivePrefix = "arXiv",
    primaryClass = "hep-ph",
    doi = "10.1103/PhysRevD.84.014015",
    journal = "Phys. Rev. D",
    volume = "84",
    pages = "014015",
    year = "2011"
}

@article{Lorce:2013pza,
  author        = {Lorc{\'e}, C{\'e}dric and Pasquini, Barbara},
  title         = {Structure analysis of the generalized correlator of quark and gluon for a spin-1/2 target},
  journal       = {JHEP},
  volume        = {09},
  year          = {2013},
  pages         = {138},
  doi           = {10.1007/JHEP09(2013)138},
  eprint        = {1307.4497},
  archivePrefix = {arXiv},
  primaryClass  = {hep-ph}
}

@article{Kanazawa:2014nha,
    author = "Kanazawa, Koichi and Lorc{\'e}, C{\'e}dric and Metz, Andreas and Pasquini, Barbara and Schlegel, Marc",
    title = "{Twist-2 generalized transverse-momentum dependent parton distributions and the spin/orbital structure of the nucleon}",
    eprint = "1403.5226",
    archivePrefix = "arXiv",
    primaryClass = "hep-ph",
    doi = "10.1103/PhysRevD.90.014028",
    journal = "Phys. Rev. D",
    volume = "90",
    pages = "014028",
    year = "2014"
}

@article{Echevarria:2012js,
    author = "Echevarria, Miguel G. and Idilbi, Ahmad and Scimemi, Ignazio",
    title = "{Soft and Collinear Factorization and Transverse Momentum Dependent Parton Distribution Functions}",
    eprint = "1211.1947",
    archivePrefix = "arXiv",
    primaryClass = "hep-ph",
    doi = "10.1016/j.physletb.2013.07.014",
    journal = "Phys. Lett. B",
    volume = "726",
    pages = "795--801",
    year = "2013"
}

@article{Echevarria:2016mrc,
    author = "Echevarria, Miguel G. and Idilbi, Ahmad and Kanazawa, Koichi and Lorc{\'e}, C{\'e}dric and Metz, Andreas and Pasquini, Barbara and Schlegel, Marc",
    title = "{Proper definition and evolution of generalized transverse momentum dependent distributions}",
    eprint = "1602.06953",
    archivePrefix = "arXiv",
    primaryClass = "hep-ph",
    doi = "10.1016/j.physletb.2016.05.086",
    journal = "Phys. Lett. B",
    volume = "759",
    pages = "336--341",
    year = "2016"
}

@article{Bertone:2022nxa,
    author = "Bertone, Valerio",
    title = "{Matching generalised transverse-momentum-dependent distributions onto generalised parton distributions at one loop}",
    eprint = "2207.09526",
    archivePrefix = "arXiv",
    primaryClass = "hep-ph",
    doi = "10.1140/epjc/s10052-022-10863-3",
    journal = "Eur. Phys. J. C",
    volume = "82",
    number = "10",
    pages = "941",
    year = "2022"
}

@article{Bertone:2025gtmd,
    author = "Bertone, Valerio and Echevarria, Miguel G. and del Rio, {\'O}scar and Rodini, Simone",
    title = "{One-loop matching for leading-twist generalised transverse-momentum-dependent distributions}",
    eprint = "2502.07576",
    archivePrefix = "arXiv",
    primaryClass = "hep-ph",
    doi = "10.1007/JHEP05(2025)183",
    journal = "JHEP",
    volume = "05",
    pages = "183",
    year = "2025"
}

@article{Maldacena:1997re,
    author = "Maldacena, Juan M.",
    title = "{The Large N limit of superconformal field theories and supergravity}",
    eprint = "hep-th/9711200",
    archivePrefix = "arXiv",
    reportNumber = "HUTP-97-A097, HUTP-98-A097",
    doi = "10.1023/A:1026654312961",
    journal = "Int. J. Theor. Phys.",
    volume = "38",
    pages = "1113--1133",
    year = "1999",
    note = "[Adv. Theor. Math. Phys. 2, 231 (1998)]"
}

@article{Gubser:1998bc,
    author = "Gubser, S. S. and Klebanov, Igor R. and Polyakov, Alexander M.",
    title = "{Gauge theory correlators from noncritical string theory}",
    eprint = "hep-th/9802109",
    archivePrefix = "arXiv",
    reportNumber = "PUPT-1767",
    doi = "10.1016/S0370-2693(98)00377-3",
    journal = "Phys. Lett. B",
    volume = "428",
    pages = "105--114",
    year = "1998"
}

@article{Witten:1998qj,
    author = "Witten, Edward",
    title = "{Anti-de Sitter space and holography}",
    eprint = "hep-th/9802150",
    archivePrefix = "arXiv",
    reportNumber = "IASSNS-HEP-98-15",
    doi = "10.4310/ATMP.1998.v2.n2.a2",
    journal = "Adv. Theor. Math. Phys.",
    volume = "2",
    pages = "253--291",
    year = "1998"
}

@article{Freedman:1998tz,
  author         = "Freedman, Daniel Z. and Mathur, Samir D. and Matusis, Alec and Rastelli, Leonardo",
  title          = "{Correlation functions in the CFT(d) / AdS(d+1) correspondence}",
  journal        = "Nucl. Phys. B",
  volume         = "546",
  year           = "1999",
  pages          = "96--118",
  doi            = "10.1016/S0550-3213(99)00053-X",
  eprint         = "hep-th/9804058",
  archivePrefix  = "arXiv",
  primaryClass   = "hep-th"
}

@article{DHoker:1999jke,
  author         = "D'Hoker, Eric and Freedman, Daniel Z.",
  title          = "{Supersymmetric gauge theories and the AdS/CFT correspondence}",
  journal        = "TASI Lect. 1999",
  year           = "1999",
  eprint         = "hep-th/0201253",
  archivePrefix  = "arXiv",
  primaryClass   = "hep-th"
}

@article{Maldacena:1998im,
    author = "Maldacena, Juan M.",
    title = "{Wilson loops in large N field theories}",
    eprint = "hep-th/9803002",
    archivePrefix = "arXiv",
    reportNumber = "HUTP-98-A012",
    doi = "10.1103/PhysRevLett.80.4859",
    journal = "Phys. Rev. Lett.",
    volume = "80",
    pages = "4859--4862",
    year = "1998"
}

@article{Rey:1998ik,
  author         = "Rey, Soo-Jong and Yee, Jung-Tay",
  title          = "{Macroscopic strings as heavy quarks in large N gauge theory and anti-de Sitter supergravity}",
  journal        = "Eur. Phys. J. C",
  volume         = "22",
  year           = "2001",
  pages          = "379--394",
  doi            = "10.1007/s100520100799",
  eprint         = "hep-th/9803001",
  archivePrefix  = "arXiv",
  primaryClass   = "hep-th"
}

@article{Drukker:1999zq,
  author         = "Drukker, Nadav and Gross, David J. and Ooguri, Hirosi",
  title          = "{Wilson loops and minimal surfaces}",
  journal        = "Phys. Rev. D",
  volume         = "60",
  year           = "1999",
  pages          = "125006",
  doi            = "10.1103/PhysRevD.60.125006",
  eprint         = "hep-th/9904191",
  archivePrefix  = "arXiv",
  primaryClass   = "hep-th"
}

@article{Brandhuber:1998bs,
    author = "Brandhuber, Andreas and Itzhaki, Nissan and Sonnenschein, Jacob and Yankielowicz, Shimon",
    title = "{Wilson loops, confinement, and phase transitions in large N gauge theories from supergravity}",
    eprint = "hep-th/9803263",
    archivePrefix = "arXiv",
    reportNumber = "TAUP-2524-98, EFI-98-21",
    doi = "10.1088/1126-6708/1998/06/001",
    journal = "JHEP",
    volume = "06",
    pages = "001",
    year = "1998"
}

@article{Kinar:1998vq,
  author         = "Kinar, Y. and Schreiber, E. and Sonnenschein, J.",
  title          = "{Q anti-Q potential from strings in curved spacetime: Classical results}",
  journal        = "Nucl. Phys. B",
  volume         = "566",
  pages          = "103-125",
  year           = "2000",
  doi            = "10.1016/S0550-3213(99)00652-5",
  eprint         = "hep-th/9811192",
  archivePrefix  = "arXiv"
}

@article{Korchemsky:1987wg,
  author         = "Korchemsky, G.~P. and Radyushkin, A.~V.",
  title          = "{Renormalization of the Wilson Loops Beyond the Leading Order}",
  journal        = "Nucl. Phys. B",
  volume         = "283",
  year           = "1987",
  pages          = "342--364",
  doi            = "10.1016/0550-3213(87)90277-X"
}

@article{Kruczenski:2002fb,
  author        = {Kruczenski, Martin},
  title         = {A note on twist two operators in N=4 SYM and Wilson loops in Minkowski signature},
  journal       = {JHEP},
  volume        = {12},
  year          = {2002},
  pages         = {024},
  eprint        = {hep-th/0210115},
  archivePrefix = {arXiv},
  doi           = {10.1088/1126-6708/2002/12/024}
}

@article{Kruczenski:2007cy,
  author         = "Kruczenski, Martin and Roiban, Radu and Tirziu, Alin and Tseytlin, Arkady A.",
  title          = "{Strong-coupling expansion of cusp anomaly and gluon amplitudes from quantum open strings in AdS$_5 \times S^5$}",
  journal        = "Nucl. Phys. B",
  volume         = "791",
  year           = "2008",
  pages          = "93--124",
  doi            = "10.1016/j.nuclphysb.2007.09.014",
  eprint         = "0707.4254",
  archivePrefix  = "arXiv",
  primaryClass   = "hep-th"
}

@article{Makeenko:2006ds,
  author        = {Makeenko, Yuri},
  title         = {Light-cone Wilson loops and the string/gauge correspondence},
  journal       = {JHEP},
  volume        = {01},
  year          = {2007},
  pages         = {007},
  eprint        = {hep-th/0611049},
  archivePrefix = {arXiv},
  doi           = {10.1088/1126-6708/2007/01/007}
}

@article{Rajan:2016tlg,
    author = "Rajan, Abha and Engelhardt, Michael and Liuti, Simonetta",
    title = "{Parton transverse momentum and orbital angular momentum}",
    eprint = "1601.06117",
    archivePrefix = "arXiv",
    primaryClass = "hep-ph",
    doi = "10.1103/PhysRevD.94.034041",
    journal = "Phys. Rev. D",
    volume = "94",
    number = "3",
    pages = "034041",
    year = "2016"
}

@article{Maynard:2026gtmd,
    author        = "Maynard, Brean and Schweitzer, Peter",
    title         = "{GTMDs, orbital angular momentum, and pretzelosity}",
    eprint        = "2605.06412",
    archivePrefix = "arXiv",
    primaryClass  = "hep-ph",
    journal       = "arXiv",
    month         = "5",
    year          = "2026"
}

@article{Tan:2024zjs,
    author = "Tan, Chung-I",
    title = "{Gluon GTMDs at nonzero skewness and impact parameter dependent parton distributions}",
    eprint = "2402.17162",
    archivePrefix = "arXiv",
    primaryClass = "hep-ph",
    doi = "10.1140/epjc/s10052-024-12976-3",
    journal = "Eur. Phys. J. C",
    volume = "84",
    number = "6",
    pages = "637",
    year = "2024"
}

@article{Chakrabarti:2025gtmd,
    author = "Chakrabarti, Dipankar and Gurjar, Bheemsehan and Mukherjee, Asmita and Saha, Kauship",
    title = "{Gluon generalized transverse momentum dependent parton distributions and Wigner functions of the proton}",
    eprint = "2509.14208",
    archivePrefix = "arXiv",
    primaryClass = "hep-ph",
    doi = "10.1103/f5tn-lw8v",
    journal = "Phys. Rev. D",
    volume = "112",
    number = "11",
    pages = "114031",
    year = "2025"
}

@article{Hagiwara:2016kam,
    author = "Hagiwara, Yoshikazu and Hatta, Yoshitaka and Ueda, Takahiro",
    title = "{Wigner, Husimi, and generalized transverse momentum dependent distributions in the color glass condensate}",
    eprint = "1609.05773",
    archivePrefix = "arXiv",
    primaryClass = "hep-ph",
    doi = "10.1103/PhysRevD.94.094036",
    journal = "Phys. Rev. D",
    volume = "94",
    number = "9",
    pages = "094036",
    year = "2016"
}

@article{Zhou:2016rnt,
    author = "Zhou, Jian",
    title = "{Elliptic gluon generalized transverse-momentum-dependent distribution inside a large nucleus}",
    eprint = "1611.02397",
    archivePrefix = "arXiv",
    primaryClass = "hep-ph",
    doi = "10.1103/PhysRevD.94.114017",
    journal = "Phys. Rev. D",
    volume = "94",
    number = "11",
    pages = "114017",
    year = "2016"
}

@article{Benic:2026gtmd,
    author        = "Beni{\'c}, Sanjin and Hagiwara, Yoshikazu and {\v S}ari{\'c}, Boris and Vivoda, Eric Andreas",
    title         = "{Generalized transverse momentum distributions at small-$x$}",
    eprint        = "2603.06092",
    archivePrefix = "arXiv",
    primaryClass  = "hep-ph",
    journal       = "arXiv",
    month         = "3",
    year          = "2026"
}

@article{Brower:2006ea,
  author         = "Brower, Richard C. and Polchinski, Joseph and Strassler, Matthew J. and Tan, Chung-I",
  title          = "{The Pomeron and gauge/string duality}",
  journal        = "JHEP",
  volume         = "12",
  year           = "2007",
  pages          = "005",
  doi            = "10.1088/1126-6708/2007/12/005",
  eprint         = "hep-th/0603115",
  archivePrefix  = "arXiv",
  primaryClass   = "hep-th"
}

@article{Brower:2010wf,
  author         = "Brower, Richard C. and Djuric, Marko and Sarcevic, Ina and Tan, Chung-I",
  title          = "{String-Gauge Dual Description of Deep Inelastic Scattering at Small-x}",
  journal        = "JHEP",
  volume         = "11",
  year           = "2010",
  pages          = "051",
  doi            = "10.1007/JHEP11(2010)051",
  eprint         = "1007.2259",
  archivePrefix  = "arXiv",
  primaryClass   = "hep-ph"
}

@article{Costa:2012cb,
  author         = "Costa, Miguel S. and Goncalves, Vasco and Penedones, Joao",
  title          = "{Conformal Regge theory}",
  journal        = "JHEP",
  volume         = "12",
  year           = "2012",
  pages          = "091",
  doi            = "10.1007/JHEP12(2012)091",
  eprint         = "1209.4355",
  archivePrefix  = "arXiv",
  primaryClass   = "hep-th"
}

@article{Brower:2013,
    author = "Brower, Richard C. and Costa, Miguel S. and Djuric, Marko and Tan, Chung-I",
    title = "{Conformal Pomeron and Odderon in Strong Coupling}",
    eprint = "1302.1419",
    archivePrefix = "arXiv",
    primaryClass = "hep-th",
    doi = "10.1007/JHEP12(2013)062",
    journal = "JHEP",
    volume = "12",
    pages = "062",
    year = "2013"
}

@article{Mueller:2005ed,
    author = "Mueller, Dieter and Schafer, A.",
    title = "{Complex conformal spin partial wave expansion of generalized parton distributions and distribution amplitudes}",
    eprint = "hep-ph/0509204",
    archivePrefix = "arXiv",
    doi = "10.1016/j.nuclphysb.2006.01.019",
    journal = "Nucl. Phys. B",
    volume = "739",
    pages = "1--59",
    year = "2006"
}

@article{Mamo:2024jwp,
    author = "Mamo, Kiminad A. and Zahed, Ismail",
    title = "{Parametrization of Generalized Parton Distributions from t-Channel String Exchange in AdS Spaces}",
    eprint = "2411.04162",
    archivePrefix = "arXiv",
    primaryClass = "hep-ph",
    doi = "10.1103/PhysRevLett.133.241901",
    journal = "Phys. Rev. Lett.",
    volume = "133",
    number = "24",
    pages = "241901",
    year = "2024"
}

@article{Berenstein:1998ij,
  author         = "Berenstein, David E. and Corrado, Richard and Fischler, Willy and Maldacena, Juan M.",
  title          = "{Operator product expansion for Wilson loops and surfaces in the large N limit}",
  journal        = "Phys. Rev. D",
  volume         = "59",
  year           = "1999",
  pages          = "105023",
  doi            = "10.1103/PhysRevD.59.105023",
  eprint         = "hep-th/9809188",
  archivePrefix  = "arXiv",
  primaryClass   = "hep-th"
}

@article{Miwa:2006vv,
  author         = "Miwa, Akitsugu and Yoneya, Tamiaki",
  title          = "{Holography of Wilson-loop expectation values with local operator insertions}",
  journal        = "JHEP",
  volume         = "12",
  year           = "2006",
  pages          = "060",
  doi            = "10.1088/1126-6708/2006/12/060",
  eprint         = "hep-th/0609007",
  archivePrefix  = "arXiv",
  primaryClass   = "hep-th"
}

@article{Alday:2011ga,
  author         = "Alday, Luis F. and Tseytlin, Arkady A.",
  title          = "{On strong-coupling correlation functions of circular Wilson loops and local operators}",
  journal        = "J. Phys. A",
  volume         = "44",
  year           = "2011",
  pages          = "395401",
  doi            = "10.1088/1751-8113/44/39/395401",
  eprint         = "1105.1537",
  archivePrefix  = "arXiv",
  primaryClass   = "hep-th"
}

@article{Alday:2011pf,
  author         = "Alday, Luis F. and Buchbinder, Evgeny I. and Tseytlin, Arkady A.",
  title          = "{Correlation function of null polygonal Wilson loops with local operators}",
  journal        = "JHEP",
  volume         = "09",
  year           = "2011",
  pages          = "034",
  doi            = "10.1007/JHEP09(2011)034",
  eprint         = "1107.5702",
  archivePrefix  = "arXiv",
  primaryClass   = "hep-th"
}

@article{Buchbinder:2012vr,
  author         = "Buchbinder, Evgeny I. and Tseytlin, Arkady A.",
  title          = "{Correlation function of circular Wilson loop with two local operators and conformal invariance}",
  journal        = "Phys. Rev. D",
  volume         = "87",
  year           = "2013",
  pages          = "026006",
  doi            = "10.1103/PhysRevD.87.026006",
  eprint         = "1208.5138",
  archivePrefix  = "arXiv",
  primaryClass   = "hep-th"
}

@article{Alday:2007hr,
  author         = "Alday, Luis F. and Maldacena, Juan M.",
  title          = "{Gluon scattering amplitudes at strong coupling}",
  journal        = "JHEP",
  volume         = "06",
  year           = "2007",
  pages          = "064",
  doi            = "10.1088/1126-6708/2007/06/064",
  eprint         = "0705.0303",
  archivePrefix  = "arXiv",
  primaryClass   = "hep-th"
}

@article{Alday:2010vh,
  author         = "Alday, Luis F. and Maldacena, Juan M.",
  title          = "{Null polygonal Wilson loops and minimal surfaces in Anti-de-Sitter space}",
  journal        = "JHEP",
  volume         = "11",
  year           = "2010",
  pages          = "082",
  doi            = "10.1007/JHEP11(2010)082",
  eprint         = "0904.0663",
  archivePrefix  = "arXiv",
  primaryClass   = "hep-th"
}

@article{Costa:2011mg,
    author = "Costa, Miguel S. and Penedones, Joao and Poland, David and Rychkov, Slava",
    title = "{Spinning Conformal Correlators}",
    eprint = "1107.3554",
    archivePrefix = "arXiv",
    primaryClass = "hep-th",
    doi = "10.1007/JHEP11(2011)071",
    journal = "JHEP",
    volume = "11",
    pages = "071",
    year = "2011"
}

@article{Costa:2014kfa,
    author = "Costa, Miguel S. and Goncalves, Vasco and Penedones, Joao",
    title = "{Spinning AdS Propagators}",
    eprint = "1404.5625",
    archivePrefix = "arXiv",
    primaryClass = "hep-th",
    doi = "10.1007/JHEP09(2014)064",
    journal = "JHEP",
    volume = "09",
    pages = "064",
    year = "2014"
}

@article{Hijano:2015zsa,
  author         = "Hijano, Eliot and Kraus, Per and Perlmutter, Eric and Snively, River",
  title          = "{Witten Diagrams Revisited: The AdS Geometry of Conformal Blocks}",
  journal        = "JHEP",
  volume         = "01",
  year           = "2016",
  pages          = "146",
  doi            = "10.1007/JHEP01(2016)146",
  eprint         = "1508.00501",
  archivePrefix  = "arXiv",
  primaryClass   = "hep-th"
}

@article{Dyer:2017zef,
  author         = "Dyer, Ethan and Freedman, Daniel Z. and Sully, James and Zhou, Yi",
  title          = "{Spinning Geodesic Witten Diagrams}",
  journal        = "JHEP",
  volume         = "11",
  year           = "2017",
  pages          = "060",
  doi            = "10.1007/JHEP11(2017)060",
  eprint         = "1708.06797",
  archivePrefix  = "arXiv",
  primaryClass   = "hep-th"
}

@article{Karch:2006pv,
  author         = "Karch, Andreas and Katz, Emanuel and Son, Dam T. and Stephanov, Mikhail A.",
  title          = "{Linear confinement and AdS/QCD}",
  journal        = "Phys. Rev. D",
  volume         = "74",
  pages          = "015005",
  year           = "2006",
  doi            = "10.1103/PhysRevD.74.015005",
  eprint         = "hep-ph/0602229",
  archivePrefix  = "arXiv"
}

@article{Polchinski:2001tt,
    author = "Polchinski, Joseph and Strassler, Matthew J.",
    title = "{Hard scattering and gauge/string duality}",
    eprint = "hep-th/0109174",
    archivePrefix = "arXiv",
    reportNumber = "NSF-ITP-01-59",
    doi = "10.1103/PhysRevLett.88.031601",
    journal = "Phys. Rev. Lett.",
    volume = "88",
    pages = "031601",
    year = "2002"
}

@article{Polchinski:2002jw,
    author = "Polchinski, Joseph and Strassler, Matthew J.",
    title = "{Deep inelastic scattering and gauge/string duality}",
    eprint = "hep-th/0209211",
    archivePrefix = "arXiv",
    reportNumber = "NSF-ITP-02-62",
    doi = "10.1088/1126-6708/2003/05/012",
    journal = "JHEP",
    volume = "05",
    pages = "012",
    year = "2003"
}

@article{Brodsky:2014yha,
    author = "Brodsky, Stanley J. and de Teramond, Guy F. and Dosch, Hans Gunter and Erlich, Joshua",
    title = "{Light-Front Holographic QCD and Emerging Confinement}",
    eprint = "1407.8131",
    archivePrefix = "arXiv",
    primaryClass = "hep-ph",
    doi = "10.1016/j.physrep.2015.05.001",
    journal = "Phys. Rept.",
    volume = "584",
    pages = "1--105",
    year = "2015"
}

@article{Mamo:2019mka,
    author = "Mamo, Kiminad A. and Zahed, Ismail",
    title = "{Diffractive photoproduction of $J/\psi$ and $\Upsilon$ using holographic QCD: gravitational form factors and GPD of gluons in the proton}",
    eprint = "1910.04707",
    archivePrefix = "arXiv",
    primaryClass = "hep-ph",
    doi = "10.1103/PhysRevD.101.086003",
    journal = "Phys. Rev. D",
    volume = "101",
    number = "8",
    pages = "086003",
    year = "2020"
}

@article{Osborn:1993cr,
  author         = "Osborn, Hugh and Petkou, Anastasios C.",
  title          = "{Implications of conformal invariance in field theories for general dimensions}",
  journal        = "Annals Phys.",
  volume         = "231",
  year           = "1994",
  pages          = "311--362",
  doi            = "10.1006/aphy.1994.1045",
  eprint         = "hep-th/9307010",
  archivePrefix  = "arXiv",
  primaryClass   = "hep-th"
}

@article{Balitsky:1995ub,
  author       = {Balitsky, Ian},
  title        = {Operator expansion for high-energy scattering},
  journal      = {Nucl. Phys. B},
  volume       = {463},
  pages        = {99--160},
  year         = {1996},
  eprint       = {hep-ph/9509348},
  archivePrefix= {arXiv},
  primaryClass = {hep-ph}
}

@article{Kovchegov:1999yj,
  author       = {Kovchegov, Yuri V.},
  title        = {Small-$x$ $F_2$ structure function of a nucleus including multiple pomeron exchanges},
  journal      = {Phys. Rev. D},
  volume       = {60},
  pages        = {034008},
  year         = {1999},
  eprint       = {hep-ph/9901281},
  archivePrefix= {arXiv},
  primaryClass = {hep-ph}
}

@book{Iancu:2003xm,
  author       = {Iancu, Edmond and Venugopalan, Raju},
  title        = {The Color Glass Condensate and High Energy Scattering in QCD},
  year         = {2003},
  eprint       = {hep-ph/0303204},
  archivePrefix= {arXiv},
  primaryClass = {hep-ph},
  note         = {In *Quark-Gluon Plasma 3*, eds. Hwa and Wang}
}

@article{Weigert:2005us,
  author       = {Weigert, Heribert},
  title        = {Evolution at small $x_{bj}$: The Color Glass Condensate},
  journal      = {Prog. Part. Nucl. Phys.},
  volume       = {55},
  pages        = {461--565},
  year         = {2005},
  eprint       = {hep-ph/0501087},
  archivePrefix= {arXiv},
  primaryClass = {hep-ph}
}

@article{Dominguez:2011wm,
  author       = {Dominguez, Fabio and Marquet, Cyril and Xiao, Bo-Wen and Yuan, Feng},
  title        = {Universality of Unintegrated Gluon Distributions at small $x$},
  journal      = {Phys. Rev. D},
  volume       = {83},
  pages        = {105005},
  year         = {2011},
  eprint       = {1101.0715},
  archivePrefix= {arXiv},
  primaryClass = {hep-ph}
}

@article{Hatta:2016dxp,
    author = "Hatta, Yoshitaka and Xiao, Bo-Wen and Yuan, Feng",
    title = "{Probing the Small- x Gluon Tomography in Correlated Hard Diffractive Dijet Production in Deep Inelastic Scattering}",
    eprint = "1601.01585",
    archivePrefix = "arXiv",
    primaryClass = "hep-ph",
    reportNumber = "YITP-16-1",
    doi = "10.1103/PhysRevLett.116.202301",
    journal = "Phys. Rev. Lett.",
    volume = "116",
    number = "20",
    pages = "202301",
    year = "2016"
}

@article{Ji:1996ek,
  author        = {Ji, Xiangdong},
  title         = {Deeply Virtual Compton Scattering},
  journal       = {Phys. Rev. D},
  volume        = {55},
  year          = {1997},
  pages         = {7114--7125},
  doi           = {10.1103/PhysRevD.55.7114},
  eprint        = {hep-ph/9609381},
  archivePrefix = {arXiv},
  primaryClass  = {hep-ph}
}

@article{Radyushkin:1996ru,
  author        = {Radyushkin, A. V.},
  title         = {Scaling limit of deeply virtual Compton scattering},
  journal       = {Phys. Lett. B},
  volume        = {380},
  year          = {1996},
  pages         = {417--425},
  doi           = {10.1016/0370-2693(96)00528-X},
  eprint        = {hep-ph/9604317},
  archivePrefix = {arXiv},
  primaryClass  = {hep-ph}
}

@article{Ji:2003ak,
  author        = {Ji, Xiangdong},
  title         = {Viewing the proton through ``color'' filters},
  journal       = {Phys. Rev. Lett.},
  volume        = {91},
  year          = {2003},
  pages         = {062001},
  doi           = {10.1103/PhysRevLett.91.062001},
  eprint        = {hep-ph/0304037},
  archivePrefix = {arXiv},
  primaryClass  = {hep-ph}
}

@article{Bhattacharya:2026qnd,
  author        = {Bhattacharya, Shohini and DeAngelo, David and Yang, Lei and Zheng, Duxin and Zhou, Jian},
  title         = {{Gluon Generalized TMD signatures at the EIC from exclusive heavy (axial-)vector meson production}},
  journal       = {arXiv},
  year          = {2026},
  eprint        = {2601.17506},
  archivePrefix = {arXiv},
  primaryClass  = {hep-ph}
}

@article{Bhattacharya:2018lgm,
    author = "Bhattacharya, Shohini and Metz, Andreas and Ojha, Vikash Kumar and Tsai, Jeng-Yuan and Zhou, Jian",
    title = "{Exclusive double quarkonium production and generalized TMDs of gluons}",
    eprint = "1802.10550",
    archivePrefix = "arXiv",
    primaryClass = "hep-ph",
    doi = "10.1016/j.physletb.2022.137383",
    journal = "Phys. Lett. B",
    volume = "833",
    pages = "137383",
    year = "2022"
}

@article{Bhattacharya:2024pi0,
    author = "Bhattacharya, Shohini and Zheng, Duxin and Zhou, Jian",
    title = "{Accessing the gluon GTMD $F_{1,4}$ in exclusive $\pi^0$ production in $ep$ collisions}",
    eprint = "2304.05784",
    archivePrefix = "arXiv",
    primaryClass = "hep-ph",
    doi = "10.1103/PhysRevD.109.096029",
    journal = "Phys. Rev. D",
    volume = "109",
    number = "9",
    pages = "096029",
    year = "2024"
}

@article{Mamo:2024vjh,
    author = "Mamo, Kiminad A. and Zahed, Ismail",
    title = "{String-based parametrization of nucleon GPDs at any skewness: A comparison to lattice QCD}",
    eprint = "2404.13245",
    archivePrefix = "arXiv",
    primaryClass = "hep-ph",
    doi = "10.1103/PhysRevD.110.114016",
    journal = "Phys. Rev. D",
    volume = "110",
    number = "11",
    pages = "114016",
    year = "2024"
}

@misc{Gimenez-Grau:2023fcy,
    author = "Gimenez-Grau, Aleix",
    title = "{The Witten Diagram Bootstrap for Holographic Defects}",
    eprint = "2306.11896",
    archivePrefix = "arXiv",
    primaryClass = "hep-th",
    month = "6",
    year = "2023"
}

@misc{carmi2026aspectswittendiagramsholographic,
      title={Aspects of Witten Diagrams for Holographic Defects}, 
      author={Dean Carmi and Sudip Ghosh and Trakshu Sharma},
      year={2026},
      eprint={2606.17719},
      archivePrefix={arXiv},
      primaryClass={hep-th},
      url={https://arxiv.org/abs/2606.17719}, 
}

@misc{Mamo:2026vuq,
    author = "Mamo, Kiminad A.",
    title = "{From Vacuum to Nucleon: Fixed-$j$ Kernel Matching of Holographic Current Correlators to QCD}",
    eprint = "2604.12037",
    archivePrefix = "arXiv",
    primaryClass = "hep-th",
    month = "4",
    year = "2026"
}

@misc{Mamo:2026fjh,
    author = "Mamo, Kiminad A.",
    title = "{Holographic Open/Closed Exchange in Double Deeply Virtual Compton Scattering: Fixed-$j$ Structural Matching to the $\pm$-Basis Wilson Kernels}",
    eprint = "2604.12038",
    archivePrefix = "arXiv",
    primaryClass = "hep-th",
    month = "4",
    year = "2026"
}

\end{document}